\documentclass[aps,twocolumn,superscriptaddress]{revtex4-1}
\usepackage{CJK}
\usepackage{physics}
\usepackage{amsmath, amsthm, amssymb, amsfonts}
\usepackage{nccmath}
\usepackage{graphicx}
\usepackage{float}
\usepackage{longtable}
\usepackage{array}
\usepackage{verbatim}
\usepackage[margin=1in]{geometry}
\usepackage{color,soul}
\usepackage{fancyhdr}
\usepackage{geometry}
\usepackage{bm}
\usepackage{hyperref}
\hypersetup{colorlinks=true, linkcolor=blue,filecolor=blue,urlcolor=blue, citecolor=blue }
\usepackage[document]{ragged2e}
\usepackage{enumitem}
\usepackage{mathtools}
\usepackage{fancyref}
\usepackage{titlesec}
\usepackage{fix-cm}
\usepackage{cleveref}
\usepackage{esvect}
\usepackage{bbold}
\usepackage{multirow}
\usepackage{array}
\usepackage{braket}
\usepackage{appendix}
\usepackage{comment}
\definecolor{rossos}{rgb}{0.8,0.2,0.3}

\usepackage{import}

\begin{document}
\begin{CJK*}{}{}
\title{Impact of shell model interactions on nuclear responses to WIMP elastic scattering}
\author{R. Abdel Khaleq}\email{raghda.abdelkhaleq@anu.edu.au}
\affiliation{Department of Fundamental and Theoretical Physics, Research School of Physics, Australian National University, ACT, 2601, Australia}
\affiliation{Department of Nuclear Physics and Accelerator Applications, Research School of Physics, Australian National University, ACT, 2601, Australia}
\affiliation{ARC Centre of Excellence for Dark Matter Particle Physics, Australia}
\author{G. Busoni}\email{giorgio.busoni@anu.edu.au}
\affiliation{Department of Fundamental and Theoretical Physics, Research School of Physics, Australian National University, ACT, 2601, Australia}
\affiliation{ARC Centre of Excellence for Dark Matter Particle Physics, Australia}
\author{C. Simenel}\email{cedric.simenel@anu.edu.au}
\affiliation{Department of Fundamental and Theoretical Physics, Research School of Physics, Australian National University, ACT, 2601, Australia}
\affiliation{Department of Nuclear Physics and Accelerator Applications, Research School of Physics, Australian National University, ACT, 2601, Australia}
\affiliation{ARC Centre of Excellence for Dark Matter Particle Physics, Australia}
\author{A. E. Stuchbery}\email{andrew.stuchbery@anu.edu.au}
\affiliation{Department of Nuclear Physics and Accelerator Applications, Research School of Physics, Australian National University, ACT, 2601, Australia}
\affiliation{ARC Centre of Excellence for Dark Matter Particle Physics, Australia}

\date{\today}

\begin{abstract}
\edef\oldrightskip{\the\rightskip}
\justify
\begin{description}
\rightskip\oldrightskip\relax
\setlength{\parskip}{0pt}
\item[Background] Nuclear recoil from scattering with weakly interacting massive particles (WIMPs) is a signature searched for in direct detection of dark matter. The underlying WIMP-nucleon interactions could be spin and/or orbital angular momentum (in)dependent. Evaluation of nuclear recoil rates through these interactions requires accounting for nuclear structure, e.g., through shell model calculations. 
\item[Purpose] To evaluate nuclear response functions induced by these interactions for $^{19}$F, $^{23}$Na, $^{28, 29, 30}$Si,  $^{40}$Ar, $^{70,72,73,74,76}$Ge,  $^{127}$I,  and $^{128, 129, 130, 131, 132, 134, 136}$Xe nuclei that are relevant to current direct detection experiments, and to estimate their sensitivity to shell model interactions. 
\item[Methods] Shell model calculations are performed with the NuShellX solver. Nuclear response functions from non-relativistic effective field theory (NREFT) are evaluated and integrated over transferred momentum for quantitative comparisons.  
\item[Results] Although the standard spin independent response is barely sensitive to the structure of the nuclei, large variations with the shell model interaction are often observed for the other channels.
\item[Conclusions] Significant uncertainties may arise from the nuclear components of WIMP-nucleus scattering amplitudes due to nuclear structure theory and modelling. These uncertainties should be accounted for in analyses of direct detection experiments.  
\end{description}
\end{abstract}
\maketitle
\end{CJK*}

\justify
\section{Introduction}
The detection of Dark Matter (DM) remains one of the most heavily pursued goals in physics today, as its exact nature continues to elude our understanding \cite{Bertone:2016nfn,Feng:2010gw,MarrodanUndagoitia:2015veg,Liu:2017drf}. Among the potential DM candidates that have been proposed, Weakly Interacting Massive Particles (WIMPs) \cite{Jungman:1995df,Schumann:2019eaa,Roszkowski:2017nbc}, that are new elementary particles, not included in the Standard Model (SM) of particle physics, with a mass and interaction strength close to the ones of the Electroweak interaction, have attracted significant interest. This class of candidates can naturally thermally produce DM in the early universe in the right amount to match the observed DM density, assuming they have a self-annihilation cross section of a similar order to that arising from the weak force - this has been termed the ``WIMP miracle".

WIMP searches include production in colliders \cite{Arcadi:2017kky}, identification of their annihilation or decay in our galaxy \cite{Gaskins:2016cha}, as well as Direct Detection (DD) of scattering between DM and SM particles through nuclear recoil or through ionisation of target atoms \cite{MarrodanUndagoitia:2015veg,Schumann:2019eaa}. However, all DD experiments so far have reported a null result \cite{DImperio:2018guh,Amare:2021yyu,COSINE-100:2021xqn}, with just one longstanding exception (DAMA/LIBRA \cite{Bernabei:2018yyw}) that still requires independent confirmation.

Evaluating WIMP-nucleus interaction rates for direct detection experiments requires detailed knowledge of the astrophysical DM halo velocity distribution, Beyond Standard Model (BSM) inputs, as well as a microscopic description of the nuclear structure properties of the target nuclei. In particular, nuclear structure properties could have a  strong impact on scattering rates \cite{Engel:1992bf}. This motivated recent studies within the context of non-relativistic effective field theory (NREFT) \cite{Fitzpatrick:2012ix,Anand:2013yka}  and chiral effective field theory (ChEFT) \cite{Hoferichter:2015ipa,Hoferichter:2016nvd, Hoferichter:2018acd, Vietze:2014vsa,Klos:2013rwa, Menendez:2012tm}.

The standard characterisation of the WIMP-nucleus  cross-section  involves both a spin-independent (SI) term and a spin-dependent (SD) one. Different DM direct detection experiments with varying targets offer the possibility of probing different WIMP-nucleus interaction channels. The SI (respectively SD) response functions can be obtained from scalar and/or vector (axial-vector) effective field theory (EFT) relativistic Lagrangians. On the one hand, the SI cross-section is proportional to $\sim A^2$, where $A$ is the atomic mass number, making SI cross-sections   larger for heavier target nuclei. On the other hand, the SD response is probed by nuclei/isotopes which have unpaired nucleons  \cite{Engel:1992bf}. It is expected to be strongly hindered for spin saturated  nuclei (i.e., with even numbers of protons and neutrons) due to the Pauli exclusion principle.

Traditionally, the SI and SD responses are obtained assuming momentum independent interactions between a WIMP and nucleons. The underlying justification is that the WIMP is expected to be non-relativistic with typical velocity $\sim10^{-3}c$ while the target nucleus is at rest. However, Fitzpatrick and collaborators \cite{Fitzpatrick:2012ix} argued that the relative velocity between the WIMPs and nucleons is dominated by the internal velocity of the nucleons in the nucleus, which is of the order of $\sim10^{-1}c$. As a result, momentum-dependent operators should be considered, opening additional interaction channels involving the nucleon orbital angular momentum $L$. In addition to the standard SI and SD responses, one should then also consider an orbital angular momentum dependent (LD) response, as well as a response that depends both on spin and $L$ (LSD). Note that these responses are evaluated within the NREFT for one-body currents \cite{Fan:2010gt,Fitzpatrick:2012ix,Anand:2013yka}, while ChEFT also predicts possible significant contributions from two-body currents \cite{Hoferichter:2018acd}, e.g., induced by WIMP scattering off virtual pions exchanged between two nucleons \cite{Hoferichter:2015ipa}. Corrections are also expected to the one-body currents due to the large Lorentz scalar and vector mean-fields present in the nucleus~\cite{Wang:2020zgp}.

Despite progress in {\it ab initio} methods \cite{Hergert:2020bxy} and their application to DM direct detection \cite{Gazda:2016mrp,Korber:2017ery,Andreoli:2018etf,Hu:2021awl}, nuclear ground-state wave-functions entering the evaluation of nuclear responses from scattering off WIMPs are traditionally computed with the nuclear shell model within restricted valence spaces. Shell model calculations usually lead to good reproductions of low-energy nuclear levels and transition amplitudes \cite{Stuchbery:2022ioh}. However, the reliability of shell model calculations for evaluating operators that are relevant to WIMP-nucleus scattering needs to be evaluated. Indeed, the uncertainty on the NREFT couplings induced by nuclear structure inputs has been recently evaluated for xenon isotopes by comparing two shell model interactions, leading to an uncertainty of up to $\sim50\%$ in some channels \cite{Heimsoth:2023jgl}.

Here, we perform a systematic study of the sensitivity of the WIMP-nucleus elastic scattering amplitude to nuclear structure for $^{19}$F, $^{23}$Na, $^{28-30}$Si, $^{40}$Ar, $^{70,72-74,76}$Ge, $^{127}$I and $^{128-132, 134, 136}$Xe target nuclei relevant to direct detection experiments. Shell model calculations are performed with various shell model nuclear interactions to obtain nuclear response functions. Variations of the magnitude of these nuclear response functions with the nuclear interaction are used to  quantify the level of uncertainty that arises purely from the nuclear components of WIMP-nucleus interaction. This is important for mappings between theory and experiment in the context of DM direct detection. 

In section \ref{background and methods} we provide an overview of the NREFT formalism adopted from \cite{Fitzpatrick:2012ix,Anand:2013yka} and the nuclear operators which are considered in the current work. Details of the nuclear shell model calculations are also presented and linked to the DM-nucleus scattering formalism. In section \ref{results} we discuss shell model predictions for typical nuclear structure observables in $^{19}$F, $^{23}$Na and $^{127}$I. Nuclear response functions calculated using different shell model interactions are presented in section~\ref{responsefunctions} for all nuclei  in consideration. We summarise the main results and conclude  in section \ref{Discussion}. Appendices contain further technical details. Analytical expressions for the nuclear response functions, together with additional results relevant to DM direct detection studies are provided in supplemental material.

\section{Theoretical  Background \& Methods} \label{background and methods}

\subsection{DM-nucleus Elastic Scattering Formalism}

The NREFT approach to WIMP-nucleus scattering  was adopted by Fitzpatrick and collaborators in \cite{Fitzpatrick:2012ix,Anand:2013yka}. This work included LD and LSD nuclear interaction responses, in addition to the standard SI and SD  ones. These provide additional avenues for a DM particle to interact with a nucleus during a scattering process, and may be comparatively significant in magnitude relative to the SI and SD responses for particular isotopes. We provide the relevant NREFT elastic scattering formalism. For brevity we only state the most important information - additional expressions and definitions can be found in Appendix \ref{formalism appendix} and in \cite{Fitzpatrick:2012ix,Anand:2013yka}.

The EFT interaction Lagrangian consists of four-field operators of the form 
\begin{equation}
    \mathcal{L}_{\text{int}} = \sum_{N=n, p} \sum_{i} c_i^{(N)} \mathcal{O}_i \ \chi^+ \chi^- N^+ N^-, \label{eq:Lint}
\end{equation}
where $\chi$ represents the dark matter field and $N$ a nucleon field. In the non-relativistic regime, only operators with terms up to second order in momentum transfer $\vec{q}= \vec{p} \ '-\vec{p}$ are included, where $\vec{p} \ '$ is the outgoing $\chi$ momentum and $\vec{p}$ is the incoming counterpart. Most of the operators considered arise from the exchange of mediators of spin-1 or less (which are at most quadratic in either $\vec{S}$ (spin operator) or $\vec{v}$ ($\equiv \vec{v}_{\chi, \text{in}} -\vec{v}_{N, \text{in}}$)), while some operators employed do not arise from this traditional exchange. The Hamiltonian is Hermitian if it is constructed from the following operators
\begin{equation}
    i \vec{q}, \hspace{4mm} \vec{v}^\perp \equiv \vec{v} + \frac{\vec{q}}{2\mu_N}, \hspace{4mm} \vec{S}_\chi, \hspace{4mm}  \vec{S}_N,
\end{equation}
where $\mu_N= m_N m_\chi/(m_N +m_\chi)$ is the reduced mass for the DM-nucleon system. Hence, the list of possible non-relativistic operators is 
\begin{widetext}
\begin{equation}
 \begin{split}
    & \hspace{3mm} \mathcal{O}_1= \mathbf{1}, \hspace{8mm} \mathcal{O}_3= i \vec{S}_N \cdot (\vec{q} \times \vec{v}^\perp), \hspace{8mm} \mathcal{O}_4= \vec{S}_\chi \cdot \vec{S}_N, \hspace{8mm} \mathcal{O}_5=  i \vec{S}_\chi \cdot (\vec{q} \times \vec{v}^\perp), \\
    \hspace{4mm} & \mathcal{O}_6=  (\vec{S}_N \cdot \vec{q}) (\vec{S}_\chi \cdot \vec{q}), \hspace{8mm} \mathcal{O}_7 =\vec{S}_N \cdot \vec{v}^\perp, \hspace{8mm} \mathcal{O}_8=  \vec{S}_\chi \cdot  \vec{v}^\perp, \hspace{8mm} \mathcal{O}_9=  i \vec{S}_\chi \cdot (\vec{S}_N \times \vec{q}),\\
    \hspace{8mm} \mathcal{O}_{10} = & \ i \vec{S}_N \cdot \vec{q}, \hspace{8mm} \mathcal{O}_{11} = i \vec{S}_\chi \cdot \vec{q}, \hspace{8mm} \mathcal{O}_{12} = \vec{S}_\chi \cdot \left(\vec{S}_N  \times \vec{v}^\perp \right), \hspace{8mm} \mathcal{O}_{13} = i \left(\vec{S}_N \cdot \vec{q}\right)  \left( \vec{S}_\chi \cdot  \vec{v}^\perp \right),\\
    & \hspace{8mm} \mathcal{O}_{14} = i \left(\vec{S}_\chi \cdot \vec{q}\right) \left(\vec{S}_N \cdot \vec{v}^\perp\right), \hspace{8mm} \mathcal{O}_{15} = - \left(\vec{S}_\chi \cdot \vec{q}\right)  \left( (\vec{S}_N \times \vec{v}^\perp) \cdot \vec{q} \right).
 \end{split}   
\end{equation}
\end{widetext}

The operator $\mathcal{O}_2= (v^\perp)^2$ is neglected as it is not obtained from the leading-order non-relativistic reduction of the relativistic four-field $\mathcal{L}_{\text{int}}$ terms in consideration. An additional operator $\mathcal{O}_{16}=- \left((\vec{S}_\chi \times \vec{v}^\perp) \cdot \vec{q} \right) \cdot (\vec{S}_N \cdot \vec{q})$ is also neglected as it is linearly dependent on $\mathcal{O}_{12}$ and $\mathcal{O}_{15}$. From the list of non-relativistic interaction operators above, one can show that the DM-nucleus elastic scattering amplitude is written as a sum of the amplitudes of various nuclear operators, of the form 
\begin{widetext}
\begin{equation} \label{General transition amplitude formula simple}
\begin{split}
       \frac{1}{2J_i+1} \sum_{M_i, M_f} \big| \langle J_i M_f  | \sum_{m=1}^A \mathcal{H}_{\text{int}} (\vec{x}_m) |  J_i M_i \rangle \big|^2= & \frac{4\pi}{2J_i+1} \Bigg[ \sum_{\{j, X\}} \sum_{J}^{\infty} |\langle J_i || l_j \ X_J (q)|| J_i \rangle |^2  \\
       + \sum_{\substack{\{j, X\}; \{k, Y\}; \{X, Y\} \\ X \neq Y}} & \sum_{J}^{\infty} \text{Re} \big[  \langle J_i || l_j \ X_J (q) || J_i \rangle \langle J_i || l_k \ Y_J (q) || J_i \rangle^* \big]\Bigg], 
\end{split}
\end{equation}
\end{widetext}

where $\mathcal{H}_{\text{int}}$ is the interaction Hamiltonian, $J_i$ is the nuclear ground state angular momentum, $A$ is the mass number, and $M_i \ (M_f)$ is the initial (final) angular momentum projection. Here, $X$ and $Y$ are one of six nuclear operators traditionally written as $M_{JM}, \ \Sigma''_{JM}, \ \Sigma'_{JM}, \ \Delta_{JM}, \ \Phi''_{JM}$ and $\tilde{\Phi}'_{JM}$. The treatment of these nuclear multipole operators here is familiar from work on semileptonic weak and electromagnetic interactions with nuclei, such as electron scattering \cite{DeForest:1966ycn, Donnelly:1975ze, Serot:1979yk}, as well as neutrino reactions, charged lepton capture, and $\beta$ decay \cite{OConnell:1972edu,Donnelly:1975ze}, where a  harmonic oscillator wave function basis was specifically employed to evaluate the single-particle matrix elements in \cite{Donnelly:1979ezn}. The long-wavelength limit ($q \rightarrow 0$) gauges the type of interaction the operators are sensitive to (see Table~\ref{long wavelength limit}). $M_{JM}$ is a SI operator, $\Sigma''_{JM}$ and $\Sigma'_{JM}$ are SD, and the remainder are $l$-dependent (LD), as well as  $\vec{\sigma}.\vec{l}$- and tensor-dependent (LSD) operators, respectively, where $\vec{l}$ is the orbital angular momentum and $\vec{\sigma}$ is the spin. The four DM scattering amplitudes $l_j,l_k\equiv l_{0,E,M,5}$, each associated with a specific nuclear operator $X$, are encoded with the DM and nuclear target physics alongside linear combinations of effective theory couplings. The cross terms in Eq.~(\ref{General transition amplitude formula simple}) exist only for two sets of operators, $M_{JM}, \ \Phi''_{JM}$ and $\Sigma'_{JM}, \ \Delta_{JM}$. The initial nuclear spins have been averaged over and final ones summed over, and the matrix element is written in reduced matrix element form using the Wigner-Eckart theorem (see Appendix \ref{formalism appendix}). The full form of Eq.~(\ref{General transition amplitude formula simple}) in terms of the operators is also provided in Eq.~(\ref{General transition amplitude formula}).

{\centering
\begin{table*}[htbp] 
\caption{Leading order terms of the nuclear operators in the long-wavelength limit $q \rightarrow 0$ \cite{Fitzpatrick:2012ix}. 
\label{long wavelength limit}}
\begin{tabular}{ |p{5cm}|p{2.5cm}|p{5.5cm}|}
 \hline
 \hfil Response Type & \centering{Leading Multipole}  & \hfil Long-wavelength Limit  \\
 \hline 
\hfil  $M_{JM}$ : Charge  & \hfil $M_{00} (q \vv{x}_m)$    & \hfil $\frac{1}{\sqrt{4\pi}} 1(m)$ \\
 \hfil  $L^5_{JM}$ : Axial Longitudinal   &  \hfil $\Sigma''_{1M} (q \vv{x}_m)$  & \hfil $\frac{1}{2\sqrt{3\pi}} \sigma_{1M} (m)$ \\
  \hfil $T^{\text{el5}}_{JM}$ : Axial Transverse Electric  & \hfil $\Sigma'_{1M} (q \vv{x}_m)$ & \hfil $\frac{1}{\sqrt{6\pi}} \sigma_{1M} (m)$ \\
 \hfil   $T^{\text{mag}}_{JM}$ : Transverse Magnetic  & \hfil $\frac{q}{m_N} \Delta_{1M} (q \vv{x}_m)$ & \hfil $-\frac{q}{2 m_N \sqrt{6\pi}} l_{1M} (m)$ \\
 \hfil $L_{JM}$ : Longitudinal & \hfil  $\frac{q}{m_N} \Phi''_{00} (q \vv{x}_m)$  & \hfil $-\frac{q}{3 m_N \sqrt{4\pi}} \vv{\sigma} (m). \vv{l} (m)$ \\
   & \hfil  $\frac{q}{m_N} \Phi''_{2M} (q \vv{x}_m)$  & \hfil  $- \frac{q}{m_N} \frac{1}{\sqrt{30 \pi}} \left[x_m \otimes \left(\vv{\sigma} (m) \times \frac{\vv{\nabla}}{i} \right)_1 \right]_{2M}$   \\
 \hfil {\centering $T^{\text{el}}_{JM}$ :  Transverse Electric}  &  \hfil $\frac{q}{m_N} \tilde{\Phi}'_{2M} (q \vv{x}_m)$  & \hfil $- \frac{q}{m_N} \frac{1}{\sqrt{20 \pi}} \left[x_m \otimes \left(\vv{\sigma} (m) \times \frac{\vv{\nabla}}{i} \right)_1 \right]_{2M}$ \\
 \hline
\end{tabular}
\label{long wavelength limit2}
\end{table*}}

In the context of elastic scattering theory we are only interested in the ground state nuclear wave function $| J_i \rangle$. The nuclear matrix elements can be written as a product of single-nucleon matrix elements and One-Body Density Matrix Elements (OBDMEs) $\Psi^{J;\tau}_{|\alpha|, |\beta|}$ in the following way
\begin{widetext}
\begin{equation} \label{reduced nuclear matrix element}
\begin{split}
     \langle J_i; T M_T \ ||  \sum_{m=1}^{A} \hat{O}_{J, \tau} (q \vec{x}_m) || & \ J_i ; T M_T \rangle  \\
    & \hspace{-15mm} = (-1)^{T-M_T}  \begin{pmatrix}
         \vspace{2mm} T  &  \tau  &  T \\
         \vspace{2mm} -M_T &  0 &  M_T  \hspace{2mm}
         \end{pmatrix}  
         \langle J_i; T \ \vdots \vdots \sum_{m=1}^{A} \hat{O}_{J, \tau} (q \vec{x}_m) \vdots \vdots \ J_i ; T \rangle \\
    & \hspace{-15mm} =    (-1)^{T-M_T}  \begin{pmatrix}
         \vspace{2mm} T  &  \tau  &  T \\
         \vspace{2mm} -M_T &  0 &  M_T  \hspace{2mm}
         \end{pmatrix}
         \sum_{|\alpha|, |\beta|} \Psi^{J;\tau}_{|\alpha|, |\beta|}  \langle |\alpha| \ \vdots \vdots \hat{O}_{J, \tau} (q \vec{x}) \vdots \vdots \ |\beta| \rangle,
\end{split}
\end{equation}
\end{widetext}
where $\alpha$ and $\beta$ are single-nucleon states given by the usual quantum numbers $\beta = \{n_\beta, l_\beta, j_\beta, m_{j_\beta}, m_{t_\beta} \}$, with the reduced state notation $|\beta|= \{n_\beta, l_\beta, j_\beta\}$. The nucleon isospin state $t_\beta=t_\alpha=1/2$ is implicit in the notation. The nuclear isospin is denoted by $T$ and its projection $M_T$. The notation $\vdots \vdots$ denotes a matrix element reduced in both angular momentum and isospin using the Wigner-Eckart theorem. Additionally, $\tau=\{0, 1\}$ with $ \hat{O}_{J, \tau} =  \hat{O}_{J} \ \tau_3^\tau$ and $\tau_3$ being the nucleon isospin operator. Hence, the $\tau=0$ term corresponds to the isospin-independent component of the single-nucleon operator, with $\tau=1$ corresponding to isospin-dependent counterpart. 

The OBDMEs  have the form
\begin{equation}
\begin{split}
    \Psi^{J;\tau}_{|\alpha|, |\beta|} & \equiv \frac{\langle J_i; T \ \vdots \vdots \left[a^\dagger_{|\alpha|} \otimes \tilde{a}_{|\beta|} \right]_{J;\tau} \vdots \vdots \ J_i ; T \rangle}{\sqrt{(2J+1)(2\tau+1)}},
\end{split}
\end{equation}
where $\tilde{a}_{\beta }=(-1)^{j_\beta -m_{j_\beta}+1/2-m_{t_\beta}} \ a _{|\beta|;-m_{j_\beta}, -m_{t_\beta}}$ and $\otimes$ denotes a tensor product. The OBDMEs contain all of the relevant information about the nuclear ground state for each isotope in consideration. Varying aspects of the nuclear model may lead to different OBDME values, and hence to different values of the nuclear matrix elements. Gauging the sensitivity of these matrix elements to nuclear structure is the goal of this  work.

\subsection{Nuclear Structure}

The nuclear shell model is a configuration interaction approach to the nuclear many-body problem that is widely used to calculate eigenstates of the nuclear hamiltonian as well as nuclear observables \cite{Stuchbery:2022ioh}. In practice, shell model calculations are performed assuming a filled inert core and a valence space of few single-nucleon orbitals above this core. Here, nuclear shell model calculations are performed with NuShellX \cite{Brown2014TheNuShellXMSU}. Among the program's user-inputs is the valence (model) space as well as a possible valence space truncation. For each model space, a range of nuclear shell model interactions are provided, each pre-developed based on constrained fits to certain nuclear data. 

For each isotope, changing the interaction used for a particular valence space truncation may provide different OBDMEs, which may impact the values of the nuclear matrix elements - this is the investigation that we undertake in the current work. We employ interactions which differ from those used in \cite{Fitzpatrick:2012ix,Anand:2013yka} and compare the two sets of results. 
 
These $\Psi^{J;\tau}_{|\alpha|, |\beta|}$ values are then inserted into a pre-developed Mathematica package \cite{Anand:2013yka}, which calculates the relevant observables associated with DM-nucleus scattering. The OBDME values used to perform the calculations in this work can be found in the supplementary material.

\subsection{Nuclear response functions}

The aforementioned Mathematica package \cite{Anand:2013yka} can be used to calculate nuclear response functions 
\begin{eqnarray}\label{FF eqn}
    F^{(N,N')}_{X,Y} (q^2) & \equiv &\frac{4\pi}{2J_i+1} \nonumber\\
    &&\sum_{J=0}^{2J_i} \langle J_i || X_J^{(N)} || J_i \rangle \langle J_i || Y^{(N')}_J || J_i \rangle
\end{eqnarray}
where $N, N'=\{p, n\}$. We also define $F^{(N,N')}_X (q^2)  \equiv  F^{(N,N')}_{X,X} (q^2)$. These response functions single out the nuclear aspect of the scattering amplitude and can be used to carry out an investigation of the effect of nuclear structure on WIMP-nucleus elastic scattering. The proton and neutron nuclear operators are given by $X_J^{(p)}= \frac{1+\tau_3}{2} \ X_J$ and $X_J^{(n)}= \frac{1-\tau_3}{2} \ X_J$. We have $F^{(p,n)}_{X,Y} (q^2)= F^{(n,p)}_{X,Y} (q^2)$ only for the non-interference responses with $X=Y$.

Using a harmonic oscillator single-particle basis, the response functions take on expressions of the form $e^{-2y}p(y)$, where $p(y)$ is a polynomial with $y=(qb/2)^2$ and $b= {1}/{\sqrt{m_N \omega}} \approx \sqrt{41.467/(45A^{-1/3} - 25A^{-2/3})}$~fm  is the harmonic oscillator length parameter ($m_N$ is the nucleon mass and $\omega$ is the oscillator frequency). Following \cite{Fitzpatrick:2012ix}, we define proton and neutron Integrated Form Factor (IFF) values as
\begin{equation}\label{IFF expressions}
  \int_{0}^{\text{100\text{ MeV}}} \frac{q \,dq}{2} F^{(N, N)}_{X(,Y)} (q^2),
\end{equation}
in units of $\text{MeV}^2$. These IFF values are a proxy for the strength of each of the nuclear interaction channels, which depend on the unique nuclear structure of the isotope in consideration. The effect of increasing the upper limit of this integral on the IFF values of $^{127}$I is touched upon in section \ref{sec:127I}.

\section{Shell model calculations} \label{results}

Before investigating nuclear responses, we compare shell model predictions of energy spectra and electromagnetic transitions against experimental values for $^{19}$F, $^{23}$Na and $^{127}$I. 

\subsection{$^{19}$F and $^{23}$Na \label{sec:sd}}

We use an unrestricted $sd$ model space with single particle levels $1d_{5/2}, \ 2s_{1/2},\ 1d_{3/2}$ for both protons and neutrons. The work of \cite{Fitzpatrick:2012ix,Anand:2013yka} uses the USD \cite{Brown1988StatusModel, Wildenthal1984EmpiricalNuclei}  nuclear shell model interaction, whereas here we use the  USDB interaction. The latter has been fitted to the energies of 608 (states) of 77 nuclei with $21\leq A\leq40$ \cite{Brown2006NewShell}. 

\begin{figure}[htbp]
\centering 
\includegraphics[width=0.95\linewidth]{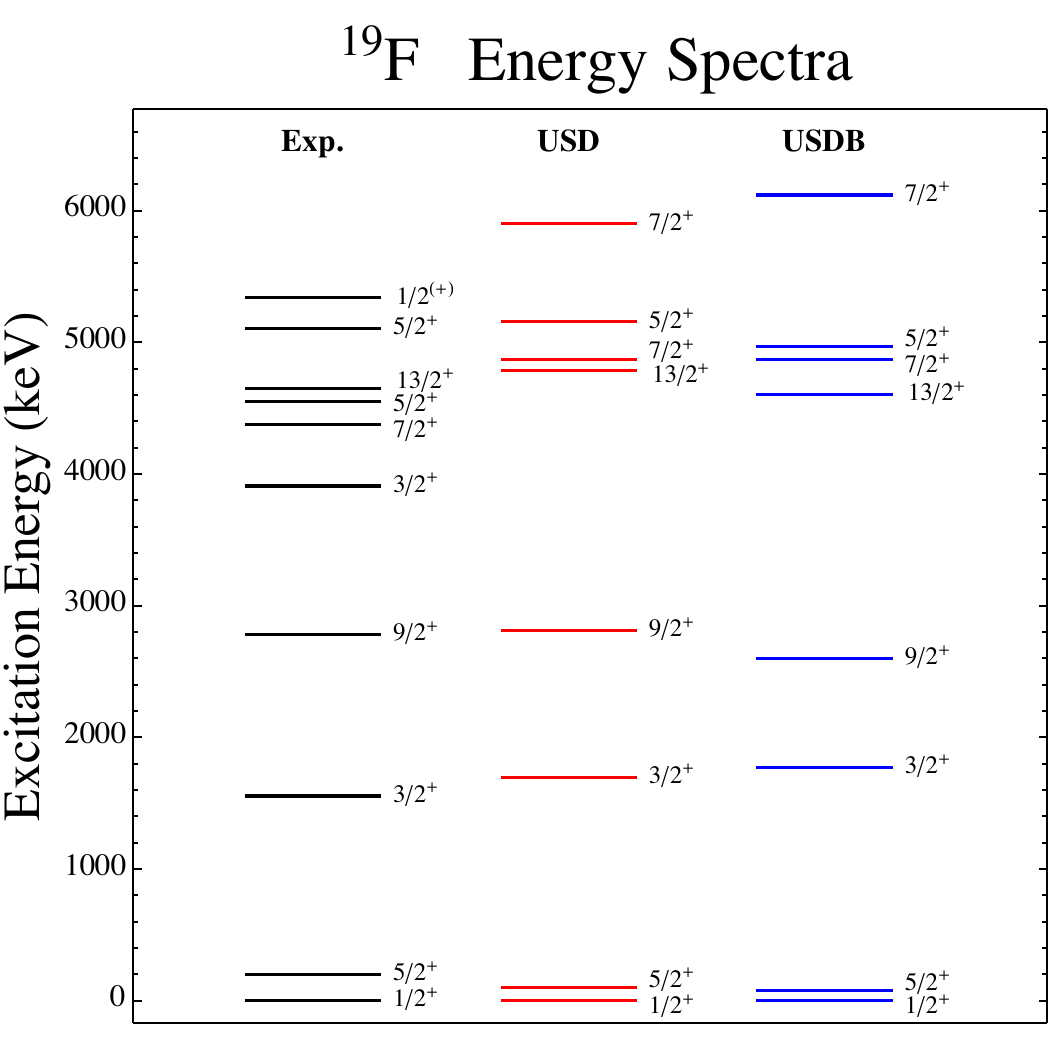}
\caption{$^{19}$F energy levels (keV) from USD and USDB interactions in the  $sd$ model space are compared with experiment (only positive parity levels are shown). \label{fig:19Flevels}}
\centering
\end{figure}

\begin{figure}[htbp]
\centering 
\includegraphics[width=0.95\linewidth]{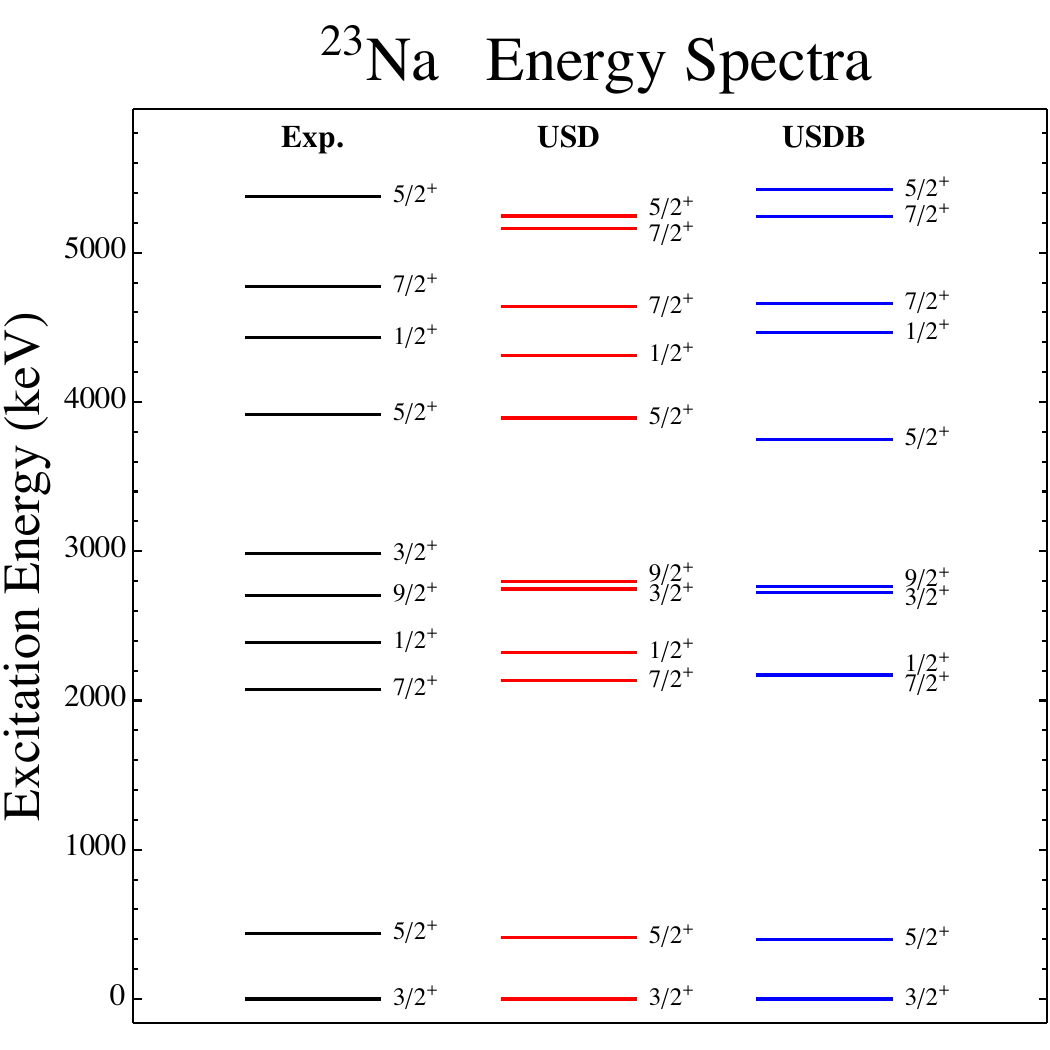}
\caption{Same as Fig.~\ref{fig:19Flevels} for $^{23}$Na.\label{fig:23Nalevels}}
\centering
\end{figure}

The $^{19}$F and $^{23}$Na experimental energy spectra are compared with the USD and USDB energy levels in Figs.~\ref{fig:19Flevels} and~\ref{fig:23Nalevels}, respectively. 
Only positive parity states are possible for the chosen model space, hence we do not include the experimental states with negative parity.
The overall agreement between experiment and theory is good up to $\sim3$~MeV in $^{19}$F and up to $\sim5$~MeV in $^{23}$Na. 

As a test of the calculated nuclear matrix elements, Table~\ref{tab:transitions1} provides a comparison between theoretical and experimental electric quadrupole [B(E2)] and magnetic dipole [B(M1)] transitions between low-lying states. We note that the USD and USDB interactions were not fitted to such data. The default values of the effective charges and parameters of the $M1$ and $E2$ operators were used in the calculations. For $^{19}$F and $^{23}$Na, the proton and neutron effective charges are $e_p=1.36$ and $e_n=0.45$, respectively, with the effective $g$~factors being those in the third row of table I of Ref.~\cite{Richter2008} for the USD interaction and the last row of the same table for USDB. 

Overall, both interactions lead to similar predictions and agree well with experimental data. As USDB is an extension of USD, the two interactions are expected to be similar for these stable isotopes. Both interactions are expected to produce realistic ground state wave functions and could then be used to predict nuclear responses to WIMP-nucleon elastic scattering.

{\tiny 
\begin{table*}[htbp]
\caption{Magnetic dipole and electric quadrupole moments, and electric quadrupole [B(E2)] and magnetic dipole [B(M1)] transitions, between low-lying states of $^{19}$F and $^{23}$Na. Experimental transition values from \cite{TILLEY1995,BASUNIA2021}, while moments taken from \cite{NSR2019STZV,INDC0816,indc.nds.0833}. \label{tab:transitions1}}
\hspace*{-10mm} \begin{tabular}{lcccccccccccccccc}
\hline \hline 
&   &   \multicolumn{3}{c}{Q [$e$fm$^2$]} & & \multicolumn{3}{c}{$\mu$ [n.m.]}  & & \multicolumn{3}{c}{B(E2) [$e^2$fm$^4$]}  & & \multicolumn{3}{c}{B(M1) [n.m.$^2$]}\\
\cline{3-5}
\cline{7-9}
\cline{11-13}
\cline{15-17}
Nucleus & State/Transition & USD & USDB & Exp. & & USD & USDB & Exp. & & USD  
& USDB & Exp. &  & USD & USDB & Exp. \\
\hline
$^{19}$F & $1/2_{\text{gs}}^+$ & & & & & +2.650 & +2.681 & +2.628 \tablenotemark[1] &  &  &  & & & &  &  \\
 & $5/2_1^+$ & -9.53 & -9.47  & -9.42(9) &  & +3.504 & +3.424 & +3.605(8)  &  &  &  & & & &  &  \\
 & $5/2_1^+ \rightarrow 1/2_{\text{gs}}^+$ & & & & & &  & & &  19.22 & 19.44 & 20.93(24) \\
&$3/2_1^+ \rightarrow 5/2_1^+$ &  &  &  & & & & & & 8.068 & 7.986 & &  & 3.03 & 3.19 & 4.1(25) \\
&$9/2_1^+ \rightarrow 5/2_1^+$ &  &  &  & & & & & & 18.82 & 19.32 & 24.7(27) &  & & &\\
\hline 
$^{23}$Na & $3/2_{\text{gs}}^+ $ & +11.0 & +10.7 & +10.4(1) &  & +2.194 & +2.128 & +2.218 \tablenotemark[1] &  &  & & & & & & \\
& $5/2_1^+ \rightarrow 3/2_{\text{gs}}^+$ & & & & & &  & & & 109.2 & 109.1 & 124(23) &  & 0.361 & 0.357 & 0.403(25)\\
&$7/2_1^+ \rightarrow 5/2_1^+$ & & & & &  & & &  & 61.72 & 57.03 & 56.7(85) &  & 0.262 &  0.238 & 0.294(34) \\
& $1/2_1^+ \rightarrow 5/2_1^+$ & & & & & & & &  & 10.31 & 14.48 & 11.3(27) &  & & & \\
\hline \hline
\end{tabular}
\tablenotetext[1]{The uncertainty is much less than $\pm 0.001$}
\end{table*}
}

\begin{table*}[htbp]
\caption{Magnetic dipole and electric quadrupole moments of low-lying states of $^{127}$I. Experimental moments taken from \cite{NSR2019STZV,INDC0816,indc.nds.0833}.
\label{tab:moments1}}
\centering
\begin{tabular}{lcccccccc}
\hline \hline 
  &  \multicolumn{3}{c}{ $\mu$ [n.m.] } &  & \multicolumn{3}{c}{$Q$ [efm$^2$]}\\
\cline{2-4}
\cline{6-8}
State  & GCN5082 & SN100PN & Exp. & & GCN5082 & SN100PN & Exp.  \\
\hline
$5/2_1^+$ & +2.8920 & +2.6046  & +2.8087(14) &  &   -49.85 & -58.96 & -68.8(10) \\
$7/2_1^+$ & +2.29 & +2.38  & +2.54(5) & & -63.83  & -50.40 & -61.7(11) \\
$3/2_1^+$ & +1.45 & +1.35  & +0.97(7) & &  +30.19 & +42.87 & \\
\hline \hline
\end{tabular}
\end{table*}

\subsection{$^{127}$I} \label{sec:SM127I}

The work of \cite{Fitzpatrick:2012ix,Anand:2013yka} utilises a nuclear shell model interaction developed by Baldridge and Dalton \cite{Baldridge1978Shell-modelCases}, which we refer to as ``B\&D". This interaction is used in the model space which includes all proton and neutron orbits in the major shell between magic numbers 50 and 82, with single particle levels $1g_{7/2}$, $2d_{5/2}$, $2d_{3/2}$, $3s_{1/2}$ and $1h_{11/2}$. The valence space restriction employed here involves fixing the occupation number of $1h_{11/2}$ to the minimum allowed nucleon number.

Here, we  consider the SN100PN \cite{Brown2005Magnetic132Sn} and GCN5082 \cite{Menendez2009DisassemblingDecay} interactions. Both begin with renormalization of the $G$ matrix based on a nucleon-nucleon potential. The GCN5082 interaction began with the Bonn-C potential and was then fitted to about 400 low-lying energy levels of 80 nuclei with $50\leq Z,N\leq82$ by varying various combinations of two-body matrix elements. The SN100PN interaction was based on the CD-Bonn nucleon-nucleon interaction, with the renormalization of the $G$ matrix carried to third order \cite{Machleidt:2000ge}. No fitting was performed, but the single-particle energies were set by reference to the energy levels of $^{133}$Sb and $^{131}$Sn. Differences between these two interactions represent the theoretical uncertainty in our shell model calculations.

Our shell model calculations are performed in the same model space as the B\&D calculations of \cite{Fitzpatrick:2012ix,Anand:2013yka}, but with different restrictions. We keep the proton valence space levels $1g_{7/2}$ and $2d_{5/2}$ unrestricted, with a maximum of 2 protons in the rest. The neutron valence space is unrestricted for the $2d_{3/2}$, $3s_{1/2}$ and $1h_{11/2}$ levels, with a full $2d_{5/2}$ level and a minimum of 6 neutrons in $1g_{7/2}$. Restrictions were needed to keep the basis space to a workable size.

\begin{figure}[htbp]
\centering 
\includegraphics[width=0.95\linewidth]{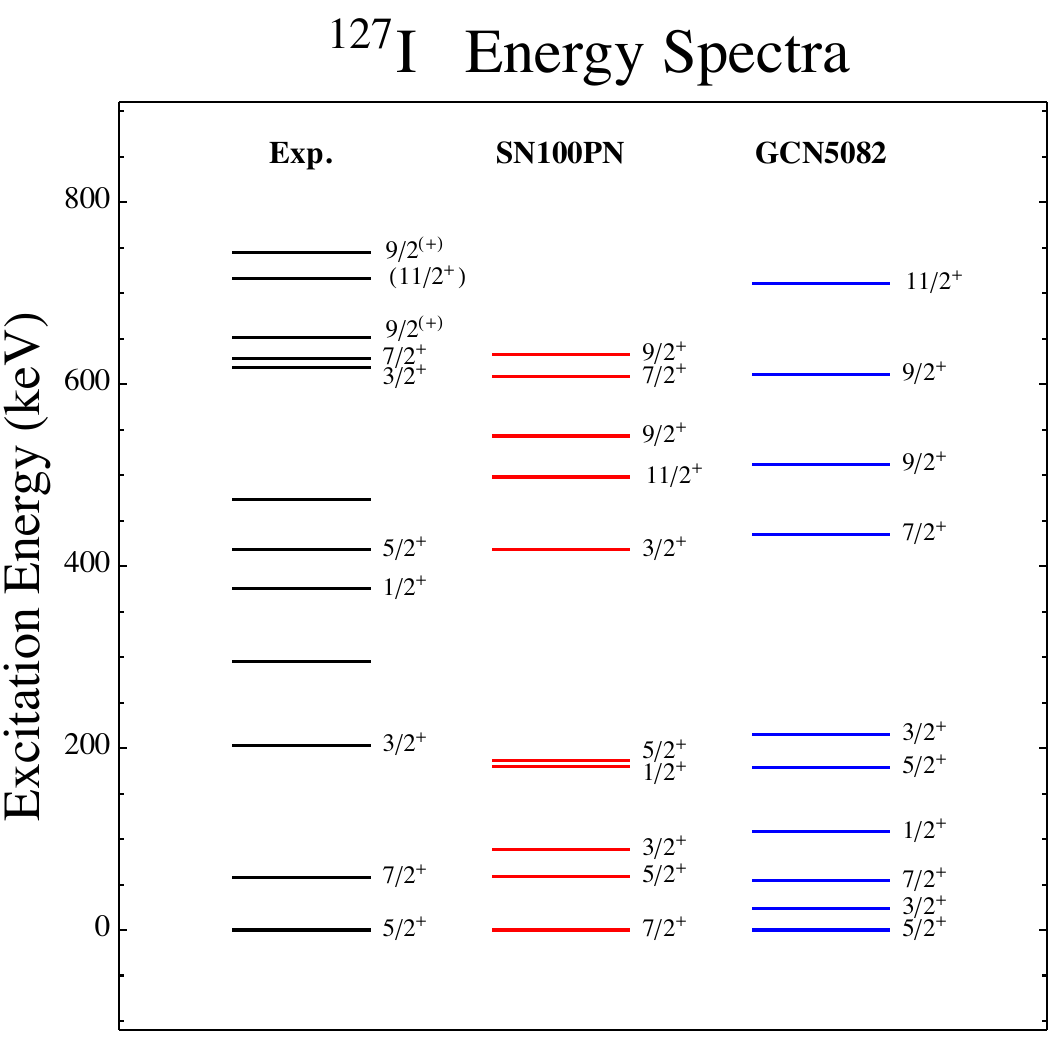}
\caption{$^{127}$I energy spectra (keV) for experimental data, as well as the SN100PN and GCN5082 interactions in a restricted model space.\label{fig:127Ilevels}}
\centering
\end{figure}

The $^{127}$I energy levels predicted by the SN100PN and GCN5082 interactions are plotted in Fig.~\ref{fig:127Ilevels} and compared with experimental data. The ordering of the first three experimental levels is not reproduced by either interaction. The calculations also compress the energy spectrum compared with what is observed experimentally. This compression (prediction of excited states at too low an excitation energy) is often a consequence of the truncation of the basis space, where the omission of many small components of the wave function can lead to the prediction of the ground state and low-lying states at too high an energy (on an absolute scale). However, the lowest few states have the same spin-parity in theory and experiment, which gives confidence that the calculated wave functions in the restricted basis at least pick up the main components of a calculation without truncation.

Table~\ref{tab:moments1} provides a comparison between theoretical and experimental magnetic dipole and electric quadrupole moments for low-lying states. The effective charges and $g$~factors used here are as adopted in recent studies of neighboring nuclei. See for example  \cite{Hicks:2022wkz}. Both interactions are in relatively good agreement with experiment. The spin-parity of the ground state being experimentally assigned to $5/2^+$, it is appropriate to use the first $5/2^+$ predicted by the shell model calculations in the computation of nuclear response in WIMP-nucleus elastic scattering. GCN5082 correctly predicts this state to be the ground state, while with SN100PN it is the first excited state. The two interactions appear to be performing with a similar level of accuracy as one another, with respect to the experimental expectations. The differences are useful in our study as it is anticipated that they may translate into different nuclear responses to WIMP-nucleus elastic scattering. Hu and collaborators have indeed found large theoretical uncertainties in $^{127}$I structure factors for SD scattering \cite{Hu:2021awl}.

\section{Nuclear response functions} \label{responsefunctions}

It is clear from the previous discussion that strong variations of nuclear response functions with the underlying shell model interactions are expected in the case of $^{127}$I. We then focus first on this isotope before presenting systematic studies of integrated form factors.  Note that, in the following, the $\tilde\Phi'$ response is included for completeness. It is however not discussed in detail as it is an exotic response induced purely by unusual couplings of WIMPs \cite{Fitzpatrick:2012ix}.

\subsection{$^{127}$I\label{sec:127I}}

\begin{figure*}[htbp]
\centering 
\includegraphics[width=\linewidth]{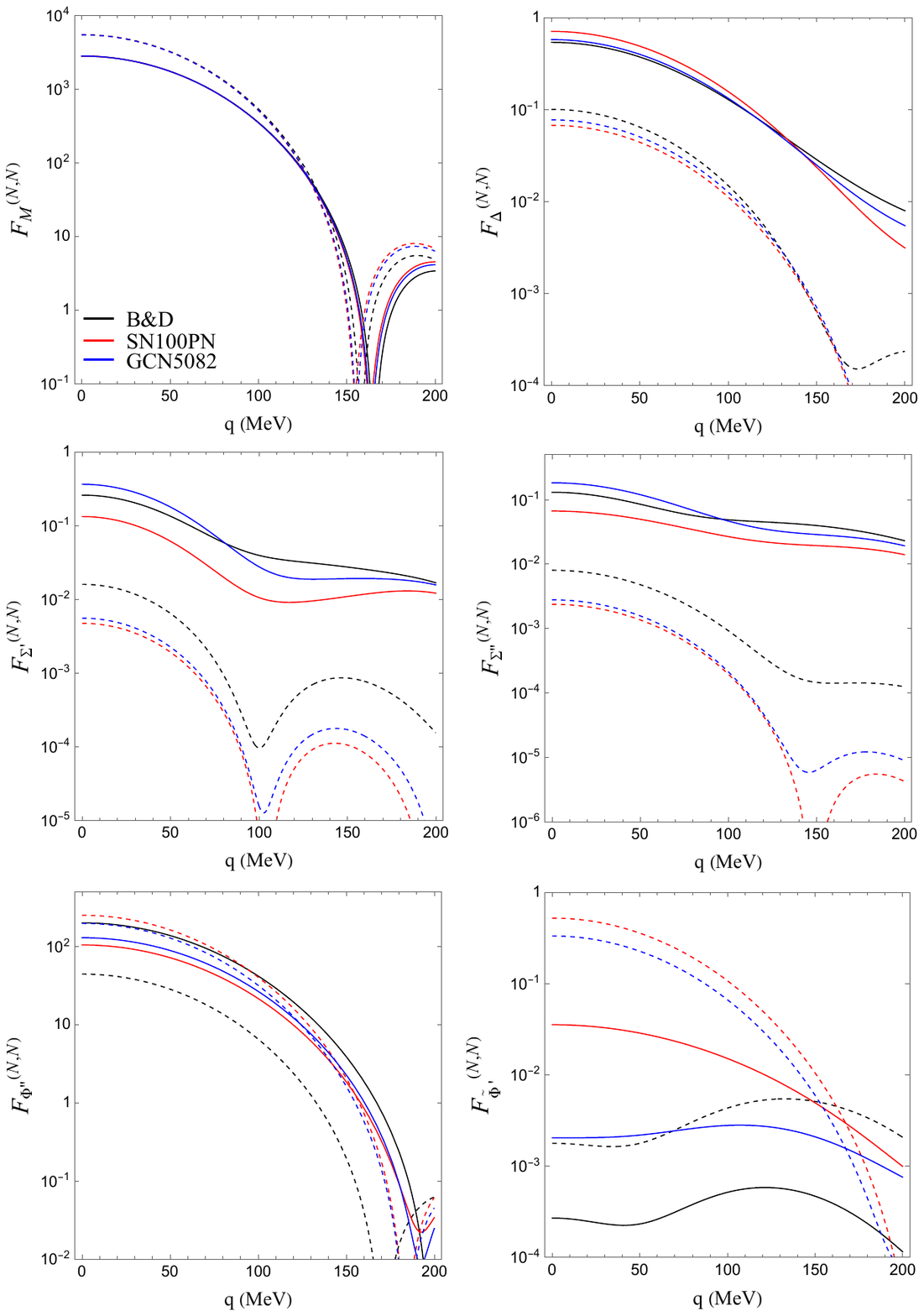}
\caption{$^{127}$I proton-proton (solid lines) and neutron-neutron (dashed lines) nuclear response functions $F^{(N,N)}_X (q^2)$ obtained with three different shell model interactions. The B\&D results are from \cite{Anand:2013yka}. \label{fig:responses}}
\centering
\end{figure*}

\begin{figure}[htbp]
\centering 
\includegraphics[width=1\linewidth]{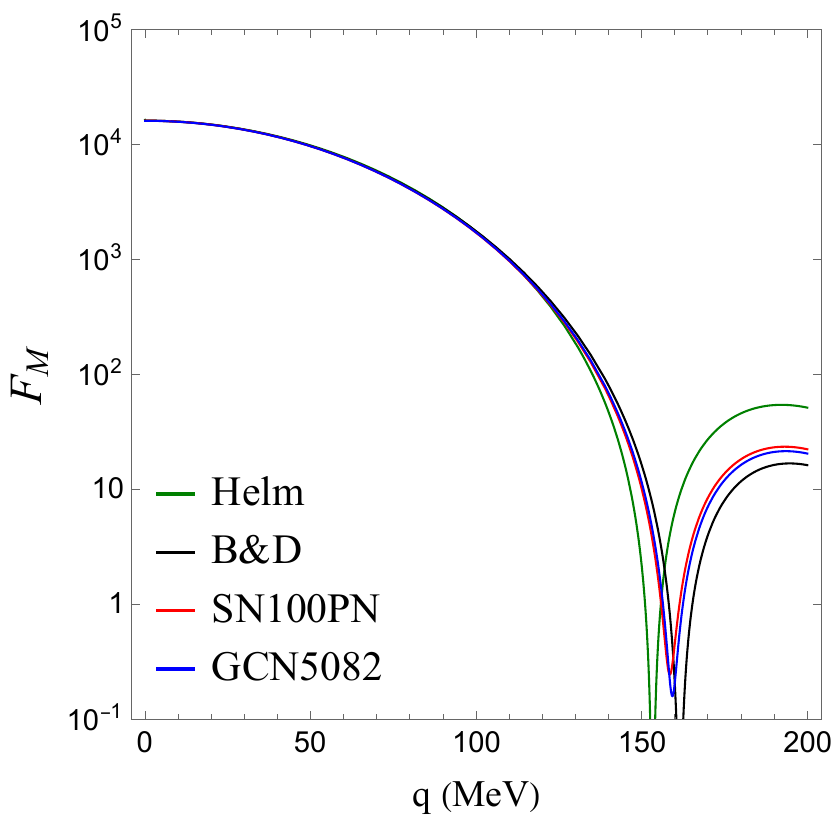}
\caption{$^{127}$I SI nuclear response function $F_M (q^2)$ obtained with three different shell model interactions, with the Helm response plotted for comparison. \label{fig:127IResponseHelm}}
\centering
\end{figure}

The response functions $F^{(p,p)}_X(q^2)$ (solid lines) and $F^{(n,n)}_X(q^2)$ (dashed lines) are shown in Fig.~\ref{fig:responses} for the nuclear operators $X=M$, $\Delta$, $\Sigma'$, $\Sigma''$, $\Phi''$ and $\tilde{\Phi}'$. The response functions obtained from the Mathematica notebook of \cite{Anand:2013yka} (i.e., not the approximated ones published in their manuscript) with the B\&D interaction are reported together with the SN100PN and GCN5082 results from the present work (see Section~\ref{sec:SM127I} for details on the valence space truncation). The Helm response \cite{Lewin:1995rx} is also plotted for comparison with the SI responses in Fig.~\ref{fig:127IResponseHelm}, which has the form $F(q)=3e^{-(qs)^2/2}(\sin(qr_n)-qr_n\cos(qr_n))/(qr_n)^3$, with nuclear skin thickness $s\approx 0.9$~fm and nuclear radius $r_n^2=\left((1.23A^{1/3}-0.6)^2 +(7/3)\pi^2(0.52)^2-5(s/\text{fm})^2 \right)$~fm$^2$. All calculations agree up to $q\sim 120$ MeV, beyond which they differ in value.

The spin independent response $F^{(N,N)}_M (q^2)$ is barely sensitive to the choice of interaction. Although all interactions predict similar shapes for the orbital angular momentum dependent responses $F^{(N,N)}_\Delta(q^2)$, significant variations in magnitude are observed, of the order of 25 - 45\% for the larger differences at low~$q$. These variations are significantly larger for the other channels involving a spin dependence ($\Sigma'$, $\Sigma''$, $\Phi''$, $\tilde{\Phi}'$).

In almost all cases, however, all interactions agree on the respective importance of proton and neutron contributions. At low $q$, the spin independent responses $F_M$ are simply proportional to the square of the number of protons and neutrons. $F_\Delta$, $F_{\Sigma'}$ and $F_{\Sigma''}$ responses are largely dominated by protons. Indeed, $^{127}$I has an odd number of protons and an even number of neutrons. Spin and orbital angular momentum are likely to be cancelled in nucleon pairs, leaving these contributions dominated by the unpaired proton. 

In the long-wavelength approximation $\Phi''$ reduces in part to the spin-orbit operator (see Table~\ref{long wavelength limit}). The latter can lead to coherent contributions that are maximised when one spin-orbit partner is fully occupied and the other one empty. In $^{127}$I for SN100PN and GCN5082, these contributions are dominated by the proton $1g_{9/2,7/2}$ and neutron $1h_{11/2,9/2}$ spin-orbit partners. Indeed, $1g_{9/2}$ belongs to the core and is thus fully occupied, with at most three protons in $1g_{7/2}$, while $1h_{9/2}$ is above the valence space and thus empty, with at least four neutrons in $1h_{11/2}$. These coherent effects can amplify differences coming from the various approximations used in the shell model calculations. In particular, we see that the calculations of \cite{Anand:2013yka} with the B\&D interaction predict a very small neutron contribution to the $\Phi''$ operator. This is likely due to the fact that they fix the occupation of $1h_{11/2}$ to its minimum (four neutrons), while in our calculations this value can be larger, thus leading to stronger neutron contribution in our case.

\begin{figure}[htbp]
\centering 
\includegraphics[width=1\linewidth]{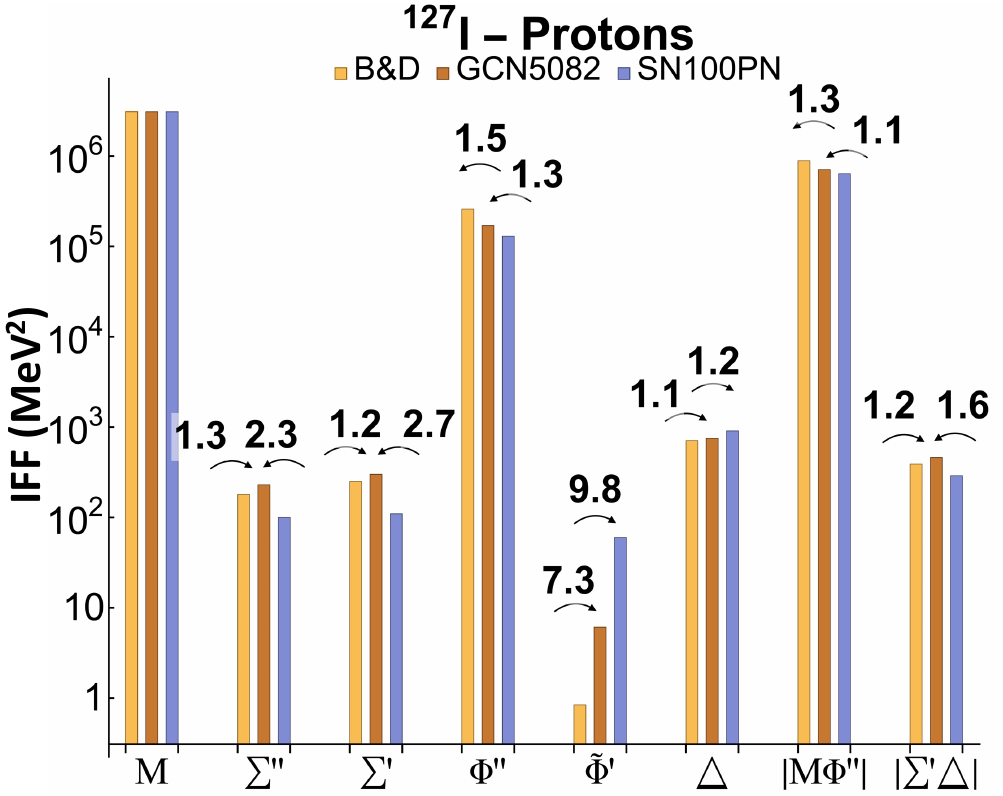}
\caption{$^{127}$I proton IFF values in units of $(\text{MeV})^2$, evaluated using Eq. (\ref{IFF expressions}).}
\centering
\label{127I all IFF pp}
\end{figure}

\begin{figure}[htbp]
\centering 
\includegraphics[width=1\linewidth]{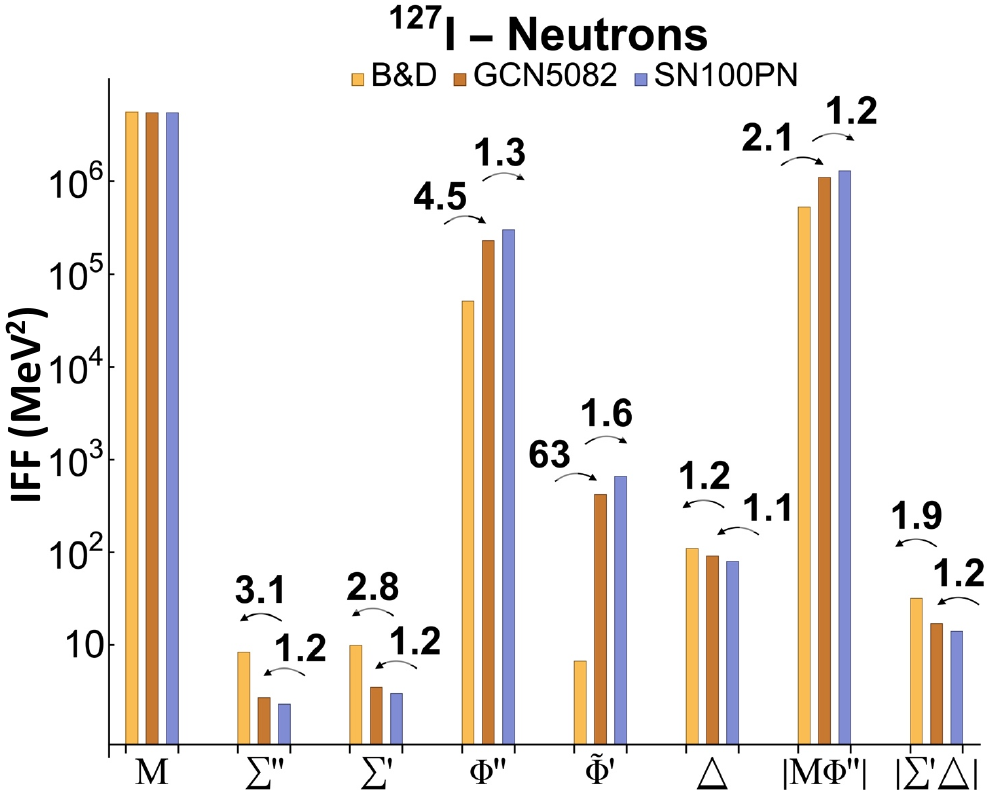}
\caption{Same as Fig.~\ref{127I all IFF pp} for $^{127}$I neutron IFF.}
\centering
\label{127I all IFF nn}
\end{figure}

Figures \ref{127I all IFF pp}  and \ref{127I all IFF nn} show the proton and neutron IFF values [see Eq.~(\ref{IFF expressions})], respectively, for the three shell model interactions. Although we choose to represent all IFFs for either protons or neutrons on the same figure, one should refrain from using them to compare the relative importance of each response. Indeed, each IFF only reflects the ability of the nucleus to interact through a specific channel. Whether WIMPs are themselves able to probe this channel depends on the particle physics model that translates into the coefficients $c_i$ in Eq.~(\ref{eq:Lint}). Nevertheless, for simplicity, and assuming that these $c_i$ coefficients are of a similar order, the SI ($M$) channels will be considered to be the leading responses, while subleading responses include the LSD ($\Phi''$) channel (due to the coherent contribution from partially occupied spin-orbit partners) as well as the SD ($\Sigma'$ and $\Sigma''$) and LD ($\Delta$) channels in nuclei with odd protons or neutron numbers. 

Large variations are observed in almost all sub-leading channels, e.g., SD proton ($\Sigma_p',\Sigma_p''$), LD proton ($\Delta_p$), LSD ($\Phi''_{p,n}$), and their interferences, including with the SI response ($M_{p,n}$). The largest of these variations reach a factor $\sim5$ between $B\&D$ and the other interactions for the $\Phi_n''$ response.

\begin{figure}[htbp]
\centering 
\includegraphics[width=1\linewidth]{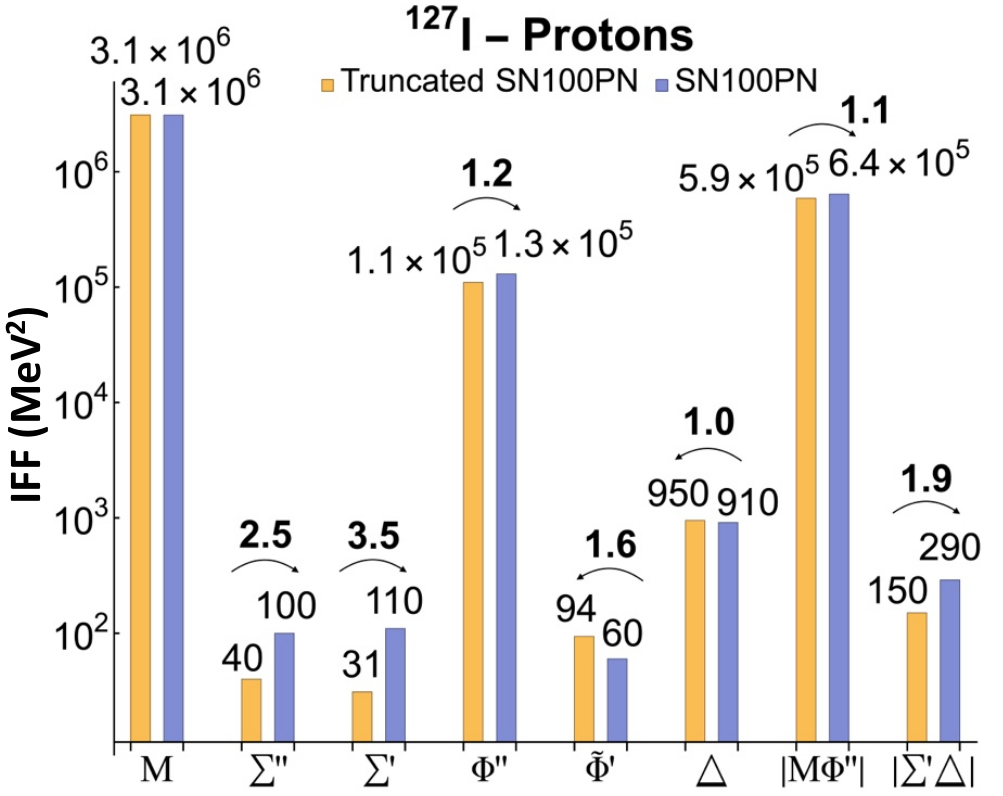}
\caption{Same as Fig.~\ref{127I all IFF pp} for different SN100PN valence space truncation.}
\centering
\label{127I truncated pp}
\end{figure}

\begin{figure}[htbp]
\centering 
\includegraphics[width=1\linewidth]{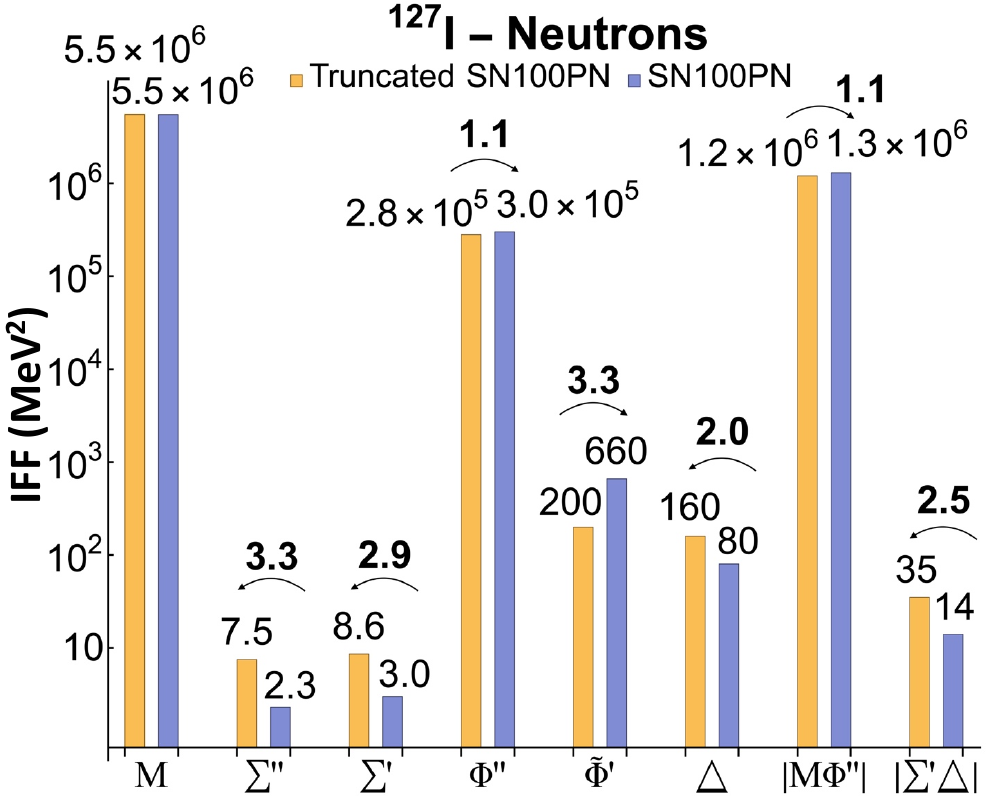}
\caption{Same as Fig.~\ref{127I all IFF nn} for different SN100PN valence space truncation.}
\centering
\label{127I truncated nn}
\end{figure}

To evaluate the effect of valence space truncation on the nuclear IFF values, we repeated our shell model calculation for  $^{127}$I with the SN100PN interaction  using a stricter valence space truncation for neutrons, with no further restriction imposed on the proton valence space. The neutron valence space is still unrestricted for the $2d_{3/2}$ and $3s_{1/2}$ levels, and the $2d_{5/2}$ level is still fully occupied. However, the $1g_{7/2}$ is now assumed to be full (instead of having a minimum of 6 neutrons), and $1h_{11/2}$ is restricted to a maximum of 8 neutrons (instead of being unrestricted). There is an overall of 6 neutron single particle states (2 occupied and 4 empty) that have been restricted compared to the previous calculation.

The resulting IFFs are plotted in Figs.~\ref{127I truncated pp} and~\ref{127I truncated nn} for protons and neutrons, respectively. 
The effect remains relatively small for some sub-dominant responses such as $\Phi_{p,n}''$ (less than $20\%$ variation), whereas some other channels exhibit larger differences, such as the proton SD channels, although they are much smaller in magnitude. When possible, the less restricted valence space should be used to evaluate nuclear response functions and limit uncertainties on the sub-leading channels. 

The IFF values change by a factor of 1.1-2.5 with $q_{max}=200$ MeV. Although the IFF variations with the shell model interaction remain of the same order, the magnitude of the SD variation sometimes depends on $q_{max}$. Hence, quantitative studies for a specific experiment should account for the $q$ range adapted to that experiment.

\subsection{Systematic study of integrated form factors}

Nuclear responses were computed for a range of isotopes relevant to DM direct detection: $^{19}$F, $^{23}$Na, $^{28, 29, 30}$Si, $^{40}$Ar, $^{70,72,73,74,76}$Ge, $^{127}$I and $^{128, 129, 130, 131, 132, 134, 136}$Xe. Analytic expressions for the momentum dependent response functions are provided in supplementary material. In the following, IFFs are used to evaluate quantitative variations induced by the choice of the nuclear interaction for each specific model space.  For elements with more than one stable isotope, IFFs have been weighted by the natural abundances of each isotope. Note that isotopes with even numbers of protons and neutrons only probe the $M$ (SI) and $\Phi''$ (LSD) responses as their ground states have spin-parity $0^+$. The results are grouped and discussed with respect to valence space (and therefore interactions) used for the shell model calculations. In all cases the SI ($M_{p,n}$) IFF values are not affected by the choice of interaction and thus we focus the discussion on the subleading channels. Only non-zero IFF values are shown in the figures. All isotopes probe the $M$ and $\Phi''$ channels (as well as their interference), however only isotopes with ground state $J_i \geq 1/2$ can additionally probe the $\Sigma''$, $\Sigma'$ and $\Delta$ channels, with $J_i \geq 1$ isotopes probing $\Tilde{\Phi}'$.

\subsubsection{$^{19}$F, $^{23}$Na, and $^{28,29,30}$Si}

\begin{figure}[htbp]
\centering 
\includegraphics[width=1.\linewidth]{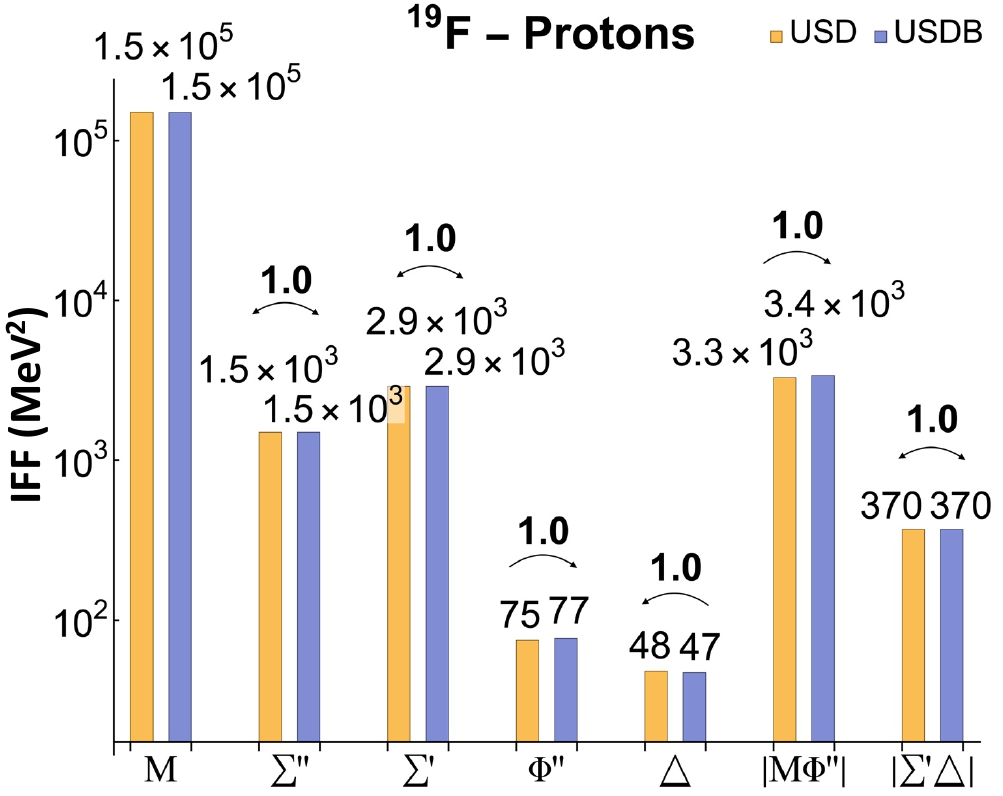}
\caption{Same as Fig.~\ref{127I all IFF pp} for $^{19}$F proton IFF.\label{fig:IFF_19Fp}}
\centering
\end{figure}

\begin{figure}[htbp]
\centering 
\includegraphics[width=1.\linewidth]{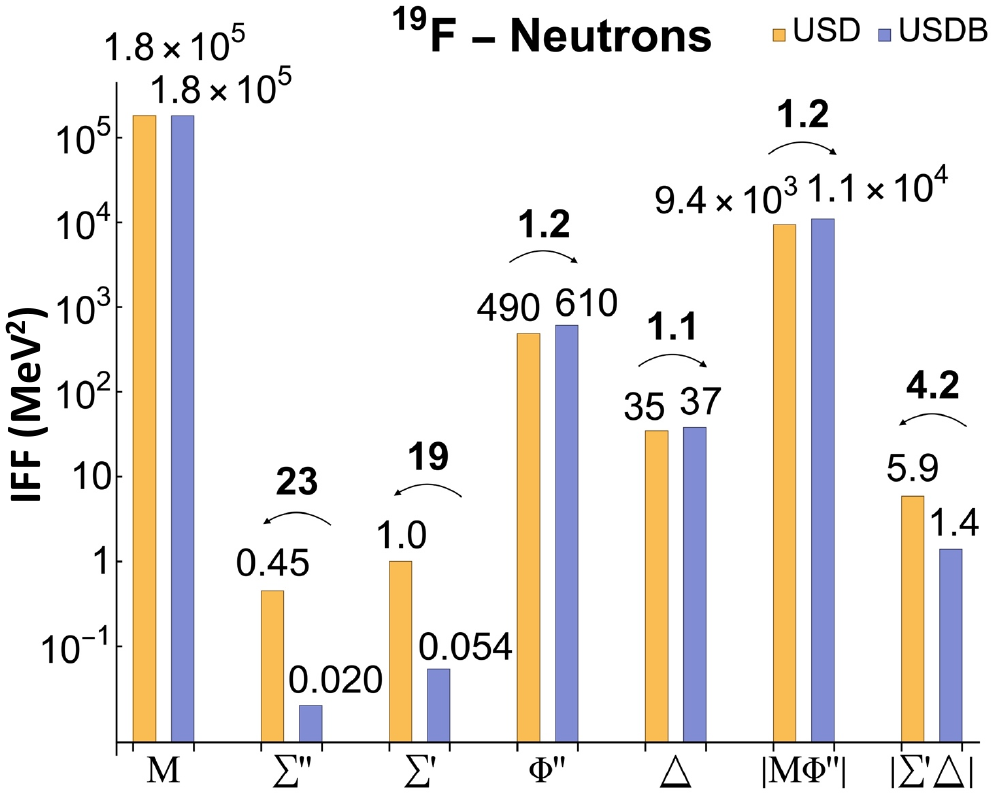}
\caption{Same as Fig.~\ref{127I all IFF pp} for $^{19}$F neutron IFF.\label{fig:IFF_19Fn}}
\centering
\end{figure}

\begin{figure}[htbp]
\centering 
\includegraphics[width=1.\linewidth]{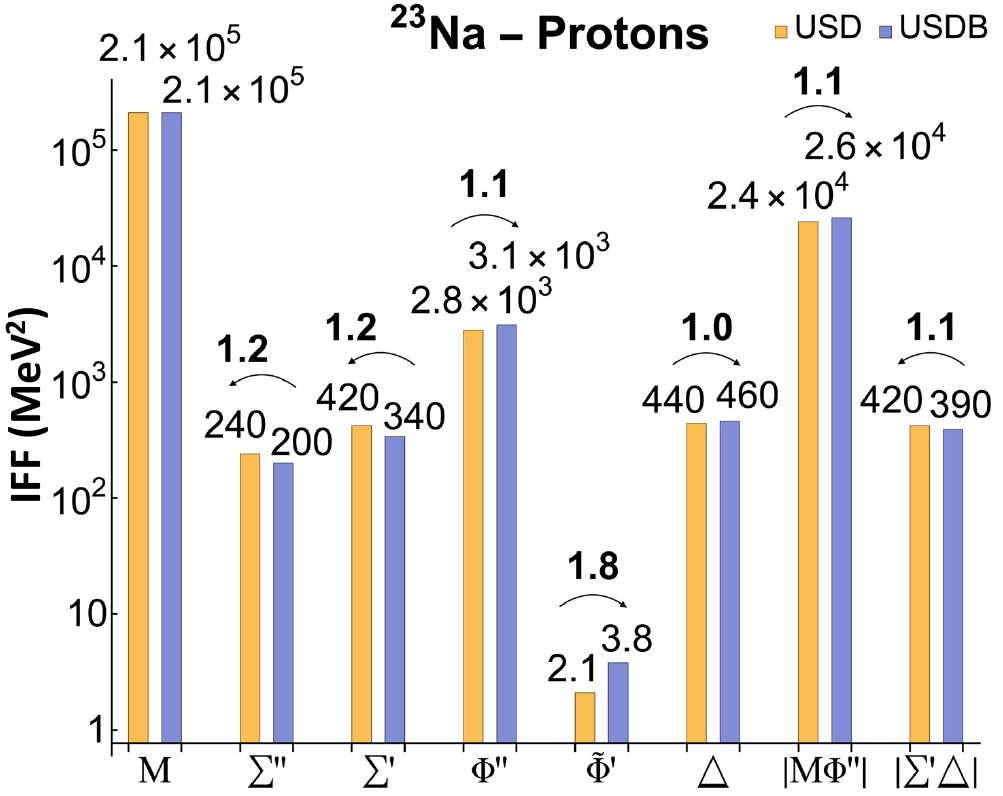}
\caption{Same as Fig.~\ref{127I all IFF pp} for $^{23}$Na proton IFF.\label{fig:IFF_23Nap}}
\centering
\end{figure}

\begin{figure}[htbp]
\centering 
\includegraphics[width=1.\linewidth]{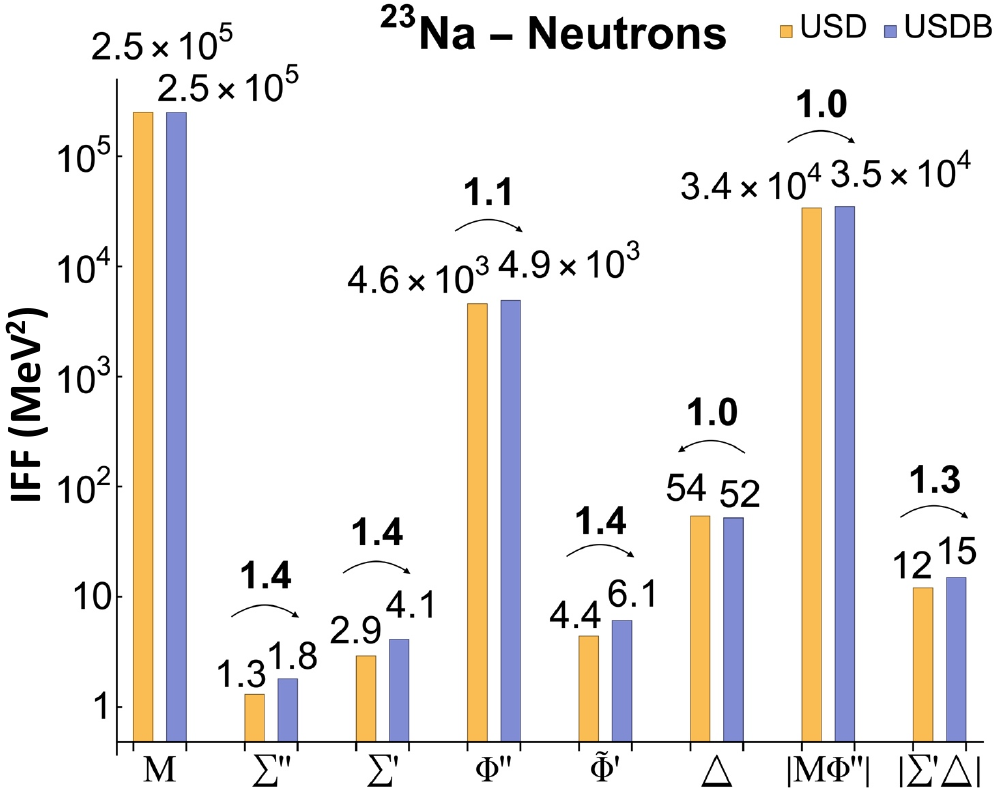}
\caption{Same as Fig.~\ref{127I all IFF pp} for $^{23}$Na neutron IFF.\label{fig:IFF_23Nan}}
\centering
\end{figure}

We consider the same $sd$ model space  for $^{19}$F, $^{23}$Na, and silicon isotopes as described in Sec.~\ref{sec:sd}. Proton and neutron IFF values are shown in Figs.~\ref{fig:IFF_19Fp}, \ref{fig:IFF_19Fn}, \ref{fig:IFF_23Nap} and \ref{fig:IFF_23Nan} for $^{19}$F and $^{23}$Na, respectively. Both $^{19}$F and $^{23}$Na have a subleading spin-dependent proton response due essentially to their unpaired proton, while their spin-dependent neutron responses are orders of magnitude smaller. 
Although both USD and USDB predict similar responses for $\Sigma_p'$ and $\Sigma_p''$ in $^{19}$F, about 20\% differences are found in $^{23}$Na. 

The orbital angular momentum dependent (LD) response $\Delta$ in $^{19}$F and $^{23}$Na is also hindered for neutrons, though this hindrance is not as strong as in the SD case. The IFF for $\Delta$ is maximum for the $^{23}$Na proton response, indicating that this nucleus could be a good candidate to probe this channel. The results for this response are also stable with respect to the interaction, with a variation of only few~\% between USD and USDB predictions. 

The LSD response in this valence space is dominated by the spin-orbit partners $1d_{5/2,3/2}$ and increases with increasing occupation of $1d_{5/2}$ as long as $1d_{3/2}$ remains comparatively lowly occupied (see attached supplementary material). 
As a result, the IFF value of $\Phi''$ in $^{19}$F and  $^{23}$Na increases with the number of valence nucleons (one proton and two neutrons in $^{19}$F, and three protons and four neutrons in $^{23}$Na). 
Here, the impact of the interaction is quite significant, with a maximum variation of $\sim25$\% between USD and USDB in the $^{19}$F neutron response. 

\begin{figure}[htbp]
\centering 
\includegraphics[width=1.0\linewidth]{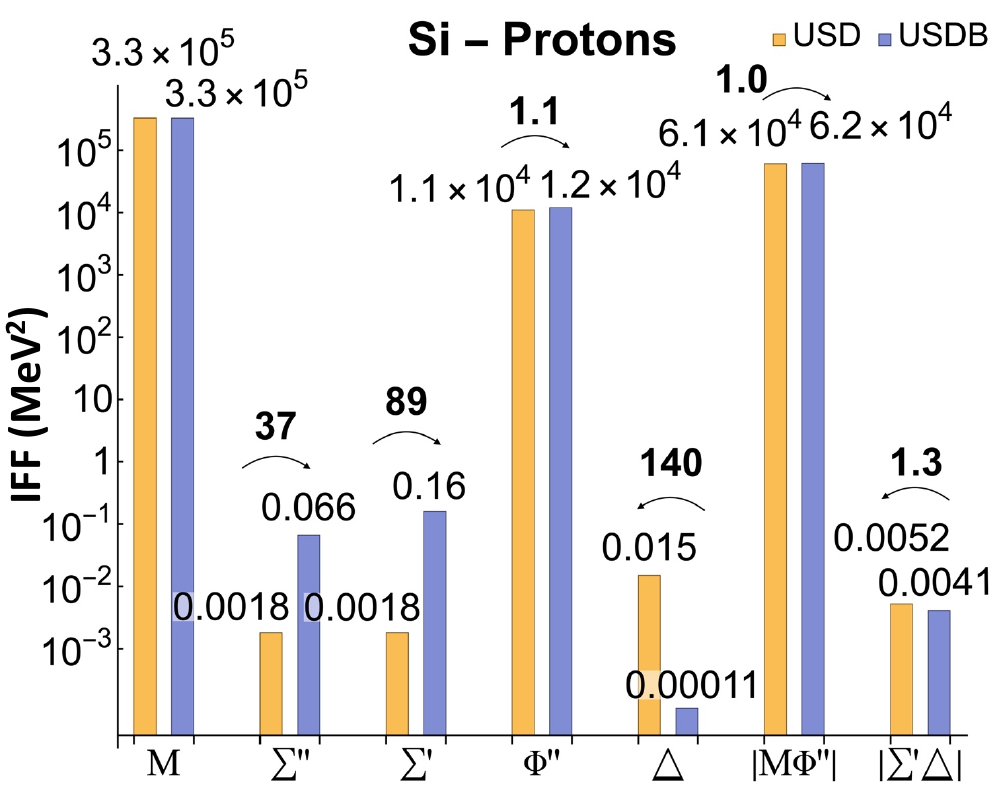}
\caption{Same as Fig.~\ref{127I all IFF pp} for silicon proton IFF.\label{fig:IFF_Sip}}
\centering
\end{figure}

\begin{figure}[htbp]
\centering 
\includegraphics[width=1.0\linewidth]{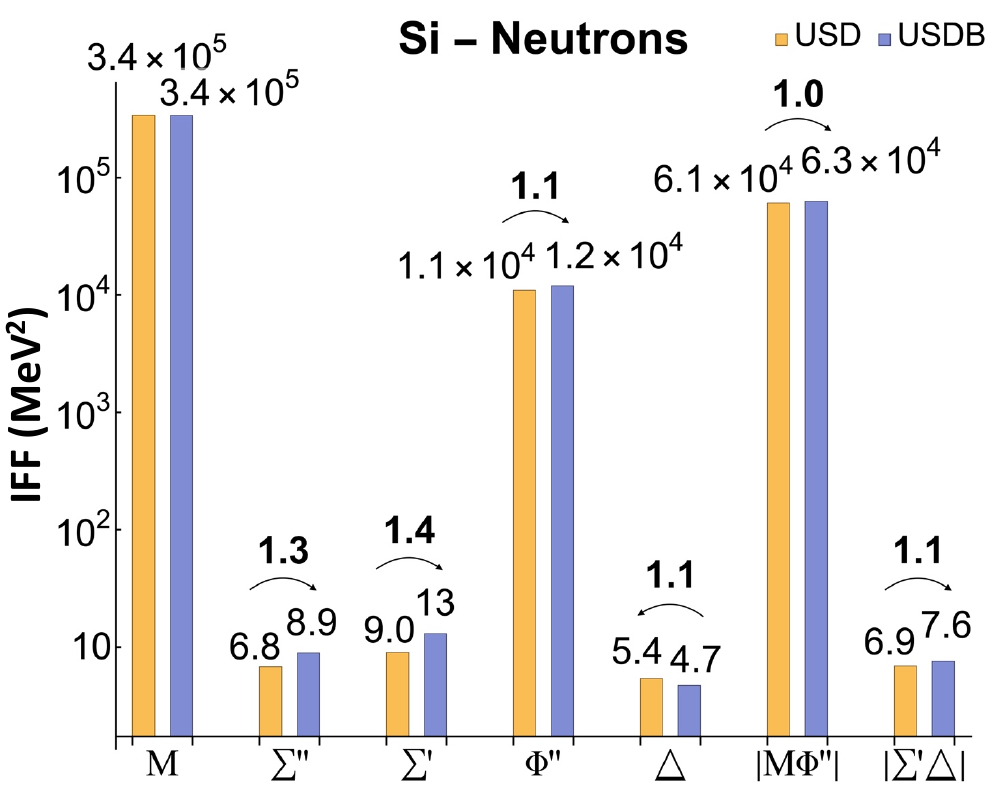}
\caption{Same as Fig.~\ref{127I all IFF pp} for silicon neutron IFF.\label{fig:IFF_Sin}}
\centering
\end{figure}

The IFF values for silicon are shown in Figs.~\ref{fig:IFF_Sip} and~\ref{fig:IFF_Sin} for protons and neutrons, respectively. We note that since $^{28}$Si and $^{30}$Si have ground states $0^+$, they only probe the $M$, $\Phi''$ and $|M\Phi''|$ channels, whereas $^{29}$Si ($1/2^+$) probes all except for $\tilde{\Phi}'$. The only subleading channel is the LSD response $\Phi''$ and its interference with the SI channel $M$. Both interactions predict corresponding IFF values within few \%. These IFF values are similar for both protons and neutrons as the most abundant isotope, $^{28}$Si ($\sim92\%$), has the same number of protons and neutrons. The $\Phi''_{p,n}$ IFF are also larger than in $^{19}$F and $^{23}$Na as  $^{28}$Si has 6 protons and 6 neutrons in the $sd$ valance space, allowing for a configuration in the ground state wave-function with more full occupation of $1d_{5/2}$ while $1d_{3/2}$ remains comparatively poorly occupied (see attached supplementary material), thus maximising the spin-orbit contribution. Silicon detectors should then be optimum for the LSD response in this mass region. Note also that the SD and LD responses are orders of magnitude larger for neutrons than for protons due to the unpaired neutron in $^{29}$Si. However, its   abundance ($\sim 4.7\%$) is too small for these channels to be subleading.

\subsubsection{$^{40}$Ar}

\begin{figure}[htbp]
\includegraphics[width=1\linewidth]{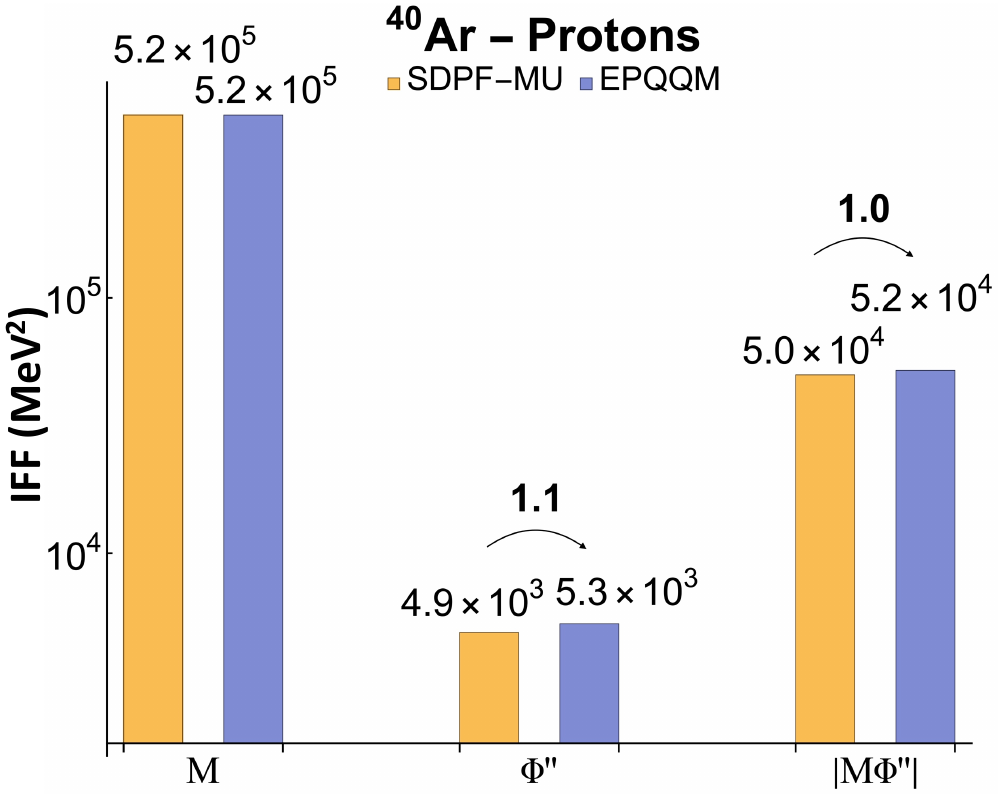}
\includegraphics[width=1\linewidth]{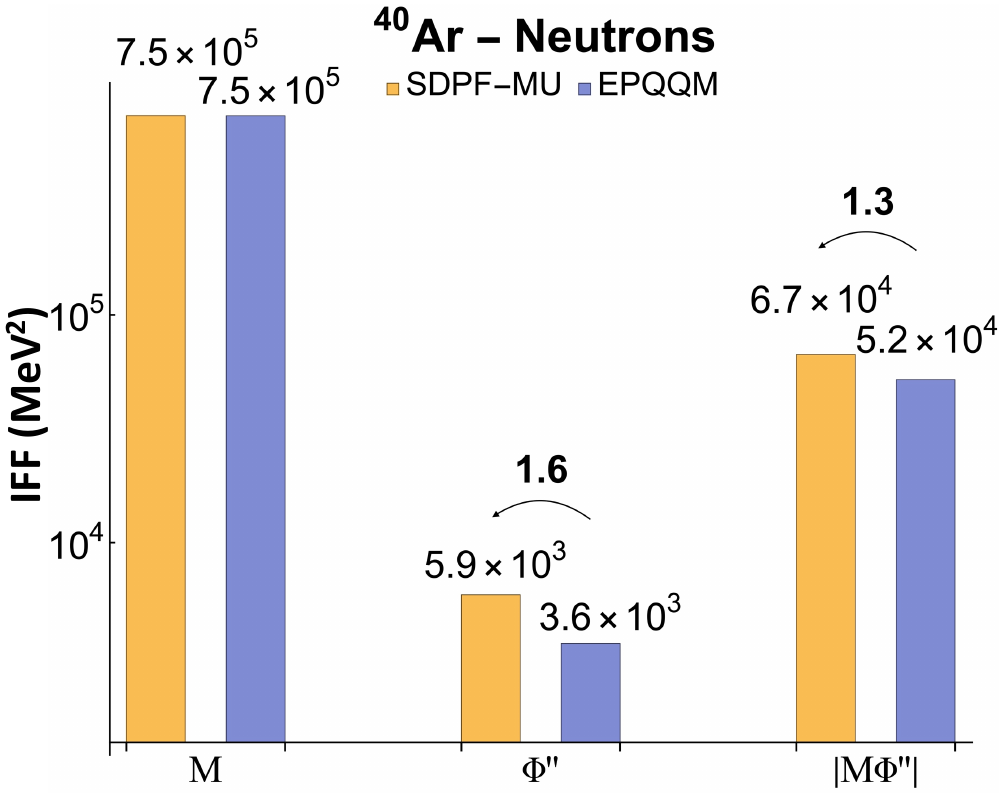}
\caption{Proton (top) and neutron (bottom) IFF values for $^{40}$Ar with the SDPF-MU and EPQQM interactions in the restricted $sdpf$ valence space.}
\label{40Ar pp and nn IFF Plots}
\end{figure}

Shell model calculations for $^{40}$Ar have been performed in the $sdpf$ valence space with single particle levels $1d_{5/2}$, $2s_{1/2}$, $1d_{3/2}$, $1f_{7/2}$, $2p_{3/2}$, $1f_{5/2}$ and $2p_{1/2}$. A valence space truncation has been employed  where the protons are unrestricted in the $sd$ shell and blocked from entering the $pf$ shell, while neutrons fill the $sd$ shell and are unrestricted in the $pf$ shell. Several interactions are available for this model space, including SDPF-NR \cite{Nummela2001SpectroscopyStates}, SDPF-U \cite{Nowacki2009NewSpace}, EPQQM \cite{Kaneko2011Shell-modelNuclei} and SDPF-MU \cite{Utsuno2012ShapeEffect}. Although we performed shell model calculations with all four interactions, we only provide the IFF values for the EPQQM and SDPF-MU interactions in Fig.~\ref{40Ar pp and nn IFF Plots} as they display the largest differences. This isotope was not considered in the work of \cite{Fitzpatrick:2012ix,Anand:2013yka}.

The LSD proton and neutron subleading responses $\Phi''$ are of the same order as that of $^{23}$Na, and about half that of silicon. As for these nuclei, the proton $\Phi''$ response is due to the $1d_{5/2,3/2}$ spin-orbit partners. However, the proton contribution to the ground-state in $^{40}$Ar is now likely to be dominated by a configuration with $1d_{5/2}$ fully occupied and $1d_{3/2}$ only half empty (see attached supplementary material for occupations of all valence levels). The unavoidable partial occupation of $1d_{3/2}$ reduces the $\Phi''$ response as compared to the optimum situation offered by silicon isotopes. For neutrons, both $1d_{5/2,3/2}$ levels are fully occupied and thus no contribution to $\Phi''$ is expected from them. The latter is expected to come from the configuration with neutrons mostly distributed to the $1f_{7/2}$ level with its spin-orbit partner $1f_{5/2}$ comparatively unoccupied.

The absence of SD and LD responses implies that this target could be a good choice to isolate the SI and LSD responses. 
However, large variations are observed for the $\Phi''$ subleading neutron response as well as in its interference with the SI response. 
In the proton case, the differences for all operators between all four interactions are much smaller as they do not exceed $10\%$.

\subsubsection{$^{70, 72, 73, 74, 76}$Ge}

Shell model calculations for germanium isotopes were performed in the 
$f_5pg_9$ model space composed of the single-particle levels $2p_{3/2}$,  $1f_{5/2}$, $2p_{1/2}$ and $1g_{9/2}$. The GCN2850 interaction \cite{Menendez2009DisassemblingDecay} was used in \cite{Fitzpatrick:2012ix,Anand:2013yka} with a valence space truncation that consisted of limiting the occupation number of the $1g_{9/2}$ level to no more than two nucleons above the minimum occupation for all isotopes. In this work, we consider an unrestricted $f_5pg_9$ model space 
and employ the JUN45 \cite{Honma2009NewNuclei} and jj44b \cite{Mukhopadhyay2017NuclearCalculations} interactions. 

\begin{figure}[htbp]
\centering 
\includegraphics[width=1\linewidth]{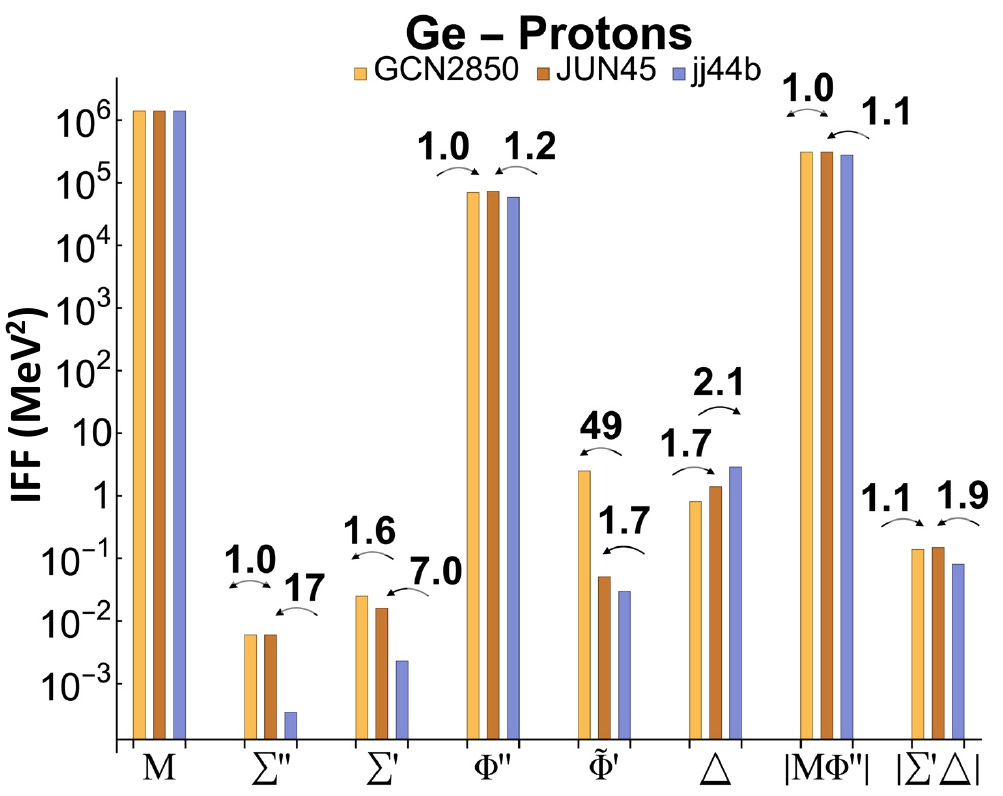}
\caption{Same as Fig.~\ref{127I all IFF pp} for germanium proton IFF.}
\label{Gep}
\centering
\end{figure}

\begin{figure}[htbp]
\centering 
\includegraphics[width=1\linewidth]{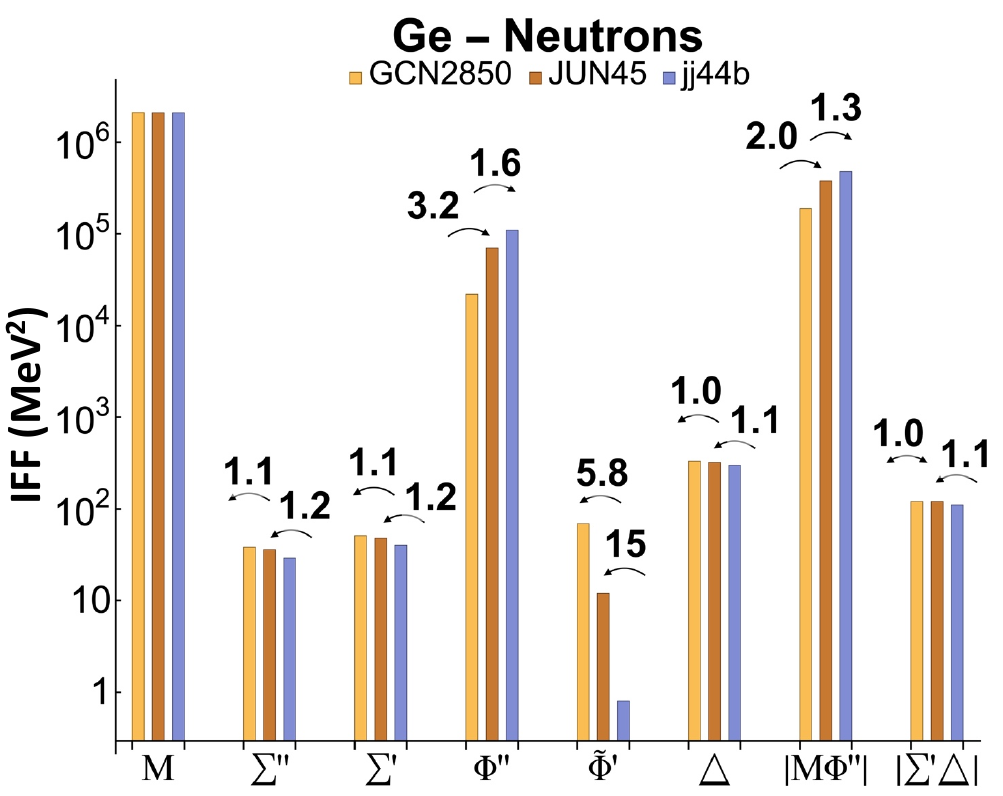}
\caption{Same as Fig.~\ref{127I all IFF pp} for germanium neutron IFF.}
\label{Gen}
\centering
\end{figure}

The isotopic IFF values shown in Figs.~\ref{Gep} and~\ref{Gen} for protons and neutrons, respectively, are weighted according to natural abundance. $^{73}$Ge is the only  stable odd  isotope and thus the only one to contribute to the $\Sigma'$, $\Sigma''$, $\Delta$, and $\tilde{\Phi}'$  responses. Due to its small abundance ($7.8\%$), these responses remain relatively small, except for the neutron LD response $\Delta_n$ whose IFF is of  the same order as, e.g., $\Delta_p$ in $^{23}$Na. The large $\Delta_n$ response in $^{73}$Ge is likely to be due to a strong contribution of the configuration with a partially occupied $1g_{9/2}$ level (with an orbital angular momentum of $4\hbar$) in the ground state (see attached supplementary material for occupations of all valence levels). 

The subleading  proton LSD response is likely to be due to the $1f_{7/2,5/2}$ spin-orbit partners, with $1f_{7/2}$ fully occupied and $1f_{5/2}$ with a small occupation in the ground states. The interpretation of the neutron LSD response is more complicated as several isotopes contribute with similar abundances. In a single-particle picture, one would expect a small contribution as $1f_{5/2}$ would be full, the $2p_{1/2,3/2}$ spin-orbit partners only have a small angular momentum, and $1g_{9/2}$ only starts getting populated from $^{73}$Ge. However, the single-particle picture is a crude approximation for mid-shell nuclei whose ground-states are expected to be composed of mixed configurations. 
In particular, configurations in which $1f_{5/2}$ is not full or in which $1g_{9/2}$ has a non-zero occupation (while its spin-orbit partner $1g_{11/2}$ remains empty as it lies outside of the valence space) all contribute to $\Phi''$.

As a result, the neutron LSD response is of the same order of magnitude as the proton one. However, uncertainties are much larger for neutrons (about a factor 5 between GCN2850 and jj44b) than for protons where IFF predictions from different interactions vary by about $20\%$.

\subsubsection{$^{127}$I and $^{128, 129, 130, 131, 132, 134, 136}$Xe}

In addition to $^{127}$I that is discussed in detail in Sec.~\ref{sec:127I}, shell model calculations have been performed with the  SN100PN interaction  \cite{Brown2005Magnetic132Sn} and within the same model space (see Sec.~\ref{sec:SM127I}) for stable xenon isotopes with isotopic abundance greater than $1\%$. These isotopes have been studied in \cite{Fitzpatrick:2012ix,Anand:2013yka} with the B\&D interaction \cite{Baldridge1978Shell-modelCases}. Their work considered unrestricted $^{134}$Xe and $^{136}$Xe calculations, and their truncation for $^{128,130,132}$Xe was identical to that of their $^{127}$I calculation. Their truncation for $^{129}$Xe and $^{131}$Xe  was further restricted by limiting the valence protons to the $2d_{5/2}$ and $1g_{7/2}$ levels whilst requiring neutrons to fully occupy these levels. 

Here, we performed unrestricted calculations for $^{131, 132, 134, 136}$Xe, with $^{129}$Xe and $^{130}$Xe employing the same truncation as our original $^{127}$I SN100PN calculation in Sec.~\ref{sec:SM127I}. To make the $^{128}$Xe calculation more feasible we further restrict the neutron valence space, by keeping the $2d_{3/2}$ and $3s_{1/2}$ levels unrestricted whilst completely filling the $1g_{7/2}$ and $2d_{5/2}$ levels, with a maximum of 8 neutrons in $1h_{11/2}$.

\begin{figure}[htbp]
\centering 
\includegraphics[width=1\linewidth]{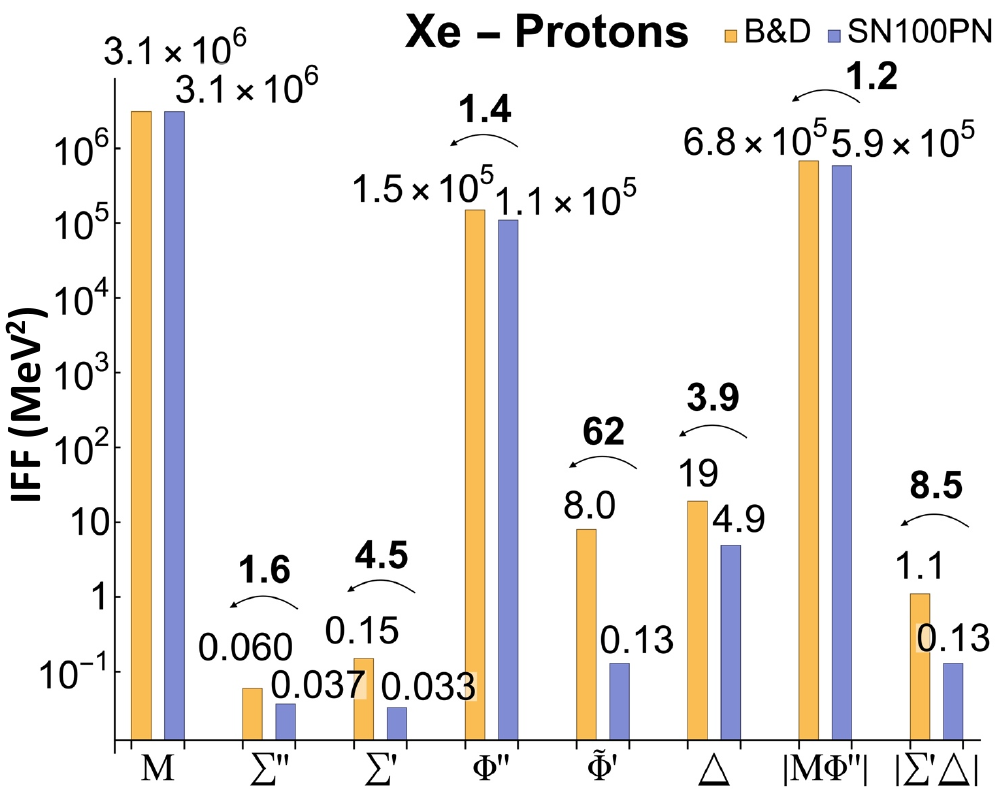}
\caption{{Same as Fig.~\ref{127I all IFF pp} for xenon proton IFF.}}
\label{Xep}
\centering
\end{figure}

\begin{figure}[htbp]
\centering 
\includegraphics[width=1\linewidth]{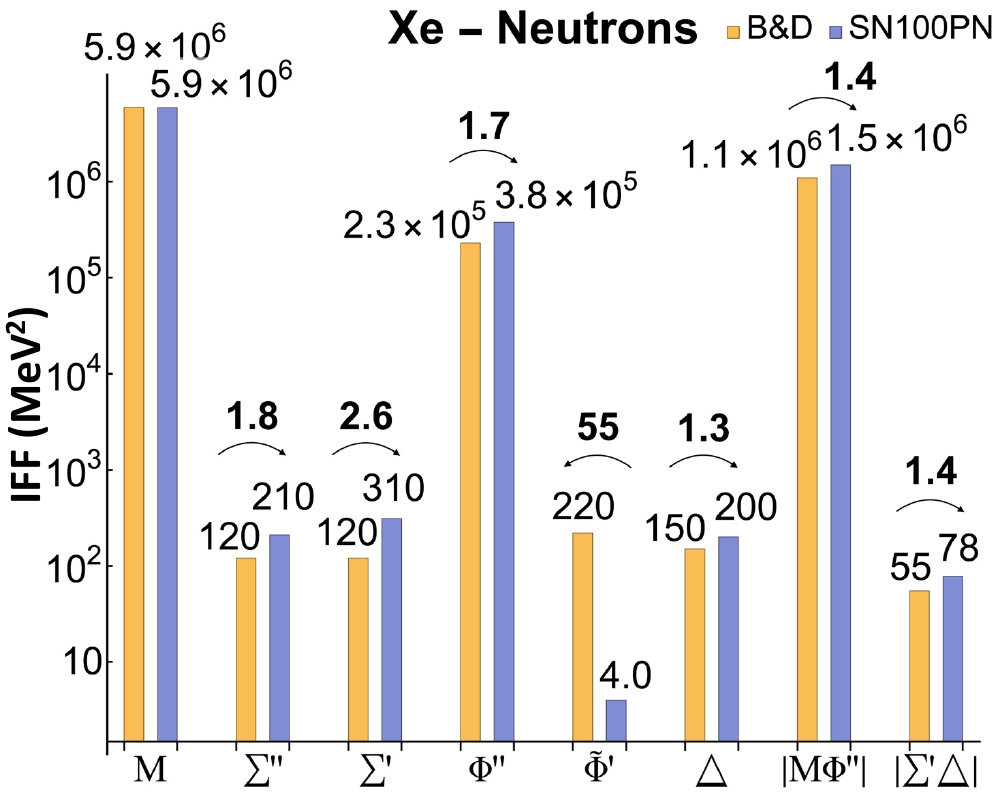}
\caption{Same as Fig.~\ref{127I all IFF pp} for xenon neutron IFF.}
\label{Xen}
\centering
\end{figure}

The xenon IFF values are shown in Fig.~\ref{Xep} and~\ref{Xen} for protons and neutrons, respectively. Comparing with $^{127}$I IFF in Figs.~\ref{127I all IFF pp} and~\ref{127I all IFF nn}, we see that SI ($M$) and LSD ($\Phi''$) responses are of the same order due to their proximity in the nuclear chart. However, SD ($\Sigma'$ and $\Sigma''$) and LD ($\Delta$) responses are significantly different between both elements. Indeed, in $^{127}$I they are dominated by the unpaired proton, while for xenon they are essentially induced by the unpaired neutrons in odd isotopes. Detectors with iodine and those with xenon  are then complementary to investigate the proton and neutron SD and LD cross-sections.

However, strong variations are observed with respect to the interactions for all but the SI responses. The $\Phi''$ proton responses vary by a factor $\sim2$ in $^{127}$I ($\sim40\%$ in xenon) between B\&D and SN100PN predictions. These variations are even larger for $\Phi''$ neutron responses (factor $\sim6$ in $^{127}$I and $\sim65\%$ in xenon). SD proton responses in $^{127}$I vary by factors $\sim2-3$ with the interaction. Similar variations are observed in SD neutron responses in xenon. Variations for LD responses are somewhat smaller ($\sim20-40\%$ except for $\Delta_p$ in xenon which has a comparatively small IFF value).

\section{Discussion and Conclusion} \label{Discussion}

Nuclear shell model calculations have been performed with the NuShellX solver for isotopes relevant to direct detection experiments. The effect of nuclear structure on WIMP-nucleus elastic scattering is studied within the non-relativistic effective field theory (NREFT) regime following \cite{Fitzpatrick:2012ix,Anand:2013yka}. Integrated form factor (IFF) values were used as a proxy to evaluate nuclear response strength and the ability of the nuclei to interact via the six NREFT operators. 

Except for the leading standard spin-independent (SI) response, WIMP-nucleus scattering may be very sensitive to nuclear structure. 
The momentum and spin dependent (LSD) $\Phi''$ response is related to spin-orbit structures in the long wavelength limit and is maximum when one spin-orbit partner is fully occupied while the other is empty. 
It is expected to be more significant in heavier nuclei due to large spin-orbit splitting, larger orbital angular momentum $l$ values, and higher degeneracies of the spin-orbit partners allowing for more nucleons to contribute coherently. The $\Sigma''$ and $\Sigma'$ operators lead to the usual spin-dependent (SD) responses, and hence are more sensitive to isotopes with an odd number of protons (such as $^{19}$F, $^{23}$Na and $^{127}$I) or neutrons (e.g., odd isotopes of germanium and xenon). The $\Delta$ operator is $l$-dependent (in the long wavelength limit) and is larger for isotopes with unpaired nucleons in higher $l$ orbits, such as $^{23}$Na and $^{127}$I for protons, and odd isotopes of germanium and xenon for neutrons. This dependence of the strength of the nuclear responses on the ground-state structure of each isotope means that different experimental efforts with various detection materials can probe different aspects of the WIMP-nucleus interaction.

The range of studied nuclei spans several standard valence spaces used in shell model calculations. These include $sd$ (F, Na, Si), $sdpf$ (Ar), $f_5pg_9$ (Ge) valence spaces, as well as the valence space with orbits in the major shell between 50 and 82 magic numbers for iodine and xenon. Several interactions are available in each valence space that are usually obtained from different fitting protocols. In addition, valence space restrictions are sometimes used to make the shell model calculations more feasible. Various nuclear shell model interactions were then used for each valence space considered to evaluate uncertainties from ground state wave-functions on nuclear responses. These sometimes lead to significant variations in subleading nuclear responses, usually increasing with the number of nucleons. Although the variations for the LD ($\Delta$) subleading responses remain relatively small (from $\lesssim 10\%$ in $sd$ and $f_5pg_9$ nuclei to $\sim30\%$ in $^{127}$I and Xe), these are more significant for SD responses ($\Sigma'$ and $\Sigma''$), going from $\lesssim20\%$ in $sd$ nuclei up to a factor $\sim3$ in $^{127}$I and Xe. Even larger variations are found in LSD ($\Phi''$) subleading responses, going from $\lesssim25\%$ ($sd$) and  $\lesssim65\%$ ($sdpf$), up to factors $\lesssim5$ ($f_5pg_9$) and $\lesssim6$ ($^{127}$I and Xe). These uncertainties should be taken into account when determining possible parameter spaces of NREFT operators from comparison with experiment (see, e.g., \cite{Cirelli:2013ufw}). \\

\begin{acknowledgements}
Valuable discussions with J. Newstead are acknowledged. This research was supported by the Australian Government through the Australian Research Council Centre of Excellence for Dark Matter Particle Physics (CDM, CE200100008). 
\end{acknowledgements}

\section*{Appendices}
 
\appendix

\section{EFT \& DM-Nucleus Elastic Scattering Formalism}\label{formalism appendix}

The general expression for the DM-nucleus scattering amplitude is given by \cite{Fitzpatrick:2012ix,Anand:2013yka}

\begin{widetext}
\begin{equation} \label{General transition amplitude formula}
\begin{split}
      & \frac{1}{2J_i+1} \sum_{M_i, M_f} \big| \langle J_i M_f  | \mathcal{H}_{\text{int}} |  J_i M_i \rangle \big|^2= \frac{4\pi}{2J_i+1} \Bigg[  \sum_{J=1, 3, ...}^{\infty} |\langle J_i || \ \vec{l}_5 \cdot \hat{q} \ \Sigma''_J (q) \ || J_i \rangle |^2 \\
      & \hspace{8mm} + \sum_{J=0,2,...}^{\infty} \bigg\{|\langle J_i || \ l_0 \ M_J (q) \ || J_i \rangle |^2 + |\langle J_i || \ \vec{l}_E \cdot \hat{q} \ \frac{q}{m_N} \Phi''_J (q) \ || J_i \rangle |^2 \\
      & \hspace{10mm} + 2 \text{Re} \left[  \langle J_i || \ \vec{l}_E \cdot \hat{q} \ \frac{q}{m_N} \Phi''_J (q) \ || J_i \rangle \langle J_i || \ l_0 \ M_J (q) \ || J_i \rangle^* \right] \bigg\}\\
      & \hspace{-5mm} + \frac{q^2}{2m_N^2}  \sum_{J=2, 4,...}^{\infty} \left( \langle J_i || \ \vec{l}_E \ \tilde{\Phi}_J' (q) \ || J_i \rangle \cdot \langle J_i || \ \vec{l}_E \ \tilde{\Phi}_J' (q) \ || J_i \rangle^*  - |\langle J_i || \ \vec{l}_E \cdot \hat{q} \ \tilde{\Phi}_J' (q) \ || J_i \rangle|^2 \right)\\
      &  \hspace{-5mm} + \sum_{J=1,3,...}^\infty \bigg\{ \frac{q^2}{2m_N^2} \left( \langle J_i || \ \vec{l}_M \ \Delta_J (q) \ || J_i \rangle \cdot \langle J_i || \ \vec{l}_M \ \Delta_J (q) \ || J_i \rangle^*  - |\langle J_i || \ \vec{l}_M \cdot \hat{q} \ \Delta_J (q) \ || J_i \rangle|^2 \right) \\
      &  + \frac{1}{2} \left( \langle J_i || \ \vec{l}_5 \ \Sigma'_J (q) \ || J_i \rangle \cdot \langle J_i || \ \vec{l}_5 \  \Sigma'_J (q) \ || J_i \rangle^*  - |\langle J_i || \ \vec{l}_5 \cdot \hat{q} \ \Sigma'_J (q) \ || J_i \rangle|^2 \right) \\
      & \hspace{8mm}+ \text{Re} \left[ i\hat{q} \cdot \langle J_i || \ \vec{l}_M \frac{q}{m_N} \Delta_J (q) \ || J_i \rangle \times \langle J_i || \ \vec{l}_5 \ \Sigma'_J (q) \ || J_i \rangle^*  \right]  \bigg\} \Bigg],
\end{split}
\end{equation}
\end{widetext}

where $m_N$ is the nucleon mass, we have averaged over initial nuclear spins and summed over final ones, and the four DM scattering amplitudes $l_j \equiv l_{0,E,M,5}$ are given by

\begin{widetext}
\begin{equation}
\begin{split}
        l_0 &= (c_1^0 +c_1^1 \tau_3) - i(\vec{q} \times \vec{S}_\chi) \cdot \vec{v}_T^\perp (c_5^0 +c_5^1 \tau_3) + \vec{S}_\chi \cdot \vec{v}_T^\perp (c_8^0 +c_8^1 \tau_3) +i\vec{q}\cdot\vec{S}_\chi (c_{11}^0 +c_{11}^1 \tau_3) \\
        \vec{l}_5 &= \frac{1}{2} \bigg[ i\vec{q}\times\vec{v}_T^\perp (c_3^0 +c_3^1 \tau_3) +\vec{S}_\chi (c_4^0 +c_4^1 \tau_3) + \vec{S}_\chi\cdot\vec{q} \ \vec{q} \ (c_6^0 +c_6^1 \tau_3) + \vec{v}_T^\perp (c_7^0 +c_7^1 \tau_3)  \\
        & + i\vec{q} \times \vec{S}_\chi (c_9^0 +c_9^1 \tau_3) + i\vec{q} \ (c_{10}^0 +c_{10}^1 \tau_3) + \vec{v}_T^\perp \times \vec{S}_\chi (c_{12}^0 +c_{12}^1 \tau_3) \\
        & + i\vec{q} \ \vec{v}_T^\perp \cdot\vec{S}_\chi (c_{13}^0 +c_{13}^1 \tau_3) + i\vec{v}_T^\perp \  \vec{q}\cdot\vec{S}_\chi (c_{14}^0 +c_{14}^1 \tau_3) + \vec{q}\times\vec{v}_T^\perp \ \vec{q}\cdot\vec{S}_\chi (c_{15}^0 +c_{15}^1 \tau_3) \bigg]  \\
        \vec{l}_M &= i\vec{q}\times\vec{S}_\chi (c_5^0 +c_5^1 \tau_3) - \vec{S}_\chi (c_8^0 +c_8^1 \tau_3) \\
        \vec{l}_E &= \frac{1}{2} \bigg[ \vec{q} \ (c_3^0 +c_3^1 \tau_3) + i\vec{S}_\chi (c_{12}^0 +c_{12}^1 \tau_3) - \vec{q} \times \vec{S}_\chi (c_{13}^0 +c_{13}^1 \tau_3) - i \vec{q} \ \Vec{q}\cdot \vec{S}_\chi (c_{15}^0 +c_{15}^1 \tau_3) \bigg]. 
\end{split}
\end{equation}
\end{widetext}

Here, $\vec{v}_T^\perp = \frac{1}{2} (\vec{v}_{\chi, \text{in}} + \vec{v}_{\chi, \text{out}} - \vec{v}_{T, \text{in}} - \vec{v}_{T, \text{out}}) = \vec{v}_T+\frac{\vec{q}}{2\mu_T}$, and $\vec{v}_{T, \text{in(out)}} = \frac{1}{A} \sum_{j=1}^A \vec{v}_{N, \text{in(out)}} (j)$. \\ \\

The Wigner-Eckart theorem has been used to express matrix elements in reduced form, using the convention 

\begin{equation}
\begin{split}
    & \langle j' m' | T_{JM}  | j m \rangle =  \\
        & (-1)^{j'-m'} \begin{pmatrix}
         \vspace{2mm} j'  &  J  &  j \\
         \vspace{2mm} -m' &  M &  m  \hspace{2mm}
         \end{pmatrix} \langle j' || T_J || j \rangle.
\end{split}
\end{equation}

The nuclear operators present in Eq.~(\ref{General transition amplitude formula}) have the form 
\begin{widetext}
\begin{equation}
\begin{split}
    \Delta_{JM} (q\vec{x}) & \equiv \vec{M}_{J J}^M (q \vec{x}) \cdot  \frac{1}{q} \vec{\nabla}, \\
    \Sigma'_{JM} (q \vec{x}) & \equiv -i \left[ \frac{\vec{\nabla}}{q} \times \vec{M}^M_{J J} (q \vec{x}) \right] \cdot \vec{\sigma}_N = [J]^{-1} \left[-\sqrt{J} \ \vec{M}^M_{J J+1} (q \vec{x}) +\sqrt{J+1} \ \vec{M}^M_{J J-1} (q \vec{x}) \right] \cdot \vec{\sigma}_N ,\\
    \Sigma''_{JM} (q \vec{x}) & \equiv \left[ \frac{\vec{\nabla}}{q} M_{JM} (q \vec{x}) \right] \cdot \vec{\sigma}_N = [J]^{-1} \left[\sqrt{J+1} \ \vec{M}^M_{J J+1} (q \vec{x}) +\sqrt{J} \ \vec{M}^M_{J J-1} (q \vec{x}) \right]\cdot \vec{\sigma}_N,\\
    \tilde{\Phi}'_{JM} (q \vec{x}) & \equiv \left( \frac{\vec{\nabla}}{q} \times \vec{M}_{JJ}^{M} (q \vec{x}) \right)\cdot \left(\vec{\sigma}_N \times \frac{1}{q} \vec{\nabla} \right) + \frac{1}{2} \vec{M}_{JJ}^{M} (q \vec{x}) \cdot \vec{\sigma}_N, \\
    \Phi''_{JM} (q \vec{x}) & \equiv i \left(\frac{\vec{\nabla}}{q} M_{JM} (q\vec{x}) \right)\cdot \left( \vec{\sigma}_N \times \frac{1}{q} \vec{\nabla} \right), 
\end{split}
\end{equation} 
\end{widetext}
where $M_{JM} (q\vec{x}) \equiv j_J(qx) Y_{JM} (\Omega_x)$, $\hspace{1mm}$ $\vec{M}_{JL}^{M} \equiv j_L(qx) \vec{Y}_{JLM}$, $[J]=\sqrt{2J+1}$, and $\vec{\sigma}_N$ is the nucleon spin operator. Here, $j_J(qx)$ is the spherical Bessel function, $Y_{JM}$ is the spherical harmonic, and $ \vec{Y}_{JLM}$ is the vector spherical harmonic. The following spherical harmonic and vector spherical harmonic identities are also employed in the derivation 

\begin{widetext}
\begin{equation}\label{partial wave expansions}
\begin{split}
     e^{i \vec{q}. \vec{x}_i} & = \sum_{J=0}^\infty \sqrt{4\pi} \ [J] \ (i)^J j_J (qx_i) Y_{J0} (\Omega_{x_i}), \\
      \hat{e}_{\lambda} e^{i \vec{q}. \vec{x}_i} & =
     \begin{cases}
        \displaystyle \sum_{J =0}^\infty \sqrt{4\pi} \ [J] \ (i)^{J-1} \ \frac{\vec{\nabla}_i}{q} \left(j_{J} (qx_i) Y_{J0} (\Omega_{x_i}) \right), \hspace{5mm} \text{for} \ \lambda= 0\\
       \hspace{5mm}\\
        \displaystyle \sum_{J \geq 1}^\infty \sqrt{2\pi} \ [J] \ (i)^{J-2}  \left[\lambda j_{J} (qx_i) \vec{Y}_{JJ1}^\lambda (\Omega_{x_i}) + \frac{\vec{\nabla}_i}{q} \times \left(j_{J} (qx_i) \vec{Y}_{JJ1}^\lambda (\Omega_{x_i}) \right) \right], \hspace{5mm} \text{for} \ \lambda= \pm 1.
    \end{cases}  
\end{split}
\end{equation}
\end{widetext}

\section{NuShellX and OBDMEs}\label{nushellx appendix}

\subsection{Valence OBDMEs}

The OBDME values provided by NuShellX are in proton-neutron (pn) formalism, and have the form 
\begin{equation}
    \frac{\langle J_f ||\left[ a^\dagger _{|\alpha| N} \ \bar{a}_{|\beta| N} \right]_{J} || J_i \rangle }{\sqrt{2J+1}} \equiv \frac{a(NN)}{\sqrt{2J+1}},
\end{equation}
where $\bar{a}_{\beta}=(-1)^{j_\beta -m_{j_\beta}} \ a _{|\beta|;-m_{j_\beta}}$ and $N=\{p, n\}$. This is converted to isospin formalism to be used with the previous expressions through 
\begin{widetext}
\begin{equation}
\begin{split}
    \Psi^{J;\tau}_{|\alpha|, |\beta|}  & \equiv  \frac{\langle J_i; T \ \vdots \vdots \left[a^\dagger_{|\alpha|} \otimes \tilde{a}_{|\beta|} \right]_{J;\tau} \vdots \vdots \ J_i ; T \rangle}{\sqrt{(2J+1)(2\tau+1)}}  \\
    & = \frac{(-1)^{-\tau -2J} \sqrt{2T+1}}{\sqrt{(2J+1)(2\tau+1)} C^{T M_T}_{\tau 0; T M_T}} \bigg[ C^{\tau 0}_{1/2 \ 1/2; 1/2 \ -1/2}  \ a(pp) -  C^{\tau 0}_{1/2 \ -1/2; 1/2 \ 1/2}  \ a(nn) \bigg],
\end{split}
\end{equation}
\end{widetext}
where $C^{j' m'}_{k \ q; j \ m} \equiv \langle k q \ j m  | j' m' \rangle$ are Clebsch-Gordan (CG) coefficients. 

For $\tau=0$ and $\tau=1$ the conversion becomes 
\begin{equation}
\begin{split}
    \Psi^{J;0}_{|\alpha|, |\beta|} & = \frac{\sqrt{2T+1} \ \left(a(pp) + a(nn) \right)}{\sqrt{2(2J+1)} \ C^{T M_T}_{0 0; T M_T}} ,
\end{split}
\end{equation}
and 
\begin{equation}
\begin{split}
    \Psi^{J;1}_{|\alpha|, |\beta|} & = \frac{\sqrt{2T+1} \  \left(-  a(pp) +  a(nn) \right)}{\sqrt{6(2J+1)} \ C^{T M_T}_{1 0; T M_T}}.
\end{split}
\end{equation}
The coefficients $ \sqrt{2T+1}/ C^{T M_T}_{0 0; T M_T}$ and $\sqrt{2T+1}/C^{T M_T}_{1 0; T M_T}$ are both isospin-dependent, and hence must be calculated separately for each isotope considered. \\

\subsection{Core OBDMEs}

The above OBDMEs provided by NuShellX only describe the valence single-particle orbitals. Equation (\ref{reduced nuclear matrix element}) also needs to be evaluated for orbitals within the filled core. It can be shown that the OBDME expression for core states has the form

\begin{widetext}
\begin{equation}
    \Psi^{J;\tau}_{|\beta|, |\beta|}  = \sum_{M_i,m_{ j_\beta}, m_{t_\beta}}  
          \frac{(-1)^{j_\beta -m_{j_\beta} +t_\beta-m_{t_\beta}} (-1)^{\tau +J} 
 C^{J0}_{j_\beta m_{j_\beta}; j_\beta -m_{j_\beta}}  C^{\tau 0}_{t_\beta m_{t_\beta}; t_\beta -m_{t_\beta}}  \sqrt{2T+1}}{C_{\tau 0; T M_T}^{T M_T} C_{J0; J_i M_i}^{J_i M_i} \sqrt{(2J+1)(2\tau+1)(2J_i+1)}},
\end{equation}
\end{widetext}

where $M_i$ is the projection of the angular momentum $J_i$, and $t_\beta=1/2$.

\section{Mathematica Package \& Density Matrix Syntax}\label{mathematica package appendix}
This work employs the Mathematica package provided by \cite{Anand:2013yka} to calculate the nuclear form factors presented above. Among its various functions, this script computes (in isospin formalism) the nuclear response functions
\begin{equation}\label{isospin W}
    W_{X,Y}^{\tau \tau'} (y) \equiv  \sum_J \langle J_i || X_{J;\tau} (q) || J_i \rangle \langle J_i || Y_{J;\tau'} (q) || J_i \rangle
\end{equation}
\noindent where $W_{X}^{\tau \tau'} (y) \equiv W_{X,X}^{\tau \tau'} (y)$, and $\tau=0 \ (1)$ indicates isospin-independence (dependence). Linear combinations of the functions (\ref{isospin W}) give the nuclear response functions in proton-neutron formalism required in Eq.~(\ref{FF eqn}). These nuclear response functions are given by the script function \texttt{ResponseNuclear[y,i,tau,tau2]} where $i$ takes on values from $1$ to $8$, to give $1)\ W_M$, $2)\ W_{\Sigma''}$, $3)\ W_{\Sigma'}$, $4)\ W_{\Phi''}$, $5)\ W_{\Tilde{\Phi}'}$, $6)\ W_\Delta$, $7)\ W_{M\Phi''}$ and $8)\ W_{\Sigma' \Delta}$.

The package provides these functions for the default density matrices $\Psi^{J,\tau} (|\alpha|, |\beta|)$ employed to calculate the results presented in \cite{Fitzpatrick:2012ix, Anand:2013yka}, for the nuclear isotopes $^{19}$F, $^{23}$Na, $^{28, 29, 30}$Si, $^{70,72,73,74,76}$Ge,  $^{127}$I,  and $^{128, 129, 130, 131, 132, 134, 136}$Xe. These can be called on using \texttt{SetIsotope[Z,A,bFM,filename]} with filename=``default". However, custom density matrices can be loaded into the script through a custom file, whose format must follow that provided in \cite{Anand:2013yka}. We provide these density matrix files as part of the supplementary material in the appropriate format for use with the aforementioned Mathematica package, with density matrix values obtained from our own nuclear shell model calculations for
$^{19}$F, $^{23}$Na, $^{28, 29, 30}$Si,  $^{40}$Ar, $^{70,72,73,74,76}$Ge,  $^{127}$I,  and $^{128, 129, 130, 131, 132, 134, 136}$Xe.

\bibliography{references}

\onecolumngrid

\newpage

{\LARGE
\begin{center}
Supplementary Material
\end{center}
}

We present the supplemental material to the paper, which constitutes the One-Body Density Matrix Elements (OBDMEs) (section \ref{OBDMEs}) and nuclear response functions (section \ref{nuclear responses}) employed in the calculation of the Integrated Form Factor (IFF) values in the aforementioned work, for the dark matter direct detection target nuclei $^{19}$F, $^{23}$Na, $^{28, 29, 30}$Si,  $^{40}$Ar, $^{70,72,73,74,76}$Ge,  $^{127}$I,  and $^{128, 129, 130, 131, 132, 134, 136}$Xe. We also present the proton and neutron occupation numbers of the single-nucleon valence orbitals (section \ref{occupation numbers}). All results employ the nuclear shell model interactions and model spaces mentioned in the main work (where only the original $^{127}$I SN100PN valence space truncation result is shown here). The results of \cite{Fitzpatrick:2012ix, Anand:2013yka} are omitted here, however can be found in these works. All results are for the experimental nuclear ground states of each of the considered nuclei. Files of the OBDMEs suitable for use directly with the Mathematica package \cite{Anand:2013yka} are also provided in \cite{OBDMEFiles}. \\


\section{One-Body Density Matrix Elements (OBDMEs)}\label{OBDMEs}

The OBDMEs  have the form
\begin{equation}
\begin{split}
    \Psi^{J;\tau}_{|\alpha|, |\beta|} & \equiv \frac{\langle J_i; T \ \vdots \vdots \left[a^\dagger_{|\alpha|} \otimes \tilde{a}_{|\beta|} \right]_{J;\tau} \vdots \vdots \ J_i ; T \rangle}{\sqrt{(2J+1)(2\tau+1)}},
\end{split}
\end{equation}
where $\tilde{a}_{\beta }=(-1)^{j_\beta -m_{j_\beta}+1/2-m_{t_\beta}} \ a _{|\beta|;-m_{j_\beta}, -m_{t_\beta}}$ and $\otimes$ denotes a tensor product. Here, $\alpha$ and $\beta$ are single-nucleon states given by the usual quantum numbers $\beta = \{n_\beta, l_\beta, j_\beta, m_{j_\beta}, m_{t_\beta} \}$, with the reduced state notation $|\beta|= \{n_\beta, l_\beta, j_\beta\}$. The nucleon isospin state $t_\beta=t_\alpha=1/2$ is implicit in the notation. The nuclear ground state angular momentum is given by $J_i$, with nuclear isospin denoted by $T$ and its projection $M_T$. Additionally, $\tau=\{0, 1\}$, and $0 \leq J \leq 2J_i$. The tables below are provided for each $(J, \ \tau)$ pair. 


\vspace{5mm}

\subsection{$^{19}$F}

{\centering
\begin{tabular}{|l|c|c|c|c|c|c|}
 \hline
    $J$ &   $N_{out}$ &   $2j_\alpha$ &   $N_{in}$ &   $2j_\beta$ &    $\Psi^{J;0}_{|\alpha|, |\beta|}$ (USDB) & $\Psi^{J;1}_{|\alpha|, |\beta|}$ (USDB) \\
 \hline 
    0 &  2   &   1    &   2 &    1 &     1.178 &   0.3558 \\
  &&    3  & &    3 &     0.1913 &   0.02824 \\
  &&   5 & &    5 &     0.8960 &   0.3489 \\
 \hline 
   1 &   2   &   1    &   2 &    1   &   0.4266 &   -0.39164    \\
 & &    3  & &    1   &   -0.01476  &   -0.01564    \\
 & &   1 & &    3   &   0.01476  &  0.01564    \\
 & &   3 & &    3   &   -0.05634  &   0.01600    \\
 & &   5 & &    3   &   -0.1339 &   0.1051    \\
 & &   3 & &    5   &   0.1339 &   -0.1051    \\
 & &   5 & &    5   &   0.1270 &   -0.2467    \\ 
 \hline
\end{tabular} \par}

\subsection{$^{23}$Na}

{\centering
\begin{tabular}{|l|c|c|c|c|c|c|}
 \hline
    $J$ &    $N_{out}$ &   $2j_\alpha$ &   $N_{in}$ &   $2j_\beta$   &  $\Psi^{J;0}_{|\alpha|, |\beta|}$ (USDB)&$\Psi^{J;1}_{|\alpha|, |\beta|}$ (USDB)\\
 \hline
    0 &  2   &   1    &   2 &    1   &   1.255     &  -0.01582    \\
  &&    3  & &    3   &   0.7097     &  0.2020    \\
  &&   5 & &    5   &   4.411     &  0.6607    \\
 \hline
   1&   2   &   1    &   2 &   1   &   -0.09913    &  0.1080    \\
 & &   3  & &   1   &   -0.02204    &  0.07772    \\
 & &   1 & &   3   &   0.02204    &  -0.07772    \\
 & &   3 & &   3   &   0.1026    &  -0.02887    \\
 & &   5 & &   3   &   -0.03910    &  0.07832    \\
 & &   3 & &   5   &   0.03910    &  -0.07832    \\
 & &   5 & &   5   &   0.4964    &  -0.2961    \\
 \hline
   2 &   2   &   3   &   2 &   1   &   0.1553    &  0.05253    \\
 & &   5   & &   1   &    -0.5832     &  0.09560     \\
 & &   1 & &   3   &   -0.1553    &  -0.05253    \\
 & &   3 & &   3   &   -0.006800    &  0.01220    \\
 & &   5 & &   3   &   -0.2725    &  -0.07176    \\
 & &   1 & &   5   &   -0.5832    &  0.09560    \\
 & &   3 & &   5   &   0.2725    &  0.07177    \\
 & &   5 & &   5   &   -0.5435     &  -0.1109    \\
 \hline
   3 &   2   &   5   &   2 &   1   &   0.03867 &   -0.05911    \\
 & &   3 & &   3   &   0.06285 &   -0.08675    \\
 & &   5 & &   3   &   0.03949 &  -0.07793    \\
 & &   1 & &   5   &   0.03867 &  -0.05911    \\
 & &   3 & &   5   &   -0.03949 &   0.07793    \\
 & &   5 & &   5   &   -0.5306 &   0.5344    \\
 \hline
\end{tabular}\par}

\subsection{$^{28, 29, 30}$Si}

\subsection*{$^{28}$Si}

{\centering
\begin{tabular}{|l|c|c|c|c|c|c|}
 \hline
     $J$  &  $N_{out}$ &   $2j_\alpha$ &   $N_{in}$ &   $2j_\beta$   &  $\Psi^{J;0}_{|\alpha|, |\beta|}$ (USDB)&$\Psi^{J;1}_{|\alpha|, |\beta|}$ (USDB)\\
 \hline 
    0&  2   &   1    &   2 &    1  &    0.7144    &  0    \\
  &&    3  & &    3    &   0.4432     &  0    \\
  &&   5 & &    5    &    2.690     &  0   \\
 \hline
\end{tabular}\par}

\subsection*{$^{29}$Si}

{\centering
\begin{tabular}{|l|c|c|c|c|c|c|}
 \hline
     $J$ &  $N_{out}$ &   $2j_\alpha$ &   $N_{in}$ &   $2j_\beta$   &  $\Psi^{J;0}_{|\alpha|, |\beta|}$ (USDB)&$\Psi^{J;1}_{|\alpha|, |\beta|}$ (USDB)\\
 \hline 
    0&  2   &   1    &   2 &    1   &   1.685     &  0.5045    \\
  &&    3  & &    3   &   0.7740     &  0.1142    \\
  &&   5 & &    5   &   5.901     &  0.1929    \\
 \hline
   1&   2   &   1    &   2 &   1   &   0.6051    &  0.5801    \\
 & &   3  & &   1   &   0.02398    &  0.008040    \\
 & &   1 & &   3   &   -0.02398    &  -0.008040    \\
 & &   3 & &   3   &   -0.08409    &  -0.07685    \\
 & &   5 & &   3   &   0.1772    &  0.1853    \\
 & &   3 & &   5   &   -0.1772    &  -0.1853    \\
 & &   5 & &   5   &   0.1117    &  0.1002    \\
 \hline
\end{tabular}\par}

\subsection*{$^{30}$Si}

{\centering
\begin{tabular}{|l|c|c|c|c|c|c|}
 \hline
     $J$ &  $N_{out}$ &   $2j_\alpha$ &   $N_{in}$ &   $2j_\beta$   &  $\Psi^{J;0}_{|\alpha|, |\beta|}$ (USDB)&$\Psi^{J;1}_{|\alpha|, |\beta|}$ (USDB)\\
 \hline 
    0&  2   &   1    &   2 &    1   &   1.646     &  0.5893    \\
  &&    3  & &    3   &   0.9312     &  0.3867    \\
  &&   5 & &    5   &   5.290     &  0.1605    \\
 \hline
\end{tabular}\par}

\subsection{$^{40}$Ar}

{\centering
\begin{tabular}{|l|c|c|c|c|c|c|c|c|c|}
 \hline
     $J$ &  $\tau$ &    $N_{out}$ &   $2j_\alpha$ &   $N_{in}$ &   $2j_\beta$   & \hspace{1mm} $\Psi^{J;\tau}_{|\alpha|, |\beta|}$ (SDPF-NR) & \hspace{1mm} $\Psi^{J;\tau}_{|\alpha|, |\beta|}$ (SDPF-U) & \hspace{1mm} $\Psi^{J;\tau}_{|\alpha|, |\beta|}$ \hspace{1mm} (SDPF-MU) &  $\Psi^{J;\tau}_{|\alpha|, |\beta|}$ (EPQQM) \\
 \hline 
   0&0&  2   &   1    &   2 &    1  &  4.278 &   4.272 &   4.226 &    4.306   \\
   &&  2 &    3  &   2 &    3    &   4.942 &   4.954 &   4.981 &   4.927   \\
   &&  2 &   5 &   2 &    5    &   7.696 &   7.690 &   7.694 &   7.692   \\
   &&  3   &   1    &   3 &    1  &   0.02680 &   0.02327 &   0.008617 &   0.07809   \\
   &&  3 &    3  &   3 &    3    &   0.09054 &   0.06587 &     0.04396 &   0.3165   \\
   &&  3 &   5 &   3 &    5    &   0.04715 &   0.04554 &   0.01527 &   0.04032   \\
   &&  3 &   7 &   3 &    7    &   0.9998 &   1.020 &   1.069 &   0.8203   \\
 \cline{2-10}
   &1&  2   &   1    &   2 &    1  &    0.1372 &   0.1418 &   0.1742 &   0.1177   \\
   &&  2 &    3  &   2 &    3    &   0.9776 &   0.9690 &   0.9499 &    0.9881   \\
   &&  2 &   5 &   2 &    5    &    0.03550 &   0.03978 &   0.03668 &   0.03814   \\
   &&  3   &   1    &   3 &    1  &   0.01895 &   0.01646 &   0.006093 &   0.05522    \\
   &&  3 &    3  &   3 &    3    &   0.06402 &   0.04658 &   0.03108 &   0.2238    \\
   &&  3 &   5 &   3 &    5    &   0.03334  &   0.03220 &   0.01080 &   0.02851   \\
   &&  3 &   7 &   3 &    7    &   0.7070 &   0.7215 &   0.7562 &   0.5800    \\
 \hline
\end{tabular}
}

\subsection{$^{70,72,73,74,76}$Ge}

\subsection*{$^{70}$Ge}

{\centering
\begin{tabular}{|l|c|c|c|c|c|c|c|c|}
 \hline
    $J$&$   N_{out}$ &   $2j_\alpha$ &   $N_{in}$ &   $2j_\beta$   &  $\Psi^{J;0}_{|\alpha|, |\beta|}$ (jj44b) &  $\Psi^{J;0}_{|\alpha|, |\beta|}$ (JUN45) & $\Psi^{J;1}_{|\alpha|, |\beta|}$ (jj44b) &  $\Psi^{J;1}_{|\alpha|, |\beta|}$ (JUN45) \\
 \hline 
    0 &  3   &   1    &   3 &    1 &     1.757 &   2.063 &   0.4607 &   0.5077\\
  &  3   &    3  &   3   &    3  &   4.307 &   5.141 &   0.9044 &   0.5944\\
  &  3   &   5 &   3   &    5   &   3.862 &   3.773 &   0.8867 &   1.544 \\
  &  4 &   9 &   4 &    9   &   1.753 &   1.164 &   0.8917 &   0.5606 \\
 \hline
\end{tabular}\par}

\subsection*{$^{72}$Ge}

{\centering
\begin{tabular}{|l|c|c|c|c|c|c|c|c|}
 \hline
     $J$&  $N_{out}$ &   $2j_\alpha$ &   $N_{in}$ &   $2j_\beta$ &     $\Psi^{J;0}_{|\alpha|, |\beta|}$ (jj44b) &  $\Psi^{J;0}_{|\alpha|, |\beta|}$ (JUN45)  & $\Psi^{J;1}_{|\alpha|, |\beta|}$ (jj44b) &$\Psi^{J;1}_{|\alpha|, |\beta|}$ (JUN45) \\
 \hline 
    0&  3   &   1    &   3 &    1 &     2.095 &   2.415     &   0.6460 &  0.8628   \\
  &  3   &    3  &   3   &    3 &     5.023 &   5.760    &   1.155 &  0.8589   \\
  &  3   &   5 &   3   &    5 &     4.965 &   4.983    &   1.197 &  1.796   \\
  &  4 &   9 &   4 &    9  &   2.737 &   2.123    &   1.503 &  1.134   \\
 \hline
\end{tabular}\par}

\subsection*{$^{73}$Ge}

{\centering
\begin{tabular}{|l|c|c|c|c|c|c|c|c|}
 \hline
    $J$&  $N_{out}$ &   $2j_\alpha$ &   $N_{in}$ &   $2j_\beta$    &  $\Psi^{J;0}_{|\alpha|, |\beta|}$ (jj44b)&  $\Psi^{J;0}_{|\alpha|, |\beta|}$ (JUN45)& $\Psi^{J;1}_{|\alpha|, |\beta|}$ (jj44b)&$\Psi^{J;1}_{|\alpha|, |\beta|}$ (JUN45)\\
 \hline 
    0&  3   &   1    &   3 &    1   &   7.090 &   8.187    &   2.063 &  3.026   \\
  &  3   &    3  &   3   &    3   &   16.78 &   18.99    &   4.099 &  3.384   \\
  &  3   &   5 &   3   &    5   &   17.19 &   17.66    &   4.091 &  5.944   \\
  &  4 &   9 &   4 &    9   &   10.54 &   8.386    &   6.012 &  4.637   \\
 \hline
   1 &   3   &   1    &   3 &   1 &   0.0449 &   -0.0231   &   0.008192 &  -0.004068   \\
 &   3   &   3    &   3   &   1 &   0.0273 &   -0.0485   &   0.003368 &  0.004153   \\
 &   3   &   1    &   3   &   3 &   -0.0273 &   0.0485   &   -0.003368 &  -0.004153   \\
 &   3   &   3  &   3   &   3 &   0.1594 &   0.0909   &   -0.1070 &  -0.03418   \\
 &   3   &   5  &   3   &   3 &   0.0185 &   -0.0680   &   0.07142 &  0.01100   \\
 &   3   &   3  &   3   &   5 &   -0.0185 &   0.0680   &   -0.07142 &  -0.01102   \\
 &   3   &   5 &   3   &   5 &   0.3357 &   0.2423   &   -0.05665 &  -0.04609   \\
 &   4 &   9 &   4 &   9 &   2.022 &   2.091   &   1.248 &  1.290   \\
 \hline
   2 &   3   &   3    &   3 &   1 &   0.2965 &   0.3800   &   -0.2281 &  -0.08411   \\
 &   3   &   5    &   3   &   1 &   0.2257 &   0.4936   &   -0.1994 &  -0.04444   \\
 &   3   &   1    &   3   &   3 &   -0.2965 &   -0.3800   &   0.2281 &  0.08414   \\
 &   3   &   3  &   3   &   3 &   -0.01697 &   0.5321   &   -0.1507 &  -0.1437   \\
 &   3   &   5  &   3   &   3 &   -0.07764 &   -0.3562   &   0.005852 &  0.08452   \\
 &   3   &   1  &   3   &   5 &   0.2258 &   0.4936   &   -0.1995 &  -0.04450   \\
 &   3   &   3  &   3   &   5 &   0.07764 &   0.3562   &   -0.005852 &  -0.08453   \\
 &   3   &   5 &   3   &   5 &   0.3397 &   0.7086   &   0.2994 &  0.005110   \\
 &   4 &   9 &   4 &   9 &   0.3582 &   0.8901   &   0.2387 &  0.5209   \\
 \hline
   3 &   3   &   5    &   3 &   1 &   -0.02144 &   0.01769   &   -0.01052 &  0.005295   \\
 &   3   &   3  &   3   &   3 &   -0.03613 &   0.03835   &   -0.04233 &  -0.02508   \\
 &   3   &   5  &   3   &   3 &   0.02337 &   -0.04199   &   0.02636 &  0.02612   \\
 &   3   &   1  &   3   &   5 &   -0.02144 &   0.01769   &   -0.01052 &  0.005295   \\
 &   3   &   3  &   3   &   5 &   -0.02337 &   0.04197   &   -0.02636 &  -0.02610   \\
 &   3   &   5 &   3   &   5   &   0.09816 &   0.09492   &   -0.01733 &  -0.06010   \\
 &   4 &   9 &   4 &   9   &   1.616 &   1.810   &   1.020 &  1.134   \\
 \hline
   4&   3 &   5  &   3 &   3 &   0.07645 &   -0.0604   &   -0.1517 &  -0.08442   \\
 &   3   &   3  &   3   &   5 &   -0.07647 &   0.06044   &   0.1517 &  0.08442   \\
 &   3   &   5 &   3   &   5 &   -0.01813 &   0.05313   &   0.2104 &  0.1628   \\
 &   4 &   9 &   4 &   9 &   -0.05485 &   0.4122   &   -0.02017 &  0.2783   \\
 \hline
   5 &   3 &   5 &   3 &   5 &   0.02898 &   0.01724   &   -0.006537 &  -0.006894   \\
 &   4 &   9 &   4 &   9 &   1.553 &   1.757   &   0.9894 &  1.116   \\
 \hline
   6 &   4 &   9 &   4 &   9 &   0.3232 &   0.7346   &   0.2033 &  0.4718   \\
 \hline
   7 &   4 &   9 &   4 &   9 &   1.582 &   1.782   &   1.008 &  1.135   \\
 \hline
   8 &   4 &   9 &   4 &   9 &   
 0.1087 &   0.5359   &   0.06612 &  0.3397   \\
 \hline
   9&   4 &   9 &   4 &   9 &   1.754 &   1.901   &   1.117 &  1.209   \\
 \hline
\end{tabular}\par}

\subsection*{$^{74}$Ge}

{\centering
\begin{tabular}{|l|c|c|c|c|c|c|c|c|}
 \hline
     $J$&  $N_{out}$&   $2j_\alpha$ &   $N_{in}$ &   $2j_\beta$ &    $\Psi^{J;0}_{|\alpha|, |\beta|}$ (jj44b)&  $\Psi^{J;0}_{|\alpha|, |\beta|}$ (JUN45)& $\Psi^{J;1}_{|\alpha|, |\beta|}$ (jj44b)&$\Psi^{J;1}_{|\alpha|, |\beta|}$ (JUN45)\\
 \hline 
    0&  3   &   1    &   3 &    1   &   2.627 &   2.691    &   0.9202 &  1.057   \\
  &  3   &    3  &   3   &    3   &   5.733 &   6.060    &   1.415 &  1.261   \\
  &  3   &   5 &   3   &    5   &   5.966 &   6.181    &   1.481 &  1.850   \\
  &  4 &   9 &   4 &    9   &   3.903 &   3.505    &   2.227 &  1.979   \\
 \hline
\end{tabular}\par}

\subsection*{$^{76}$Ge}

{\centering
\begin{tabular}{|l|c|c|c|c|c|c|c|c|}
 \hline
     $J$&  $N_{out}$&   $2j_\alpha$ &   $N_{in}$ &   $2j_\beta$ &     $\Psi^{J;0}_{|\alpha|, |\beta|}$ (jj44b) &  $\Psi^{J;0}_{|\alpha|, |\beta|}$ (JUN45)  & $\Psi^{J;1}_{|\alpha|, |\beta|}$ (jj44b) &$\Psi^{J;1}_{|\alpha|, |\beta|}$ (JUN45) \\
 \hline 
    0 &  3   &   1    &   3 &    1   &   3.282 &   3.063    &   1.282 &  1.301   \\
  &  3 &    3  &   3 &    3   &   6.453 &   6.538    &   1.733 &  1.556   \\
  &  3 &   5 &   3 &    5   &   7.014 &   7.372    &   1.777 &  2.122   \\
  &  4 &   9 &   4 &    9   &   5.135 &   4.901    &   2.985 &  2.820   \\
 \hline
\end{tabular}\par}

\subsection{$^{127}$I}

{\begin{center}
\begin{longtable}[c]{|l|c|c|c|c|c|c|c|c|}
 \hline
     $J$&  $N_{out}$&   $2j_\alpha$ &   $N_{in}$ &   $2j_\beta$ & $\Psi^{J;0}_{|\alpha|, |\beta|}$ (GCN5082)&  $\Psi^{J;0}_{|\alpha|, |\beta|}$ (SN100PN)& $\Psi^{J;1}_{|\alpha|, |\beta|}$ (GCN5082)&$\Psi^{J;1}_{|\alpha|, |\beta|}$ (SN100PN)\\
 \hline 
 \endhead
 \hline
 \multicolumn{9}{|r|}{\text{Continued on next page}}\\
 \hline
 \endfoot
 \endlastfoot
    0&  4   &   1    &   4 &    1 &   9.322  &   6.638      &   4.832  &  3.483     \\
  &  4 &    3  &   4 &    3 &   10.03  &   8.756      &   4.176  &  4.237     \\
  &  4 &   5 &   4 &    5 &   23.00  &   21.95      &   9.999  &  10.76     \\
  &  4 &   7 &   4 &    7 &   25.61   &   26.92     &   10.54  &  9.220     \\
  &  5 &   11 &   5 &    11 &   16.17  &   17.99      &   9.499  &  10.72     \\
 \hline
   1 &   4   &   1    &   4 &    1 &   0.1709  &   0.2808    &   0.02533  &   -0.01042     \\
 &   4   &   3    &   4 &    1 &   -0.1610  &   -0.2145     &   0.01134  &   0.02337     \\
 &   4   &   1    &   4 &    3 &   0.1610  &   0.2145     &   -0.01136  &   -0.02337     \\
 &   4 &    3  &   4 &    3 &   0.2892  &   0.2628     &   0.03553  &   0.05842     \\
 &   4 &    5  &   4 &    3 &   -0.08962  &   -0.06235     &   0.05415  &   0.03767     \\
 &   4 &    3  &   4 &    5 &   0.08962  &   0.06235     &   -0.05415  &   -0.03767    \\
 &   4 &   5 &   4 &    5 &    2.070  &   1.549     &  -1.250  &   -0.9362     \\
 &   4 &   7 &   4 &    5 &   -0.1905  &   -0.4202    &   0.1151  &   0.2539    \\
 &   4 &   5 &   4 &    7 &   0.1905  &   0.4202     &   -0.1151  &   -0.2539     \\
 &   4 &   7 &   4 &    7 &   0.2557  &   0.6440     &   -0.1081  &   -0.3348     \\
 &   5 &   11 &   5 &    11 &   0.2309  &   0.2032     &   0.1323  &   0.1188     \\
 \hline
   2&   4   &   3    &   4 &   1 &   -0.5824  &   -0.8225     &   -0.2599  &  -0.3763    \\
 &   4   &   5    &   4 &   1 &   0.1159  &   0.1279     &   -0.07006  &  -0.07727     \\
 &   4   &   1    &   4 &   3 &   0.5823  &   0.8224     &   0.2598  &  0.3763    \\
 &   4 &   3  &   4 &   3 &   1.269  &   1.008    &   0.5173  &  0.4914     \\
 &   4 &   5  &   4 &   3 &   0.2208  &   0.03380     &   -0.1334  &  -0.02042     \\
 &   4 &   7  &   4 &   3 &   0.4219  &   0.6312     &   0.1226  &  0.2071     \\
 &   4 &   1  &   4 &   5 &   0.1160  &   0.1279     &   -0.07008 &  -0.07727    \\
 &   4 &   3  &   4 &   5 &   -0.2209  &   -0.03380     &   0.1334  &  0.02042    \\
 &   4 &   5 &   4 &   5 &   1.426  &   1.112     &   -0.8617  &  -0.6720    \\
 &   4 &   7 &   4 &   5 &   -0.2198  &   -0.4135     &   0.1328  &  0.2498     \\
 &   4 &   3 &   4 &   7 &   0.4217  &   0.6312     &   0.1225  &  0.2071    \\
 &   4 &   5 &   4 &   7 &   0.2198  &   0.4135     &   -0.1328   &  -0.2498    \\
 &   4 &   7 &   4 &   7 &   1.235  &   2.086    &   -0.5624  &  -0.8618     \\
 &   5 &   11 &   5 &   11 &   1.346  &   1.702    &    0.8118 &  1.020     \\
 \hline
   3&   4   &   5    &   4 &   1 &   0.08278  &   0.1611     &   -0.05002  &  -0.09733     \\
 &   4   &   7    &   4 &   1 &   -0.06457  &   -0.1262     &   0.03533  &  0.07693     \\
 &   4 &   3  &   4 &   3 &   -0.02487  &   -0.06378     &   0.06361  &  0.06379     \\
 &   4 &   5  &   4 &   3 &   -0.02375  &   -0.01287     &   0.01435  &  0.007775     \\
 &   4 &   7  &   4 &   3 &   0.1029  &   0.1100     &   -0.07846  &  -0.07210     \\
 &   4 &   1  &   4 &   5 &   0.08282  &   0.1611     &    -0.05004 &  -0.09733    \\
 &   4 &   3  &   4 &   5 &   0.02375  &   0.01287    &   -0.01435  &  -0.007775    \\
 &   4 &   5 &   4 &   5 &   1.537  &   0.9317    &   -0.9289  &  -0.5629    \\
 &   4 &   7 &   4 &   5 &   -0.2330  &   -0.3106    &   0.1408  &  0.1877     \\
 &   4 &   1 &   4 &   7 &   0.06461  &   0.1262     &   -0.03535  &  -0.07693     \\
 &   4 &   3 &   4 &   7 &   0.1029  &   0.1100     &   -0.07846  &  -0.07210     \\
 &   4 &   5 &   4 &   7 &   0.2330  &   0.3106    &   -0.1408  &  -0.1877    \\
 &   4 &   7 &   4 &   7 &   -0.1368  &   -0.1445     &   0.08435  &  0.07956     \\
 &   5 &   11 &   5 &   11 &   0.03396  &   0.002819     &   0.02449  &  0.003106    \\
 \hline
   4&   4   &   7    &   4 &   1 &   0.1001  &   -0.08434     &   -0.09170  &  0.009318     \\
 &   4 &   5  &   4 &   3 &   -0.1192  &   0.007031    &   0.07200  &  -0.004248    \\
 &   4 &   7  &   4 &   3 &   -0.09084  &   -0.05406     &   0.09697  &  0.03908     \\
 &   4 &   3  &   4 &   5 &    0.1191 &   -0.007031     &   -0.07198  &  0.004248     \\
 &   4 &   5 &   4 &   5 &   1.104  &   0.7639     &   -0.6673  &  -0.4616    \\
 &   4 &   7 &   4 &   5 &   0.1167  &   0.02056     &   -0.07054  &  -0.01242     \\
 &   4 &   1 &   4 &   7 &   -0.1001  &   0.08434     &    0.09170 &  -0.009318     \\
 &   4 &   3 &   4 &   7 &   -0.09084  &   -0.05406     &   0.09697  &  0.03908     \\
 &   4 &   5 &   4 &   7 &   -0.1167  &   -0.02056     &   0.07054  &  0.01242     \\
 &   4 &   7 &   4 &   7 &   -0.2445  &   -0.2995     &   0.1599  &  0.1712     \\
 &   5 &   11 &   5 &   11 &   -0.1456  &   -0.09598     &   -0.07393  &  -0.05210   \\
 \hline
   5 &   4   &   7    &   4 &   3 &   -0.06086  &   -0.06418     &   0.04038  &  0.04327     \\
 &   4 &   5 &   4 &   5 &   1.873  &   1.367     &   -1.132  &  -0.8259    \\
 &   4 &   7 &   4 &   5 &   0.2773  &   0.4459     &    -0.1676 &  -0.2694     \\
 &   4 &   3 &   4 &   7 &   -0.06086  &   -0.06418     &   0.04038  &  0.04327    \\
 &   4 &   5 &   4 &   7 &   -0.2773  &   -0.4459     &   0.1676  &  0.2694     \\
 &   4 &   7 &   4 &   7 &   -0.3659  &   -0.6380    &    0.2196 &  0.3820     \\
 &   5 &   11 &   5 &   11 &   0.01068 &   -6.633 $\times 10^{-5}$    &    0.001483  &  -0.001283     \\
 \hline
\end{longtable} 
\end{center}}

\subsection{$^{128, 129, 130, 131, 132, 134, 136}$Xe}

\subsection*{$^{128}$Xe}

{\centering
\begin{tabular}{|l|c|c|c|c|c|c|}
 \hline
    $J$&  $N_{out}$&   $2j_\alpha$ &   $N_{in}$ &   $2j_\beta$  &  $\Psi^{J;0}_{|\alpha|, |\beta|}$ (SN100PN)&$\Psi^{J;1}_{|\alpha|, |\beta|}$ (SN100PN)\\
 \hline 
    0 &  4   &   1    &   4 &    1  &   2.121      &  0.7848     \\
  &  4   &    3  &   4   &    3   &   3.072     &  1.186     \\
  &  4   &   5 &   4   &    5  &   8.998     &  4.163     \\
  &  4   &   7 &   4   &    7   &   12.02      &  3.823     \\
  &  5 &   11 &   5 &    11  &   7.377     &  4.257    \\
 \hline
\end{tabular}\par}

\subsection*{$^{129}$Xe}

{\centering
\begin{tabular}{|l|c|c|c|c|c|c|}
 \hline
     $J$&  $N_{out}$&   $2j_\alpha$ &   $N_{in}$ &   $2j_\beta$   &  $\Psi^{J;0}_{|\alpha|, |\beta|}$ (SN100PN) &$\Psi^{J;1}_{|\alpha|, |\beta|}$ (SN100PN)\\
 \hline 
    0&  4   &   1    &   4 &    1   &   4.829      &  2.148    \\
  &  4   &    3  &   4   &    3   &   5.599     &  2.453     \\
  &  4   &   5 &   4   &    5  &   13.15      &  5.923     \\
  &  4   &   7 &   4   &    7   &   16.47      &  5.123    \\
  &  5 &   11 &   5 &    11   &   11.27     &  6.497     \\
 \hline
   1&   4   &   1    &   4 &   1   &   2.276      &  1.385     \\
 &   4   &   3    &   4   &   1   &   0.007827      &  -0.005130      \\
 &   4   &   1    &   4   &   3   &   -0.007827     &   0.005130     \\
 &   4   &   3  &   4   &   3    &   -0.2975     &  -0.1742      \\
 &   4   &   3  &   4   &   5 &   0.001526      &  -0.0009218      \\
 &   4   &   5  &   4   &   3  &   -0.001526      &  0.0009218      \\
 &   4   &   5 &   4   &   5    &   0.05774      &  -0.03489      \\
 &   4   &   7 &   4   &   5    &   -0.01546      &  0.009338     \\
 &   4   &   5 &   4   &   7   &   0.01546     &   -0.009338      \\
 &   4   &   7 &   4   &   7   &   0.06441      &  -0.04377      \\
 &   5 &   11 &   5 &   11    &   0.06195     &   0.03611      \\
 \hline
\end{tabular}\par}

\subsection*{$^{130}$Xe}

{\centering
\vspace{-0.2cm}\begin{tabular}{|l|c|c|c|c|c|c|}
 \hline
     $J$&  $N_{out}$&   $2j_\alpha$ &   $N_{in}$ &   $2j_\beta$   &  $\Psi^{J;0}_{|\alpha|, |\beta|}$ (SN100PN)&$\Psi^{J;1}_{|\alpha|, |\beta|}$ (SN100PN)\\
 \hline 
    0&  4   &   1    &   4 &    1   &   3.705      &  1.842     \\
  &  4   &    3  &   4   &    3  &   4.349      &  2.094    \\
  &  4   &   5 &   4   &    5   &   9.452     &  4.315     \\
  &  4   &   7 &   4   &    7   &   12.11      &  3.619     \\
  &  5 &   11 &   5 &    11   &   8.765      &  5.016     \\
 \hline
\end{tabular}\par}

\vspace{-0.5cm}\subsection*{$^{131}$Xe}

{\centering
\vspace{-0.2cm}\begin{tabular}{|l|c|c|c|c|c|c|}
 \hline
     $J$&  $N_{out}$&   $2j_\alpha$ &   $N_{in}$ &   $2j_\beta$   &  $\Psi^{J;0}_{|\alpha|, |\beta|}$ (SN100PN)   &$\Psi^{J;1}_{|\alpha|, |\beta|}$ (SN100PN) \\
 \hline 
    0&  4   &   1    &   4 &    1   &   6.968      &  3.664     \\
  &  4   &    3  &   4   &    3   &   9.075      &  4.525    \\
  &  4   &   5 &   4   &    5   &   17.59      &  8.464    \\
  &  4   &   7 &   4   &    7  &   25.71      &  7.231     \\
  &  5 &   11 &   5 &    11  &   20.42      &  11.66     \\
 \hline
   1&   4   &   1    &   4 &   1   &    -0.1837     &  -0.1010     \\
 &   4   &   3    &   4   &   1    &   0.1458      &  0.07953      \\
 &   4   &   1    &   4   &   3   &   -0.1458      &  -0.07955      \\
 &   4   &   3  &   4   &   3   &   2.703      &  1.632     \\
 &   4   &   5  &   4   &   3  &   -0.06623      &   -0.04416    \\
 &   4   &   3  &   4   &   5  &   0.06623      &  0.04416     \\
 &   4   &   5 &   4   &   5  &   0.05407      &  -0.006109     \\
 &   4   &   7 &   4   &   5   &   0.06578      &  0.04852     \\
 &   4 &   5 &   4   &   7    &   -0.06578      &   -0.04852    \\
 &   4 &   7 &   4 &   7   &   0.1769      &  0.01370      \\
 &   5 &   11 &   5 &   11    &    0.03812     &   0.02073     \\
 \hline
   2 &   4   &   3    &   4 &   1   &   -0.08030      &  -0.06080     \\
 &   4   &   1    &   4   &   3   &   0.08030      &  0.06080     \\
 &   4   &   1    &   4   &   5   &   0.1349      &  0.1534     \\
 &   4   &   3  &   4   &   3  &   -0.3794      &  -0.2110     \\
 &   4   &   3  &   4   &   5   &   -0.07230      &  -0.08105    \\
 &   4   &   3  &   4   &   7   &   0.04874      &  0.1682     \\
 &   4   &   5  &   4   &   1   &   0.1349      &  0.1534    \\
 &   4   &   5  &   4   &   3   &   0.07233      &  0.08107     \\
 &   4   &   5 &   4   &   5   &   0.02560      &  0.05820    \\
 &   4   &   7 &   4   &   3   &   0.04874      &  0.1682     \\
 &   4   &   7 &   4   &   5    &   -0.02439      &  0.007340     \\
 &   4   &   5 &   4   &   7   &   0.02439    &  -0.007340     \\
 &   4   &   7 &   4   &   7  &   0.1426     &  -0.04637     \\
 &   5 &   11 &   5 &   11   &   -0.2028    &  -0.1349     \\
 \hline
   3 &   4 &   1  &   4 &   5  &   0.008799     &  0.005296     \\
 &   4   &   1  &   4   &   7   &   0.02944     &  0.02018     \\
 &   4   &   3  &   4   &   3   &   2.730     &  1.641         \\
 &   4   &   3  &   4   &   5   &   -0.04763     &  -0.02667     \\
 &   4   &   3  &   4   &   7   &   -0.08958     &  -0.05071     \\
 &   4   &   5 &   4   &   1  &   0.008799    &  0.005296     \\
 &   4   &   5 &   4   &   3   &   0.04763    &  0.02667     \\
 &   4   &   5 &   4   &   5   &   -0.05068     &  -0.03910     \\
 &   4   &   7 &   4   &   1  &   -0.02944     &  -0.02018     \\
 &   4   &   7 &   4   &   3   &   -0.08958     &  -0.05071     \\
 &   4   &   7 &   4   &   5   &   0.05778     &  0.03478     \\
 &   4   &   5 &   4   &   7   &   -0.05778     &  -0.03478     \\
 &   4   &   7 &   4   &   7  &   0.06665     &  0.008424     \\
 &   5 &   11 &   5 &   11   &   0.01365     &  0.007507     \\
 \hline
\end{tabular}\par}

\subsection*{$^{132}$Xe}

{\centering
\begin{tabular}{|l|c|c|c|c|c|c|}
 \hline
     $J$&  $N_{out}$&   $2j_\alpha$ &   $N_{in}$ &   $2j_\beta$   &  $\Psi^{J;0}_{|\alpha|, |\beta|}$ (SN100PN)&$\Psi^{J;1}_{|\alpha|, |\beta|}$ (SN100PN) \\
 \hline 
    0&  4   &   1    &   4 &    1   &   3.983     &  2.168    \\
  &  4 &    3  &   4 &    3   &   5.263      &  2.773    \\
  &  4 &   5 &   4 &    5  &   9.027      &  4.306     \\
  &  4 &   7 &   4 &    7  &   13.17      &  3.682     \\
  &  5 &   11 &   5 &    11   &   10.85      &  6.180     \\
 \hline
\end{tabular}\par}

\subsection*{$^{134}$Xe}

{\centering
\begin{tabular}{|l|c|c|c|c|c|c|}
 \hline
     $J$&  $N_{out}$&   $2j_\alpha$ &   $N_{in}$ &   $2j_\beta$   &  $\Psi^{J;0}_{|\alpha|, |\beta|}$ (SN100PN)   &$\Psi^{J;1}_{|\alpha|, |\beta|}$ (SN100PN) \\
 \hline 
    0&  4   &   1    &   4 &    1  &   4.481      &  2.529     \\
  &  4 &    3  &   4 &    3   &   6.180     &  3.420     \\
  &  4 &   5 &   4 &    5   &   9.519     &  4.703     \\
  &  4 &   7 &   4 &    7   &   14.12      &  3.806     \\
  &  5 &   11 &   5 &    11   &   12.41      &  7.082     \\
 \hline
\end{tabular}\par}

\subsection*{$^{136}$Xe}

{\centering
\begin{tabular}{|l|c|c|c|c|c|c|}
 \hline
    $J$&  $N_{out}$&   $2j_\alpha$ &   $N_{in}$ &   $2j_\beta$  &  $\Psi^{J;0}_{|\alpha|, |\beta|}$ (SN100PN) &$\Psi^{J;1}_{|\alpha|, |\beta|}$ (SN100PN)\\
 \hline 
    0&  4   &   1    &   4 &    1   &   5.498      &  3.151     \\
  &  4 &    3  &   4 &    3   &   7.820      &  4.429     \\
  &  4 &   5 &   4 &    5   &   10.25      &  5.023     \\
  &  4 &   7 &   4 &    7   &   14.77      &  4.046    \\
  &  5 &   11 &   5 &    11   &   13.51     &  7.695    \\
 \hline
\end{tabular}\par}

\vspace{10mm}

\section{Nuclear Response Functions}\label{nuclear responses}

The nuclear response functions are given by
\begin{eqnarray}\label{FF eqn2}
    F^{(N,N')}_{X,Y} (q^2) & \equiv &\frac{4\pi}{2J_i+1}  \sum_{J=0}^{2J_i} \langle J_i || X_J^{(N)} || J_i \rangle \langle J_i || Y^{(N')}_J || J_i \rangle,
\end{eqnarray}
where $N, N'=\{p, n\}$ and $F^{(N,N')}_X (q^2)  \equiv  F^{(N,N')}_{X,X} (q^2)$. Here, $X$ and $Y$ are one of six nuclear operators traditionally written as $M_{JM}, \ \Sigma''_{JM}, \ \Sigma'_{JM}, \ \Delta_{JM}, \ \Phi''_{JM}$ and $\tilde{\Phi}'_{JM}$. The cross terms in Eq.~(\ref{FF eqn2}) exist only for two sets of operators, $\{M_{JM}, \ \Phi''_{JM}\}$ and $\{\Sigma'_{JM}, \ \Delta_{JM}\}$. A harmonic oscillator single-particle basis is employed, where the response functions take on expressions of the form $e^{-2y}p(y)$, with $p(y)$ a polynomial and $y=(qb/2)^2$. The harmonic oscillator length parameter is 
$b \approx \sqrt{41.467/(45A^{-1/3} - 25A^{-2/3})}$~fm ($A$ is the mass number). We have $F^{(p,n)}_{X,Y} (q^2)= F^{(n,p)}_{X,Y} (q^2)$ only for $X=Y$, hence we omit repetitive expressions.

\vspace{5mm}

\subsection{$^{19}$F}

$F_{M}^{(p,p)}=  e^{-2 y} (81.0-96.0 y+36.2 y^2-4.60 y^3+0.186 y^4)$

$F_{\Sigma''}^{(p,p)}=   e^{-2 y} (0.913-2.40 y+2.37 y^2-1.04 y^3+0.171 y^4) $

$F_{\Sigma'}^{(p,p)}=  e^{-2 y} (1.83-4.90 y+4.90 y^2-2.16 y^3+0.354 y^4) $

$F_{\Phi''}^{(p,p)}=   e^{-2 y} (0.0403-0.0322 y+0.00644 y^2)$

$F_{\Delta}^{(p,p)}=   e^{-2 y} (0.0247-0.0198 y+0.00396 y^2)$

$F_{M \Phi''}^{(p,p)}=  e^{-2 y} (-1.81+1.79 y-0.514 y^2+0.0346 y^3)$

$F_{\Sigma' \Delta}^{(p,p)}=   e^{-2 y} (-0.212+0.370 y-0.208 y^2+0.0374 y^3)$

\vspace{5mm}

$F_{M}^{(p,n)}=  e^{-2 y} (90.0-113. y+47.4 y^2-7.35 y^3+0.362 y^4) $

$F_{\Sigma''}^{(p,n)}= e^{-2 y} (-0.00381+0.0151 y-0.0181 y^2+0.00858 y^3-0.00138 y^4) $

$F_{\Sigma'}^{(p,n)}= e^{-2 y} (-0.00763+0.0154 y+0.00995 y^2-0.0249 y^3+0.00890 y^4) $

$F_{\Phi''}^{(p,n)}=e^{-2 y} (0.113-0.0905 y+0.0181 y^2)  $

$F_{\Delta}^{(p,n)}=  e^{-2 y} (-0.0220+0.0176 y-0.00352 y^2) $

$F_{M \Phi''}^{(p,n)}=  e^{-2 y} (-5.07+5.03 y-1.45 y^2+0.0972 y^3) $

$F_{\Sigma' \Delta}^{(p,n)}= e^{-2 y} (0.189-0.330 y+0.185 y^2-0.0334 y^3) $

\vspace{5mm}

$F_{M \Phi''}^{(n,p)}= e^{-2 y} (-2.01+2.14 y-0.704 y^2+0.0674 y^3)  $

$F_{\Sigma' \Delta}^{(n,p)}=  e^{-2 y} (0.000888-0.000952 y-0.00211 y^2+0.000940 y^3) $

\vspace{5mm}

$F_{M}^{(n,n)}=   e^{-2 y} (100.-133. y+61.2 y^2-11.2 y^3+0.706 y^4)$

$F_{\Sigma''}^{(n,n)}=   e^{-2 y} (0.0000159-0.0000846 y+0.000139 y^2-0.0000709 y^3+0.0000112 y^4)$

$F_{\Sigma'}^{(n,n)}=   e^{-2 y} (0.0000319-0.0000428 y-0.000154 y^2+0.000113 y^3+0.000223 y^4)$

$F_{\Phi''}^{(n,n)}=  e^{-2 y} (0.318-0.254 y+0.0508 y^2)$

$F_{\Delta}^{(n,n)}=   e^{-2 y} (0.0196-0.0157 y+0.00314 y^2)$

$F_{M \Phi''}^{(n,n)}=   e^{-2 y} (-5.64+6.01 y-1.98 y^2+0.189 y^3)$

$F_{\Sigma' \Delta}^{(n,n)}=   e^{-2 y} (-0.000790+0.000848 y+0.00188 y^2-0.000837 y^3)$

\vspace{5mm}

\subsection{$^{23}$Na}

$F_{M}^{(p,p)}=  e^{-2 y} (121.-176. y+87.0 y^2-16.9 y^3+1.22 y^4)$

$F_{\Sigma''}^{(p,p)}=  e^{-2 y} (0.112-0.192 y+0.386 y^2-0.0911 y^3+0.00707 y^4)$

$F_{\Sigma'}^{(p,p)}=  e^{-2 y} (0.223-0.700 y+1.08 y^2-0.449 y^3+0.0576 y^4)$

$F_{\Phi''}^{(p,p)}=  e^{-2 y} (1.63-1.32 y+0.303 y^2)$

$F_{\tilde{\Phi}'}^{(p,p)}= e^{-2 y} (0.00220-0.00358 y+0.00145 y^2-6.28\times10^{-19} y^3+6.77\times10^{-35} y^4) $

$F_{\Delta}^{(p,p)}=  e^{-2 y} (0.244-0.195 y+0.0534 y^2)$

$F_{M \Phi''}^{(p,p)}=  e^{-2 y} (-14.1+15.9 y-5.60 y^2+0.605 y^3)$

$F_{\Sigma' \Delta}^{(p,p)}=  e^{-2 y} (-0.233+0.459 y-0.287 y^2+0.0498 y^3)$

\vspace{5mm}

$F_{M}^{(p,n)}= e^{-2 y} (132.-199. y+102. y^2-20.3 y^3+1.48 y^4) $

$F_{\Sigma''}^{(p,n)}=e^{-2 y} (0.0118-0.0315 y+0.0237 y^2-0.00842 y^3+0.000184 y^4) $

$F_{\Sigma'}^{(p,n)}= e^{-2 y} (0.0236-0.0630 y+0.0452 y^2-0.0112 y^3+0.00125 y^4) $

$F_{\Phi''}^{(p,n)}= e^{-2 y} (2.06-1.67 y+0.363 y^2) $

$F_{\tilde{\Phi}'}^{(p,n)}= e^{-2 y} (0.00245-0.000942 y-0.000856 y^2+1.92\times10^{-19} y^3-1.67\times10^{-36} y^4) $

$F_{\Delta}^{(p,n)}= e^{-2 y} (0.0820-0.0656 y+0.0145 y^2) $

$F_{M \Phi''}^{(p,n)}= e^{-2 y} (-17.8+20.1 y-6.96 y^2+0.714 y^3)$

$F_{\Sigma' \Delta}^{(p,n)}=e^{-2 y} (-0.0784+0.154 y-0.0780 y^2+0.0109 y^3)$

\vspace{5mm}

$F_{M \Phi''}^{(n,p)}= e^{-2 y} (-15.3+18.1 y-6.64 y^2+0.740 y^3)$

$F_{\Sigma' \Delta}^{(n,p)}= e^{-2 y} (-0.0247+0.0370 y-0.0124 y^2+0.00130 y^3)$

\vspace{5mm}

$F_{M}^{(n,n)}= e^{-2 y} (144.-224. y+119. y^2-24.5 y^3+1.81 y^4)$

$F_{\Sigma''}^{(n,n)}=e^{-2 y} (0.00125-0.00451 y+0.00591 y^2-0.00322 y^3+0.000660 y^4)$

$F_{\Sigma'}^{(n,n)}=  e^{-2 y} (0.00250-0.00549 y+0.00352 y^2-0.000466 y^3+0.0000316 y^4)$

$F_{\Phi''}^{(n,n)}=  e^{-2 y} (2.61-2.10 y+0.443 y^2)$

$F_{\tilde{\Phi}'}^{(n,n)}= e^{-2 y} (0.00274+0.00235 y+0.000504 y^2-9.11\times10^{-21} y^3+4.12\times10^{-38} y^4)$

$F_{\Delta}^{(n,n)}=  e^{-2 y} (0.0275-0.0220 y+0.00454 y^2)$

$F_{M \Phi''}^{(n,n)}=  e^{-2 y} (-19.4+22.9 y-8.26 y^2+0.885 y^3)$

$F_{\Sigma' \Delta}^{(n,n)}=  e^{-2 y} (-0.00830+0.0124 y-0.00433 y^2+0.000344 y^3)$

\vspace{5mm}

\subsection{$^{28, 29, 30}$Si}

\subsection*{$^{28}$Si}

$F_{M}^{(p,p)}=  e^{-2 y} (196.-336. y+197. y^2-45.3 y^3+3.56 y^4)$

$F_{\Phi''}^{(p,p)}= e^{-2 y} (6.15-4.92 y+0.983 y^2) $

$F_{M \Phi''}^{(p,p)}=e^{-2 y} (-34.7+43.6 y-16.6 y^2+1.87 y^3) $

\vspace{5mm}

$F_{M}^{(p,n)}= e^{-2 y} (196.-336. y+197. y^2-45.3 y^3+3.56 y^4)$

$F_{\Phi''}^{(p,n)}= e^{-2 y} (6.15-4.92 y+0.983 y^2) $

$F_{M \Phi''}^{(p,n)}= e^{-2 y} (-34.7+43.6 y-16.6 y^2+1.87 y^3) $

\vspace{5mm}

$F_{M \Phi''}^{(n,p)}= e^{-2 y} (-34.7+43.6 y-16.6 y^2+1.87 y^3) $

\vspace{5mm}

$F_{M}^{(n,n)}= e^{-2 y} (196.-336. y+197. y^2-45.3 y^3+3.56 y^4)$

$F_{\Phi''}^{(n,n)}= e^{-2 y} (6.15-4.92 y+0.983 y^2) $

$F_{M \Phi''}^{(n,n)}= e^{-2 y} (-34.7+43.6 y-16.6 y^2+1.87 y^3) $

\vspace{5mm}

\subsection*{$^{29}$Si}

$F_{M}^{(p,p)}= e^{-2 y} (196.-336. y+195. y^2-44.1 y^3+3.37 y^4) $

$F_{\Sigma''}^{(p,p)}= e^{-2 y} (0.00102-0.00374 y+0.00484 y^2-0.00259 y^3+0.000491 y^4) $

$F_{\Sigma'}^{(p,p)}= e^{-2 y} (0.00204-0.00443 y+0.00358 y^2-0.00128 y^3+0.000170 y^4) $

$F_{\Phi''}^{(p,p)}= e^{-2 y} (8.00-6.40 y+1.28 y^2) $

$F_{\Delta}^{(p,p)}= e^{-2 y} (1.24\times10^{-6}-9.96\times10^{-7} y+1.99\times10^{-7} y^2) $

$F_{M \Phi''}^{(p,p)}= e^{-2 y} (-39.6+49.8 y-18.8 y^2+2.08 y^3) $

$F_{\Sigma' \Delta}^{(p,p)}= e^{-2 y} (-0.0000504+0.0000748 y-0.0000364 y^2+5.82\times10^{-6} y^3) $

\vspace{5mm}

$F_{M}^{(p,n)}= e^{-2 y} (210.-367. y+220. y^2-52.1 y^3+4.23 y^4) $

$F_{\Sigma''}^{(p,n)}= e^{-2 y} (0.00992-0.0206 y+0.0187 y^2-0.0152 y^3+0.00513 y^4) $

$F_{\Sigma'}^{(p,n)}=e^{-2 y} (0.0198-0.0587 y+0.0693 y^2-0.0360 y^3+0.00671 y^4) $

$F_{\Phi''}^{(p,n)}= e^{-2 y} (8.18-6.54 y+1.31 y^2) $

$F_{\Delta}^{(p,n)}= e^{-2 y} (0.000258-0.000207 y+0.0000413 y^2) $

$F_{M \Phi''}^{(p,n)}= e^{-2 y} (-40.5+50.9 y-19.2 y^2+2.12 y^3) $

$F_{\Sigma' \Delta}^{(p,n)}= e^{-2 y} (-0.0105+0.0155 y-0.00755 y^2+0.00121 y^3)$

\vspace{5mm}

$F_{M \Phi''}^{(n,p)}= e^{-2 y} (-42.4+54.7 y-21.6 y^2+2.61 y^3) $

$F_{\Sigma' \Delta}^{(n,p)}= e^{-2 y} (-0.000490+0.00111 y-0.000942 y^2+0.000230 y^3) $

\vspace{5mm}

$F_{M}^{(n,n)}= e^{-2 y} (225.-400. y+247. y^2-61.5 y^3+5.31 y^4) $

$F_{\Sigma''}^{(n,n)}=  e^{-2 y} (0.0963-0.0472 y+0.149 y^2-0.0352 y^3+0.0536 y^4) $

$F_{\Sigma'}^{(n,n)}= e^{-2 y} (0.193-0.723 y+1.13 y^2-0.848 y^3+0.265 y^4) $

$F_{\Phi''}^{(n,n)}= e^{-2 y} (8.35-6.68 y+1.34 y^2) $

$F_{\Delta}^{(n,n)}= e^{-2 y} (0.0536-0.0428 y+0.00857 y^2) $

$F_{M \Phi''}^{(n,n)}= e^{-2 y} (-43.4+55.9 y-22.1 y^2+2.66 y^3) $

$F_{\Sigma' \Delta}^{(n,n)}= e^{-2 y} (-0.102+0.231 y-0.195 y^2+0.0476 y^3) $

\vspace{5mm}

\subsection*{$^{30}$Si}

$F_{M}^{(p,p)}= e^{-2 y} (196.-336. y+195. y^2-43.5 y^3+3.29 y^4) $

$F_{\Phi''}^{(p,p)}= e^{-2 y} (9.13-7.30 y+1.46 y^2) $

$F_{M \Phi''}^{(p,p)}= e^{-2 y} (-42.3+53.2 y-20.0 y^2+2.19 y^3) $

\vspace{5mm}

$F_{M}^{(p,n)}= e^{-2 y} (224.-397. y+243. y^2-58.8 y^3+4.86 y^4) $

$F_{\Phi''}^{(p,n)}= e^{-2 y} (7.59-6.07 y+1.21 y^2) $

$F_{M \Phi''}^{(p,n)}= e^{-2 y} (-35.1+44.2 y-16.6 y^2+1.82 y^3)$

\vspace{5mm}

$F_{M \Phi''}^{(n,p)}= e^{-2 y} (-48.3+63.7 y-25.8 y^2+3.24 y^3) $

\vspace{5mm}

$F_{M}^{(n,n)}= e^{-2 y} (256.-469. y+301. y^2-78.6 y^3+7.18 y^4) $
  
$F_{\Phi''}^{(n,n)}= e^{-2 y} (6.30-5.04 y+1.01 y^2) $

$F_{M \Phi''}^{(n,n)}= e^{-2 y} (-40.2+52.9 y-21.5 y^2+2.69 y^3) $

\vspace{5mm}

\subsection{$^{40}$Ar}

\subsubsection{SDPF-NR}

$F_{M}^{(p,p)}= e^{-2 y} (324.-624. y+423. y^2-118. y^3+11.5 y^4-2.61\times10^{-6} y^5+1.48\times10^{-13} y^6) $

$F_{\Phi''}^{(p,p)}= e^{-2 y} (2.88-2.30 y+0.461 y^2-5.91\times10^{-7} y^3+1.90\times10^{-13} y^4) $

$F_{M \Phi''}^{(p,p)}= e^{-2 y} (-30.5+41.6 y-17.5 y^2+2.31 y^3-1.74\times10^{-6} y^4+1.68\times10^{-13} y^5) $

\vspace{5mm}

$F_{M}^{(p,n)}= e^{-2 y} (396.-813. y+593. y^2-183. y^3+22.4 y^4-0.607 y^5+6.88\times10^{-8} y^6) $

$F_{\Phi''}^{(p,n)}= e^{-2 y} (2.91-3.49 y+1.27 y^2-0.137 y^3+8.80\times10^{-8} y^4) $

$F_{M \Phi''}^{(p,n)}= e^{-2 y} (-30.8+54.4 y-33.2 y^2+8.16 y^3-0.687 y^4+7.78\times10^{-8} y^5)$

\vspace{5mm}

$F_{M \Phi''}^{(n,p)}= e^{-2 y} (-37.3+55.7 y-25.9 y^2+4.17 y^3-0.121 y^4+7.78\times10^{-8} y^5) $

\vspace{5mm}

$F_{M}^{(n,n)}= e^{-2 y} (484.-1060. y+826. y^2-281. y^3+41.0 y^4-2.04 y^5+0.0320 y^6) $

$F_{\Phi''}^{(n,n)}= e^{-2 y} (2.94-4.70 y+2.57 y^2-0.554 y^3+0.0409 y^4)$

$F_{M \Phi''}^{(n,n)}= e^{-2 y} (-37.7+71.3 y-47.1 y^2+13.0 y^3-1.40 y^4+0.0361 y^5) $

\vspace{5mm}

\subsubsection{SDPF-U}

$F_{M}^{(p,p)}= e^{-2 y} (324.-624. y+423. y^2-118. y^3+11.5 y^4-2.52\times10^{-6} y^5+1.37\times10^{-13} y^6) $

$F_{\Phi''}^{(p,p)}= e^{-2 y} (2.81-2.25 y+0.449 y^2-5.87\times10^{-7} y^3+1.92\times10^{-13} y^4) $

$F_{M \Phi''}^{(p,p)}= e^{-2 y} (-30.2+41.1 y-17.3 y^2+2.28 y^3-1.74\times10^{-6} y^4+1.62\times10^{-13} y^5) $

\vspace{5mm}

$F_{M}^{(p,n)}= e^{-2 y} (396.-813. y+593. y^2-183. y^3+22.2 y^4-0.585 y^5+6.38\times10^{-8} y^6) $

$F_{\Phi''}^{(p,n)}= e^{-2 y} (2.92-3.51 y+1.28 y^2-0.136 y^3+8.91\times10^{-8} y^4) $

$F_{M \Phi''}^{(p,n)}= e^{-2 y} (-31.4+55.4 y-33.8 y^2+8.27 y^3-0.691 y^4+7.54\times10^{-8} y^5)$

\vspace{5mm}

$F_{M \Phi''}^{(n,p)}= e^{-2 y} (-36.9+55.0 y-25.6 y^2+4.09 y^3-0.115 y^4+7.54\times10^{-8} y^5)$

\vspace{5mm}

$F_{M}^{(n,n)}= e^{-2 y} (484.-1060. y+825. y^2-280. y^3+40.4 y^4-1.95 y^5+0.0297 y^6) $

$F_{\Phi''}^{(n,n)}= e^{-2 y} (3.05-4.87 y+2.66 y^2-0.568 y^3+0.0414 y^4) $

$F_{M \Phi''}^{(n,n)}= e^{-2 y} (-38.4+72.6 y-47.9 y^2+13.1 y^3-1.39 y^4+0.0350 y^5) $

\vspace{5mm}

\subsubsection{SDPF-MU}

$F_{M}^{(p,p)}= e^{-2 y} (324.-624. y+422. y^2-117. y^3+11.4 y^4-2.39\times10^{-6} y^5+1.25\times10^{-13} y^6) $

$F_{\Phi''}^{(p,p)}= e^{-2 y} (2.71-2.17 y+0.433 y^2-6.25\times10^{-7} y^3+ 2.25\times10^{-13} y^4) $

$F_{M \Phi''}^{(p,p)}=  e^{-2 y} (-29.6+40.4 y-17.0 y^2+2.22 y^3-1.84\times10^{-6} y^4+1.68\times10^{-13} y^5) $

\vspace{5mm}

$F_{M}^{(p,n)}= e^{-2 y} (396. - 813. y + 592. y^2 - 182. y^3 + 21.9 y^4 - 0.556 y^5 + 5.82\times10^{-8} y^6)$

$F_{\Phi''}^{(p,n)}= e^{-2 y} (3.12 - 3.74 y + 1.36 y^2 - 0.145 y^3 + 1.05\times10^{-7} y^4) $

$F_{M \Phi''}^{(p,n)}= e^{-2 y} (-34.1 + 60.1 y - 36.6 y^2 + 8.94 y^3 - 0.745 y^4 + 7.80\times10^{-8} y^5) $

\vspace{5mm}

$F_{M \Phi''}^{(n,p)}= e^{-2 y} (-36.2 + 54.0 y - 25.1 y^2 + 3.98 y^3 - 0.108 y^4 + 7.81\times10^{-8} y^5) $

\vspace{5mm}

$F_{M}^{(n,n)}= e^{-2 y} (484. - 1060. y + 824. y^2 - 278. y^3 + 39.7 y^4 - 1.86 y^5 + 0.0270 y^6) $

$F_{\Phi''}^{(n,n)}= e^{-2 y} (3.59 - 5.74 y + 3.13 y^2 - 0.668 y^3 + 0.0486 y^4) $

$F_{M \Phi''}^{(n,n)}= e^{-2 y} (-41.7 + 78.8 y - 51.9 y^2 + 14.2 y^3 - 1.49 y^4 + 0.0362 y^5) $

\vspace{5mm}

\subsubsection{EPQQM}

$F_{M}^{(p,p)}= e^{-2 y} (324. - 624. y + 423. y^2 - 118. y^3 + 11.6 y^4 - 3.55\times10^{-6} y^5 + 2.72\times10^{-13} y^6) $

$F_{\Phi''}^{(p,p)}= e^{-2 y} (2.93 - 2.34 y + 0.469 y^2 - 5.69\times10^{-7} y^3 + 1.73\times10^{-13} y^4) $

$F_{M \Phi''}^{(p,p)}= e^{-2 y} (-30.8 + 42.0 y - 17.7 y^2 + 2.33 y^3 - 1.77\times10^{-6} y^4 + 2.17\times10^{-13} y^5) $

\vspace{5mm}

$F_{M}^{(p,n)}= e^{-2 y} (396. - 813. y + 597. y^2 - 189. y^3 + 24.3 y^4 - 0.824 y^5 + 1.26\times10^{-7} y^6) $

$F_{\Phi''}^{(p,n)}= e^{-2 y} (2.52 - 3.02 y + 1.14 y^2 - 0.132 y^3 + 8.02\times10^{-8} y^4) $

$F_{M \Phi''}^{(p,n)}= e^{-2 y} (-26.5 + 46.7 y - 28.9 y^2 + 7.35 y^3 - 0.657 y^4 + 1.01\times10^{-7} y^5) $

\vspace{5mm}

$F_{M \Phi''}^{(n,p)}= e^{-2 y} (-37.7 + 56.1 y - 26.6 y^2 + 4.46 y^3 - 0.166 y^4 + 1.01\times10^{-7} y^5) $

\vspace{5mm}

$F_{M}^{(n,n)}= e^{-2 y} (484. - 1060. y + 836. y^2 - 294. y^3 + 46.6 y^4 - 2.86 y^5 + 0.0585 y^6) $

$F_{\Phi''}^{(n,n)}= e^{-2 y} (2.16 - 3.46 y + 1.95 y^2 - 0.454 y^3 + 0.0372 y^4) $

$F_{M \Phi''}^{(n,n)}= e^{-2 y} (-32.4 + 61.2 y - 41.2 y^2 + 11.9 y^3 - 1.43 y^4 + 0.0467 y^5)$

\vspace{5mm}

\subsection{$^{70,72,73,74,76}$Ge}

\subsubsection{JUN45}

\subsection*{$^{70}$Ge}

$F_{M}^{(p,p)}= e^{-2 y} (1020.-2830. y+2950. y^2-1480. y^3+378. y^4-47.4 y^5+2.40 y^6-0.0139 y^7+0.0000214 y^8) $

$F_{\Phi''}^{(p,p)}= e^{-2 y} (56.6 - 92.7 y + 54.0 y^2 - 13.3 y^3 + 1.25 y^4 -  0.0197 y^5 + 0.0000857 y^6) $

$F_{M \Phi''}^{(p,p)}= e^{-2 y} (-241. + 529. y - 424. y^2 + 155. y^3 - 26.4 y^4 + 1.77 y^5 - 0.0188 y^6 + 0.0000429 y^7) $

\vspace{5mm}

$F_{M}^{(p,n)}= e^{-2 y} (1210.-3500. y+3840. y^2-2030. y^3+554. y^4-75.8 y^5+4.42 y^6-0.0550 y^7+0.000133 y^8) $

$F_{\Phi''}^{(p,n)}= e^{-2 y} (39.9 - 71.5 y + 46.6 y^2 - 13.5 y^3 + 1.71 y^4 -  0.0725 y^5 + 0.000531 y^6) $

$F_{M \Phi''}^{(p,n)}= e^{-2 y} (-170. + 399. y - 350. y^2 + 145. y^3 - 29.5 y^4 + 2.77 y^5 - 0.0919 y^6 + 0.000266 y^7)$

\vspace{5mm}

$F_{M \Phi''}^{(n,p)}= e^{-2 y} (-286.+663. y-562. y^2+220. y^3-40.7 y^4+3.13 y^5-0.0544 y^6+0.000266 y^7) $

\vspace{5mm}

$F_{M}^{(n,n)}= e^{-2 y} (1440.-4330. y+4970. y^2-2770. y^3+807. y^4-120. y^5+7.92 y^6-0.148 y^7+0.000823 y^8) $

$F_{\Phi''}^{(n,n)}= e^{-2 y} (28.2-54.7 y+39.6 y^2-13.2 y^3+2.10 y^4-0.141 y^5+0.00329 y^6) $

$F_{M \Phi''}^{(n,n)}= e^{-2 y} (-202.+499. y-460. y^2+202. y^3-44.4 y^4+4.60 y^5-0.183 y^6+0.00165 y^7) $

\vspace{5mm}

\subsection*{$^{72}$Ge}

$F_{M}^{(p,p)}= e^{-2 y} (1020.-2830. y+2940. y^2-1460. y^3+366. y^4-44.9 y^5+2.22 y^6-0.0133 y^7+0.0000213 y^8) $

$F_{\Phi''}^{(p,p)}= e^{-2 y} (50.8-83.3 y+48.8 y^2-12.1 y^3+1.16 y^4-0.0190 y^5+0.0000852 y^6) $

$F_{M \Phi''}^{(p,p)}=e^{-2 y} (-228.+502. y-401. y^2+146. y^3-24.7 y^4+1.65 y^5-0.0181 y^6+0.0000426 y^7) $

\vspace{5mm}

$F_{M}^{(p,n)}= e^{-2 y} (1280.-3740. y+4160. y^2-2230. y^3+617. y^4-86.2 y^5+5.28 y^6-0.0847 y^7+0.000226 y^8) $

$F_{\Phi''}^{(p,n)}= e^{-2 y} (43.9-82.1 y+56.0 y^2-17.1 y^3+2.34 y^4-0.116 y^5+0.000904 y^6) $

$F_{M \Phi''}^{(p,n)}= e^{-2 y} (-197.+479. y-434. y^2+185. y^3-39.4 y^4+3.94 y^5-0.149 y^6+0.000452 y^7) $

\vspace{5mm}

$F_{M \Phi''}^{(n,p)}= e^{-2 y} (-285.+675. y-583. y^2+234. y^3-44.9 y^4+3.66 y^5-0.0786 y^6+0.000452 y^7) $

\vspace{5mm}

$F_{M}^{(n,n)}= e^{-2 y} (1600.-4950. y+5850. y^2-3380. y^3+1020. y^4-161. y^5+11.8 y^6-0.299 y^7+0.00240 y^8) $

$F_{\Phi''}^{(n,n)}= e^{-2 y} (37.9-79.7 y+62.4 y^2-22.8 y^3+4.04 y^4-0.326 y^5+0.00959 y^6) $

$F_{M \Phi''}^{(n,n)}= e^{-2 y} (-246.+640. y-623. y^2+290. y^3-68.4 y^4+7.89 y^5-0.381 y^6+0.00480 y^7) $

\vspace{5mm}

\subsection*{$^{73}$Ge}

$F_{M}^{(p,p)}= e^{-2 y} (1020.-2820. y+2930. y^2-1440. y^3+359. y^4-43.5 y^5+2.12 y^6-0.0120 y^7+0.0000181 y^8) $

$F_{\Sigma''}^{(p,p)}= e^{-2 y} (0.0000894-0.000566 y+0.00130 y^2-0.00143 y^3+0.000884 y^4-0.000260 y^5+0.0000309 y^6- 7.52\times10^{-7} y^7+7.35\times10^{-9} y^8) $

$F_{\Sigma'}^{(p,p)}= e^{-2 y} (0.000179-0.000880 y+0.00262 y^2-0.00412 y^3+0.00414 y^4-0.00165 y^5+0.000258 y^6-0.0000104 y^7+1.30\times10^{-7} y^8) $

$F_{\Phi''}^{(p,p)}= e^{-2 y} (46.5-76.3 y+44.6 y^2-11.0 y^3+1.05 y^4-0.0165 y^5+0.0000725 y^6)$

$F_{\tilde{\Phi}'}^{(p,p)}=  e^{-2 y} (0.000503-0.00143 y+0.00122 y^2-0.000269 y^3+0.0000160 y^4+1.07\times10^{-6} y^5+4.87\times10^{-8} y^6) $

$F_{\Delta}^{(p,p)}= e^{-2 y} (0.0120-0.0216 y+0.0152 y^2-0.00486 y^3+0.000688 y^4-0.0000353 y^5+5.92\times10^{-7} y^6) $

$F_{M \Phi''}^{(p,p)}= e^{-2 y} (-218.+480. y-382. y^2+138. y^3-23.1 y^4+1.52 y^5-0.0161 y^6+0.0000362 y^7) $

$F_{\Sigma' \Delta}^{(p,p)}= e^{-2 y} (-0.00146+0.00492 y-0.00985 y^2+0.00796 y^3-0.00280 y^4+0.000414 y^5-0.0000195 y^6+2.75\times10^{-7} y^7) $

\vspace{5mm}

$F_{M}^{(p,n)}= e^{-2 y} (1310.-3860. y+4310. y^2-2320. y^3+646. y^4-91.0 y^5+5.67 y^6-0.0981 y^7+0.000253 y^8)$

$F_{\Sigma''}^{(p,n)}= e^{-2 y} (0.00553-0.0264 y+0.0489 y^2-0.0505 y^3+0.0266 y^4-0.00721 y^5+0.000953 y^6-0.0000520 y^7+1.92\times10^{-6} y^8) $

$F_{\Sigma'}^{(p,n)}= e^{-2 y} (0.0111-0.0626 y+0.163 y^2-0.232 y^3+0.157 y^4-0.0509 y^5+0.00778 y^6-0.000497 y^7+0.0000108 y^8) $

$F_{\Phi''}^{(p,n)}= e^{-2 y} (45.8-86.9 y+60.0 y^2-18.6 y^3+2.58 y^4-0.131 y^5+0.00101 y^6) $

$F_{\tilde{\Phi}'}^{(p,n)}= e^{-2 y} (-0.00732+0.0173 y-0.0126 y^2+0.00399 y^3-0.000571 y^4+0.0000260 y^5+3.42\times10^{-7} y^6) $

$F_{\Delta}^{(p,n)}= e^{-2 y} (0.186-0.389 y+0.322 y^2-0.123 y^3+0.0222 y^4-0.00172 y^5+0.0000407 y^6) $

$F_{M \Phi''}^{(p,n)}= e^{-2 y} (-215.+528. y-484. y^2+209. y^3-44.9 y^4+4.56 y^5-0.176 y^6+0.000507 y^7) $

$F_{\Sigma' \Delta}^{(p,n)}= e^{-2 y} (-0.0227+0.0830 y-0.174 y^2+0.162 y^3-0.0679 y^4+0.0129 y^5-0.000974 y^6+0.0000195 y^7) $

\vspace{5mm}

$F_{M \Phi''}^{(n,p)}= e^{-2 y} (-279.+666. y-580. y^2+234. y^3-45.3 y^4+3.76 y^5-0.0854 y^6+0.000507 y^7) $

$F_{\Sigma' \Delta}^{(n,p)}= e^{-2 y} (-0.0906+0.371 y-0.498 y^2+0.305 y^3-0.0919 y^4+0.0136 y^5-0.000898 y^6+0.0000195 y^7) $

\vspace{5mm}

$F_{M}^{(n,n)}= e^{-2 y} (1670.-5250. y+6300. y^2-3690. y^3+1140. y^4-184. y^5+14.1 y^6-0.394 y^7+0.00361 y^8) $

$F_{\Sigma''}^{(n,n)}= e^{-2 y} (0.343-1.10 y+2.16 y^2-1.91 y^3+1.03 y^4-0.315 y^5+0.0649 y^6-0.00731 y^7+0.000989 y^8)$

$F_{\Sigma'}^{(n,n)}= e^{-2 y} (0.686-4.38 y+10.9 y^2-11.8 y^3+6.60 y^4-2.01 y^5+0.349 y^6-0.0323 y^7+0.00199 y^8) $

$F_{\Phi''}^{(n,n)}= e^{-2 y} (45.2-97.5 y+78.4 y^2-29.4 y^3+5.42 y^4-0.460 y^5+0.0145 y^6) $

$F_{\tilde{\Phi}'}^{(n,n)}= e^{-2 y} (0.107-0.200 y+0.157 y^2-0.0598 y^3+0.0158 y^4-0.00220 y^5+0.000163 y^6) $

$F_{\Delta}^{(n,n)}= e^{-2 y} (2.89-6.88 y+6.65 y^2-3.00 y^3+0.693 y^4-0.0779 y^5+0.00362 y^6) $

$F_{M \Phi''}^{(n,n)}= e^{-2 y} (-275.+727. y-723. y^2+343. y^3-83.1 y^4+9.94 y^5-0.509 y^6+0.00723 y^7) $

$F_{\Sigma' \Delta}^{(n,n)}= e^{-2 y} (-1.41+6.17 y-9.56 y^2+6.77 y^3-2.48 y^4+0.481 y^5-0.0483 y^6+0.00201 y^7)$

\vspace{5mm}

\subsection*{$^{74}$Ge}

$F_{M}^{(p,p)}= e^{-2 y} (1020.-2830. y+2920. y^2-1430. y^3+350. y^4-41.6 y^5+1.98 y^6-0.0117 y^7+0.0000184 y^8) $

$F_{\Phi''}^{(p,p)}= e^{-2 y} (41.5-68.2 y+39.8 y^2-9.84 y^3+0.937 y^4-0.0158 y^5+0.0000734 y^6)$

$F_{M \Phi''}^{(p,p)}= e^{-2 y} (-206.+454. y-361. y^2+130. y^3-21.6 y^4+1.40 y^5-0.0156 y^6+0.0000367 y^7) $

\vspace{5mm}

$F_{M}^{(p,n)}= e^{-2 y} (1340.-4000. y+4500. y^2-2440. y^3+682. y^4-96.8 y^5+6.17 y^6-0.118 y^7+0.000324 y^8) $

$F_{\Phi''}^{(p,n)}= e^{-2 y} (50.8-97.8 y+68.4 y^2-21.4 y^3+3.03 y^4-0.159 y^5+0.00130 y^6) $

$F_{M \Phi''}^{(p,n)}= e^{-2 y} (-252.+627. y-579. y^2+251. y^3-54.1 y^4+5.52 y^5-0.216 y^6+0.000649 y^7) $

\vspace{5mm}

$F_{M \Phi''}^{(n,p)}= e^{-2 y} (-270.+653. y-575. y^2+235. y^3-46.2 y^4+3.96 y^5-0.101 y^6+0.000649 y^7) $

\vspace{5mm}

$F_{M}^{(n,n)}= e^{-2 y} (1760.-5620. y+6860. y^2-4090. y^3+1290. y^4-214. y^5+17.3 y^6-0.545 y^7+0.00573 y^8) $

$F_{\Phi''}^{(n,n)}= e^{-2 y} (62.3-137. y+113. y^2-43.1 y^3+8.12 y^4-0.709 y^5+0.0229 y^6) $

$F_{M \Phi''}^{(n,n)}= e^{-2 y} (-331.+893. y-904. y^2+438. y^3-108. y^4+13.4 y^5-0.723 y^6+0.0115 y^7) $

\vspace{5mm}

\subsection*{$^{76}$Ge}

$F_{M}^{(p,p)}= e^{-2 y} (1020.-2830. y+2910. y^2-1410. y^3+343. y^4-40.1 y^5+1.87 y^6-0.0105 y^7+0.0000159 y^8) $

$F_{\Phi''}^{(p,p)}=e^{-2 y} (37.4-61.4 y+35.9 y^2-8.84 y^3+0.837 y^4-0.0139 y^5+0.0000634 y^6) $

$F_{M \Phi''}^{(p,p)}= e^{-2 y} (-196.+431. y-341. y^2+122. y^3-20.1 y^4+1.29 y^5-0.0140 y^6+0.0000317 y^7) $

\vspace{5mm}

$F_{M}^{(p,n)}= e^{-2 y} (1410.-4240. y+4830. y^2-2650. y^3+748. y^4-108. y^5+7.04 y^6-0.147 y^7+0.000394 y^8)  $

$F_{\Phi''}^{(p,n)}= e^{-2 y} (55.8-109. y+77.9 y^2-24.8 y^3+3.59 y^4-0.195 y^5+0.00158 y^6) $

$F_{M \Phi''}^{(p,n)}= e^{-2 y} (-292.+736. y-689. y^2+302. y^3-65.9 y^4+6.83 y^5-0.273 y^6+0.000788 y^7) $

\vspace{5mm}

$F_{M \Phi''}^{(n,p)}=e^{-2 y} (-269.+661. y-590. y^2+245. y^3-49.0 y^4+4.32 y^5-0.119 y^6+0.000788 y^7)$

\vspace{5mm}

$F_{M}^{(n,n)}= e^{-2 y} (1940.-6320. y+7910. y^2-4840. y^3+1570. y^4-271. y^5+23.1 y^6-0.813 y^7+0.00979 y^8) $

$F_{\Phi''}^{(n,n)}= e^{-2 y} (83.2-190. y+161. y^2-63.8 y^3+12.5 y^4-1.14 y^5+0.0392 y^6) $

$F_{M \Phi''}^{(n,n)}= e^{-2 y} (-401.+1110. y-1160. y^2+578. y^3-148. y^4+19.1 y^5-1.10 y^6+0.0196 y^7) $

\vspace{5mm}

\subsubsection{jj44b}

\subsection*{$^{70}$Ge}

$F_{M}^{(p,p)}= e^{-2 y} (1020.-2830. y+2930. y^2-1430. y^3+352. y^4-42.2 y^5+2.06 y^6-0.0163 y^7+0.0000353 y^8) $

$F_{\Phi''}^{(p,p)}= e^{-2 y} (42.4-70.4 y+41.2 y^2-10.1 y^3+0.984 y^4-0.0220 y^5+0.000141 y^6) $

$F_{M \Phi''}^{(p,p)}= e^{-2 y} (-208.+461. y-367. y^2+132. y^3-22.0 y^4+1.48 y^5-0.0218 y^6+0.0000707 y^7) $

\vspace{5mm}

$F_{M}^{(p,n)}= e^{-2 y} (1210.-3520. y+3870. y^2-2040. y^3+552. y^4-75.3 y^5+4.54 y^6-0.0768 y^7+0.000263 y^8) $

$F_{\Phi''}^{(p,n)}= e^{-2 y} (47.8-87.0 y+57.4 y^2-16.8 y^3+2.18 y^4-0.103 y^5+0.00105 y^6) $

$F_{M \Phi''}^{(p,n)}= e^{-2 y} (-235.+557. y-489. y^2+201. y^3-40.5 y^4+3.78 y^5-0.132 y^6+0.000526 y^7) $

\vspace{5mm}

$F_{M \Phi''}^{(n,p)}= e^{-2 y} (-247.+581. y-497. y^2+196. y^3-37.0 y^4+3.03 y^5-0.0732 y^6+0.000526 y^7) $

\vspace{5mm}

$F_{M}^{(n,n)}= e^{-2 y} (1440.-4380. y+5080. y^2-2870. y^3+850. y^4-131. y^5+9.43 y^6-0.240 y^7+0.00196 y^8) $

$F_{\Phi''}^{(n,n)}= e^{-2 y} (53.9-107. y+78.5 y^2-26.7 y^3+4.34 y^4-0.309 y^5+0.00782 y^6) $

$F_{M \Phi''}^{(n,n)}= e^{-2 y} (-279.+699. y-655. y^2+292. y^3-65.5 y^4+7.11 y^5-0.318 y^6+0.00391 y^7)$

\vspace{5mm}

\subsection*{$^{72}$Ge}

$F_{M}^{(p,p)}= e^{-2 y} (1020.-2830. y+2920. y^2-1420. y^3+347. y^4-41.0 y^5+1.96 y^6-0.0138 y^7+0.0000265 y^8) $

$F_{\Phi''}^{(p,p)}= e^{-2 y} (39.4-65.1 y+38.0 y^2-9.31 y^3+0.891 y^4-0.0182 y^5+0.000106 y^6) $

$F_{M \Phi''}^{(p,p)}= e^{-2 y} (-201.+443. y-352. y^2+126. y^3-20.8 y^4+1.37 y^5-0.0184 y^6+0.0000529 y^7) $

\vspace{5mm}

$F_{M}^{(p,n)}= e^{-2 y} (1280.-3760. y+4190. y^2-2240. y^3+617. y^4-86.0 y^5+5.38 y^6-0.102 y^7+0.000329 y^8) $

$F_{\Phi''}^{(p,n)}= e^{-2 y} (51.3-96.0 y+65.3 y^2-19.8 y^3+2.69 y^4-0.136 y^5+0.00132 y^6) $

$F_{M \Phi''}^{(p,n)}= e^{-2 y} (-261.+634. y-571. y^2+241. y^3-50.3 y^4+4.92 y^5-0.183 y^6+0.000658 y^7)$

\vspace{5mm}

$F_{M \Phi''}^{(n,p)}= e^{-2 y} (-251.+599. y-521. y^2+210. y^3-40.4 y^4+3.42 y^5-0.0895 y^6+0.000658 y^7) $

\vspace{5mm}

$F_{M}^{(n,n)}=  e^{-2 y} (1600.-4990. y+5960. y^2-3480. y^3+1070. y^4-173. y^5+13.5 y^6-0.408 y^7+0.00409 y^8) $

$F_{\Phi''}^{(n,n)}= e^{-2 y} (66.7-140. y+109. y^2-39.6 y^3+7.00 y^4-0.561 y^5+0.0163 y^6) $

$F_{M \Phi''}^{(n,n)}= e^{-2 y} (-326.+851. y-832. y^2+389. y^3-92.4 y^4+10.8 y^5-0.548 y^6+0.00817 y^7) $

\vspace{5mm}

\subsection*{$^{73}$Ge}

$F_{M}^{(p,p)}= e^{-2 y} (1020.-2820. y+2910. y^2-1410. y^3+342. y^4-40.1 y^5+1.88 y^6-0.0113 y^7+0.0000182 y^8) $

$F_{\Sigma''}^{(p,p)}= e^{-2 y} (8.79\times10^{-6}-0.000129 y+0.000564 y^2-0.000702 y^3+0.000435 y^4-0.000116 y^5+0.0000130 y^6-3.47\times10^{-7} y^7+3.47\times10^{-9} y^8) $

$F_{\Sigma'}^{(p,p)}= e^{-2 y} (0.0000176-0.0000905 y+0.000558 y^2-0.00125 y^3+0.00306 y^4-0.00160 y^5+0.000270 y^6-9.82\times10^{-6} y^7+1.06\times10^{-7} y^8) $

$F_{\Phi''}^{(p,p)}= e^{-2 y} (36.9-60.7 y+35.2 y^2-8.48 y^3+0.789 y^4-0.0143 y^5+0.0000727 y^6) $

$F_{\tilde{\Phi}'}^{(p,p)}= e^{-2 y} (0.000354-0.00150 y+0.00208 y^2-0.000857 y^3+0.000119 y^4-9.78\times10^{-7} y^5+1.46\times10^{-8} y^6) $

$F_{\Delta}^{(p,p)}= e^{-2 y} (0.0243-0.0423 y+0.0286 y^2-0.00901 y^3+0.00123 y^4-0.0000454 y^5+5.00\times10^{-7} y^6) $

$F_{M \Phi''}^{(p,p)}= e^{-2 y} (-194.+428. y-338. y^2+120. y^3-19.5 y^4+1.25 y^5-0.0148 y^6+0.0000363 y^7) $

$F_{\Sigma' \Delta}^{(p,p)}= e^{-2 y} (-0.000654+0.00225 y-0.00976 y^2+0.00981 y^3-0.00382 y^4+0.000575 y^5-0.0000211 y^6+2.29\times10^{-7} y^7) $

\vspace{5mm}

$F_{M}^{(p,n)}= e^{-2 y} (1310.-3870. y+4330. y^2-2330. y^3+647. y^4-91.0 y^5+5.79 y^6-0.115 y^7+0.000322 y^8) $

$F_{\Sigma''}^{(p,n)}= e^{-2 y} (0.00159-0.0145 y+0.0313 y^2-0.0339 y^3+0.0173 y^4-0.00458 y^5+0.000606 y^6-0.0000333 y^7+1.00\times10^{-6} y^8) $

$F_{\Sigma'}^{(p,n)}= e^{-2 y} (0.00319-0.0183 y+0.0742 y^2-0.157 y^3+0.123 y^4-0.0422 y^5+0.00654 y^6-0.000417 y^7+8.05\times10^{-6} y^8)$

$F_{\Phi''}^{(p,n)}= e^{-2 y} (54.1-102. y+70.1 y^2-21.4 y^3+2.92 y^4-0.148 y^5+0.00129 y^6) $

$F_{\tilde{\Phi}'}^{(p,n)}= e^{-2 y} (-0.00169+0.00611 y-0.00711 y^2+0.00268 y^3-0.000388 y^4+0.0000192 y^5+1.23\times10^{-7} y^6) $

$F_{\Delta}^{(p,n)}=  e^{-2 y} (0.260-0.535 y+0.420 y^2-0.155 y^3+0.0268 y^4-0.00188 y^5+0.0000332 y^6)$

$F_{M \Phi''}^{(p,n)}= e^{-2 y} (-285.+697. y-633. y^2+269. y^3-56.7 y^4+5.61 y^5-0.211 y^6+0.000644 y^7) $

$F_{\Sigma' \Delta}^{(p,n)}= e^{-2 y} (-0.00700+0.0263 y-0.112 y^2+0.135 y^3-0.0613 y^4+0.0117 y^5-0.000855 y^6+0.0000156 y^7) $

\vspace{5mm}

$F_{M \Phi''}^{(n,p)}= e^{-2 y} (-248.+597. y-522. y^2+211. y^3-40.9 y^4+3.48 y^5-0.0924 y^6+0.000644 y^7) $

$F_{\Sigma' \Delta}^{(n,p)}=  e^{-2 y} (-0.119+0.480 y-0.610 y^2+0.357 y^3-0.105 y^4+0.0151 y^5-0.000923 y^6+0.0000156 y^7) $

\vspace{5mm}

$F_{M}^{(n,n)}= e^{-2 y} (1670.-5290. y+6400. y^2-3790. y^3+1190. y^4-196. y^5+15.9 y^6-0.519 y^7+0.00573 y^8) $

$F_{\Sigma''}^{(n,n)}= e^{-2 y} (0.289-1.02 y+1.97 y^2-1.71 y^3+0.884 y^4-0.260 y^5+0.0522 y^6-0.00581 y^7+0.000824 y^8)$

$F_{\Sigma'}^{(n,n)}= e^{-2 y} (0.578-3.68 y+9.02 y^2-9.63 y^3+5.35 y^4-1.63 y^5+0.285 y^6-0.0267 y^7+0.00168 y^8) $

$F_{\Phi''}^{(n,n)}= e^{-2 y} (79.4-169. y+135. y^2-50.1 y^3+9.10 y^4-0.755 y^5+0.0229 y^6) $

$F_{\tilde{\Phi}'}^{(n,n)}= e^{-2 y} (0.00804-0.0242 y+0.0240 y^2-0.00834 y^3+0.00222 y^4-0.000337 y^5+0.0000207 y^6) $

$F_{\Delta}^{(n,n)}=e^{-2 y} (2.78-6.60 y+6.26 y^2-2.77 y^3+0.627 y^4-0.0690 y^5+0.00313 y^6) $

$F_{M \Phi''}^{(n,n)}= e^{-2 y} (-364.+965. y-959. y^2+456. y^3-111. y^4+13.4 y^5-0.708 y^6+0.0115 y^7)$

$F_{\Sigma' \Delta}^{(n,n)}= e^{-2 y} (-1.27+5.54 y-8.43 y^2+5.89 y^3-2.13 y^4+0.412 y^5-0.0411 y^6+0.00171 y^7)$

\vspace{5mm}

\subsection*{$^{74}$Ge}

$F_{M}^{(p,p)}= e^{-2 y} (1020.-2830. y+2910. y^2-1410. y^3+343. y^4-40.2 y^5+1.88 y^6-0.0116 y^7+0.0000190 y^8) $

$F_{\Phi''}^{(p,p)}= e^{-2 y} (37.0-60.8 y+35.3 y^2-8.59 y^3+0.807 y^4-0.0148 y^5+0.0000761 y^6) $

$F_{M \Phi''}^{(p,p)}=e^{-2 y} (-195.+429. y-339. y^2+121. y^3-19.8 y^4+1.28 y^5-0.0153 y^6+0.0000381 y^7) $

\vspace{5mm}

$F_{M}^{(p,n)}= e^{-2 y} (1340.-4010. y+4520. y^2-2450. y^3+686. y^4-97.6 y^5+6.30 y^6-0.128 y^7+0.000370 y^8) $

$F_{\Phi''}^{(p,n)}= e^{-2 y} (56.0-107. y+74.2 y^2-23.0 y^3+3.20 y^4-0.167 y^5+0.00148 y^6) $

$F_{M \Phi''}^{(p,n)}= e^{-2 y} (-294.+727. y-667. y^2+286. y^3-60.9 y^4+6.12 y^5-0.236 y^6+0.000739 y^7) $

\vspace{5mm}

$F_{M \Phi''}^{(n,p)}= e^{-2 y} (-255.+619. y-547. y^2+223. y^3-43.9 y^4+3.81 y^5-0.104 y^6+0.000739 y^7)$

\vspace{5mm}

$F_{M}^{(n,n)}= e^{-2 y} (1760.-5650. y+6930. y^2-4170. y^3+1330. y^4-224. y^5+18.5 y^6-0.627 y^7+0.00718 y^8) $

$F_{\Phi''}^{(n,n)}= e^{-2 y} (84.7-185. y+150. y^2-56.6 y^3+10.5 y^4-0.904 y^5+0.0287 y^6) $

$F_{M \Phi''}^{(n,n)}= e^{-2 y} (-386.+1040. y-1050. y^2+508. y^3-126. y^4+15.6 y^5-0.853 y^6+0.0144 y^7) $

\vspace{5mm}

\subsection*{$^{76}$Ge}

$F_{M}^{(p,p)}= e^{-2 y} (1020.-2820. y+2910. y^2-1400. y^3+339. y^4-39.3 y^5+1.81 y^6-0.00953 y^7+0.0000134 y^8)$

$F_{\Phi''}^{(p,p)}= e^{-2 y} (34.5-56.6 y+32.7 y^2-7.88 y^3+0.726 y^4-0.0119 y^5+0.0000536 y^6) $

$F_{M \Phi''}^{(p,p)}= e^{-2 y} (-188.+413. y-326. y^2+116. y^3-18.7 y^4+1.18 y^5-0.0125 y^6+0.0000268 y^7) $

\vspace{5mm}

$F_{M}^{(p,n)}= e^{-2 y} (1410.-4250. y+4850. y^2-2660. y^3+755. y^4-109. y^5+7.15 y^6-0.151 y^7+0.000381 y^8) $

$F_{\Phi''}^{(p,n)}= e^{-2 y} (59.4-115. y+81.1 y^2-25.4 y^3+3.59 y^4-0.191 y^5+0.00152 y^6) $

$F_{M \Phi''}^{(p,n)}= e^{-2 y} (-323.+810. y-752. y^2+326. y^3-70.3 y^4+7.19 y^5-0.283 y^6+0.000762 y^7) $

\vspace{5mm}

$F_{M \Phi''}^{(n,p)}= e^{-2 y} (-258.+635. y-568. y^2+235. y^3-46.9 y^4+4.14 y^5-0.115 y^6+0.000762 y^7) $

\vspace{5mm}

$F_{M}^{(n,n)}= e^{-2 y} (1940.-6340. y+7970. y^2-4910. y^3+1610. y^4-280. y^5+24.3 y^6-0.876 y^7+0.0109 y^8) $

$F_{\Phi''}^{(n,n)}= e^{-2 y} (102.-230. y+192. y^2-74.5 y^3+14.3 y^4-1.29 y^5+0.0434 y^6) $

$F_{M \Phi''}^{(n,n)}= e^{-2 y} (-445.+1230. y-1270. y^2+633. y^3-161. y^4+20.8 y^5-1.20 y^6+0.0217 y^7) $

\vspace{5mm}

\subsection{$^{127}$I}

\subsubsection{SN100PN}

$F_{M}^{(p,p)}= e^{-2 y} (2810.-10000. y+13800. y^2-9430. y^3+3480. y^4-711. y^5+77.8 y^6-4.14 y^7+0.0861 y^8-0.0000932 y^9+2.68\times10^{-8} y^{10})$

$F_{\Sigma''}^{(p,p)}=  e^{-2 y} (0.0666-0.117 y+0.300 y^2-0.382 y^3+0.575 y^4-0.464 y^5+0.186 y^6-0.0368 y^7+0.00291 y^8-6.45\times10^{-8} y^9+1.20\times10^{-12} y^{10})$

$F_{\Sigma'}^{(p,p)}= e^{-2 y} (0.133-0.950 y+3.14 y^2-5.43 y^3+5.52 y^4-3.14 y^5+0.964 y^6-0.147 y^7+0.00884 y^8-6.45\times10^{-7} y^9+3.39\times10^{-11} y^{10}) $

$F_{\Phi''}^{(p,p)}=e^{-2 y} (105.-253. y+227. y^2-95.7 y^3+20.3 y^4-2.09 y^5+0.0846 y^6-0.000231 y^7+1.67\times10^{-7} y^8) $

$F_{\tilde{\Phi}'}^{(p,p)}= e^{-2 y} (0.0354-0.0182 y+0.00494 y^2-0.00995 y^3+0.00854 y^4-0.00241 y^5+0.000273 y^6+1.44\times10^{-7} y^7+9.85\times10^{-11} y^8)  $

$F_{\Delta}^{(p,p)}= e^{-2 y} (0.705-1.69 y+1.83 y^2-1.09 y^3+0.379 y^4-0.0688 y^5+0.00498 y^6-1.43\times10^{-6} y^7+2.38\times10^{-10} y^8) $

$F_{M \Phi''}^{(p,p)}= e^{-2 y} (-543.+1620. y-1830. y^2+992. y^3-282. y^4+42.2 y^5-3.06 y^6+0.0833 y^7-0.000163 y^8+6.69\times10^{-8} y^9)$

$F_{\Sigma' \Delta}^{(p,p)}= e^{-2 y} (-0.306+1.46 y-2.89 y^2+2.91 y^3-1.59 y^4+0.465 y^5-0.0670 y^6+0.00364 y^7-1.09\times10^{-6} y^8+8.96\times10^{-11} y^9) $

\vspace{5mm}

$F_{M}^{(p,n)}= e^{-2 y} (3920.-15300. y+23200. y^2-17800. y^3+7640. y^4-1880. y^5+262. y^6-19.4 y^7+0.657 y^8-0.00705 y^9+3.85\times10^{-6} y^{10})  $

$F_{\Sigma''}^{(p,n)}= e^{-2 y} (0.0126-0.0388 y+0.0459 y^2-0.0294 y^3+0.0120 y^4-0.00375 y^5+0.000931 y^6-0.000156 y^7+0.0000118 y^8+5.76\times10^{-9} y^9+2.22\times10^{-11} y^{10}) $

$F_{\Sigma'}^{(p,n)}= e^{-2 y} (0.0251-0.190 y+0.565 y^2-0.853 y^3+0.701 y^4-0.322 y^5+0.0840 y^6-0.0121 y^7+0.000839 y^8-0.0000189 y^9+1.69\times10^{-9} y^{10}) $

$F_{\Phi''}^{(p,n)}= e^{-2 y} (162.-441. y+462. y^2-238. y^3+64.9 y^4-9.35 y^5+0.656 y^6-0.0172 y^7+0.0000241 y^8) $

$F_{\tilde{\Phi}'}^{(p,n)}= e^{-2 y} (-0.136+0.203 y-0.0825 y^2-0.00934 y^3+0.0137 y^4-0.00327 y^5+0.000299 y^6-8.77\times10^{-6} y^7+1.05\times10^{-8} y^8)$

$F_{\Delta}^{(p,n)}=  e^{-2 y} (0.218-0.586 y+0.656 y^2-0.392 y^3+0.131 y^4-0.0233 y^5+0.00197 y^6-0.0000554 y^7+1.24\times10^{-8} y^8) $

$F_{M \Phi''}^{(p,n)}= e^{-2 y} (-840.+2780. y-3570. y^2+2300. y^3-813. y^4+160. y^5-16.8 y^6+0.866 y^7-0.0172 y^8+9.63\times10^{-6} y^9)$

$F_{\Sigma' \Delta}^{(p,n)}= e^{-2 y} (-0.0946+0.479 y-0.976 y^2+0.995 y^3-0.545 y^4+0.163 y^5-0.0259 y^6+0.00197 y^7-0.0000513 y^8+4.59\times10^{-9} y^9) $

\vspace{5mm}

$F_{M \Phi''}^{(n,p)}= e^{-2 y} (-757.+2510. y-3160. y^2+1980. y^3-670. y^4+126. y^5-12.5 y^6+0.571 y^7-0.00733 y^8+9.63\times10^{-6} y^9) $

$F_{\Sigma' \Delta}^{(n,p)}= e^{-2 y} (-0.0578+0.300 y-0.532 y^2+0.474 y^3-0.240 y^4+0.0694 y^5-0.0112 y^6+0.000872 y^7-0.0000205 y^8+4.59\times10^{-9} y^9) $

\vspace{5mm}

$F_{M}^{(n,n)}=e^{-2 y} (5470.-23000. y+38300. y^2-32900. y^3+16100. y^4-4650. y^5+796. y^6-77.5 y^7+3.88 y^8-0.0798 y^9+0.000556 y^{10}) $

$F_{\Sigma''}^{(n,n)}= e^{-2 y} (0.00237-0.0105 y+0.0182 y^2-0.0162 y^3+0.00823 y^4-0.00252 y^5+0.000479 y^6-0.0000558 y^7+3.73\times10^{-6} y^8-1.15\times10^{-7} y^9+1.45\times10^{-9} y^{10}) $

$F_{\Sigma'}^{(n,n)}= e^{-2 y} (0.00474-0.0378 y+0.112 y^2-0.157 y^3+0.119 y^4-0.0522 y^5+0.0138 y^6-0.00218 y^7+0.000190 y^8-7.45\times10^{-6} y^9+1.04\times10^{-7} y^{10}) $

$F_{\Phi''}^{(n,n)}= e^{-2 y} (251.-764. y+918. y^2-561. y^3+189. y^4-35.8 y^5+3.66 y^6-0.183 y^7+0.00348 y^8) $

$F_{\tilde{\Phi}'}^{(n,n)}= e^{-2 y} (0.521-1.29 y+1.27 y^2-0.646 y^3+0.186 y^4-0.0308 y^5+0.00283 y^6-0.000130 y^7+2.29\times10^{-6} y^8)$

$F_{\Delta}^{(n,n)}= e^{-2 y} (0.0672-0.201 y+0.238 y^2-0.145 y^3+0.0487 y^4-0.00911 y^5+0.000909 y^6-0.0000432 y^7+7.63\times10^{-7} y^8) $

$F_{M \Phi''}^{(n,n)}=e^{-2 y} (-1170.+4250. y-6050. y^2+4410. y^3-1810. y^4+427. y^5-57.6 y^6+4.15 y^7-0.136 y^8+0.00139 y^9) $

$F_{\Sigma' \Delta}^{(n,n)}= e^{-2 y} (-0.0179+0.0979 y-0.187 y^2+0.173 y^3-0.0884 y^4+0.0261 y^5-0.00447 y^6+0.000416 y^7-0.0000181 y^8+2.82\times10^{-7} y^9) $

\vspace{5mm}

\subsubsection{GCN5082}

$F_{M}^{(p,p)}= e^{-2 y} (2810.-10000. y+13900. y^2-9570. y^3+3600. y^4-756. y^5+86.4 y^6-4.91 y^7+0.112 y^8-0.000190 y^9+8.80\times10^{-8} y^{10}) $

$F_{\Sigma''}^{(p,p)}= e^{-2 y} (0.183-0.594 y+1.38 y^2-1.72 y^3+1.70 y^4-1.08 y^5+0.388 y^6-0.0714 y^7+0.00535 y^8-3.62\times10^{-7} y^9+1.22\times10^{-11} y^{10})$

$F_{\Sigma'}^{(p,p)}= e^{-2 y} (0.365-2.33 y+6.76 y^2-10.3 y^3+9.33 y^4-4.93 y^5+1.45 y^6-0.220 y^7+0.0134 y^8-1.75\times10^{-6} y^9+1.60\times10^{-10} y^{10}) $

$F_{\Phi''}^{(p,p)}= e^{-2 y} (130.-313. y+281. y^2-119. y^3+25.5 y^4-2.68 y^5+0.113 y^6-0.000478 y^7+5.50\times10^{-7} y^8) $

$F_{\tilde{\Phi}'}^{(p,p)}= e^{-2 y} (0.00204+0.00411 y+0.0323 y^2-0.0523 y^3+0.0316 y^4-0.00829 y^5+0.000898 y^6+1.03\times10^{-6} y^7+3.45\times10^{-10} y^8)$

$F_{\Delta}^{(p,p)}=  e^{-2 y} (0.574-1.38 y+1.64 y^2-1.10 y^3+0.438 y^4-0.0891 y^5+0.00711 y^6-2.60\times10^{-6} y^7+1.03\times10^{-9} y^8) $

$F_{M \Phi''}^{(p,p)}= e^{-2 y} (-604.+1810. y-2040. y^2+1120. y^3-325. y^4+49.8 y^5-3.77 y^6+0.111 y^7-0.000333 y^8+2.20\times10^{-7} y^9)$

$F_{\Sigma' \Delta}^{(p,p)}= e^{-2 y} (-0.458+2.01 y-3.74 y^2+3.74 y^3-2.12 y^4+0.665 y^5-0.104 y^6+0.00612 y^7-2.58\times10^{-6} y^8+4.05\times10^{-10} y^9) $

\vspace{5mm}

$F_{M}^{(p,n)}= e^{-2 y} (3910.-15200. y+23100. y^2-17900. y^3+7710. y^4-1920. y^5+273. y^6-20.6 y^7+0.711 y^8-0.00733 y^9+6.21\times10^{-6} y^{10}) $

$F_{\Sigma''}^{(p,n)}= e^{-2 y} (0.0225-0.0887 y+0.158 y^2-0.161 y^3+0.0991 y^4-0.0376 y^5+0.00873 y^6-0.00118 y^7+0.0000761 y^8-1.20\times10^{-6} y^9+3.85\times10^{-11} y^{10}) $

$F_{\Sigma'}^{(p,n)}= e^{-2 y} (0.0451-0.324 y+0.908 y^2-1.28 y^3+0.990 y^4-0.436 y^5+0.110 y^6-0.0156 y^7+0.00110 y^8-0.0000284 y^9+3.27\times10^{-9} y^{10}) $

$F_{\Phi''}^{(p,n)}= e^{-2 y} (160.-437. y+459. y^2-237. y^3+65.1 y^4-9.45 y^5+0.670 y^6-0.0180 y^7+0.0000388 y^8)$

$F_{\tilde{\Phi}'}^{(p,n)}= e^{-2 y} (-0.0260+0.00668 y+0.0498 y^2-0.0535 y^3+0.0209 y^4-0.00377 y^5+0.000292 y^6-5.28\times10^{-6} y^7+5.46\times10^{-9} y^8) $

$F_{\Delta}^{(p,n)}=  e^{-2 y} (0.210-0.570 y+0.666 y^2-0.428 y^3+0.154 y^4-0.0293 y^5+0.00263 y^6-0.0000789 y^7+2.41\times10^{-8} y^8) $

$F_{M \Phi''}^{(p,n)}= e^{-2 y} (-745.+2470. y-3180. y^2+2070. y^3-738. y^4+147. y^5-15.9 y^6+0.844 y^7-0.0175 y^8+0.0000155 y^9) $

$F_{\Sigma' \Delta}^{(p,n)}= e^{-2 y} (-0.168+0.788 y-1.50 y^2+1.45 y^3-0.768 y^4+0.227 y^5-0.0360 y^6+0.00277 y^7-0.0000758 y^8+8.89\times10^{-9} y^9) $

\vspace{5mm}

$F_{M \Phi''}^{(n,p)}= e^{-2 y} (-841.+2780. y-3510. y^2+2200. y^3-751. y^4+142. y^5-14.2 y^6+0.649 y^7-0.00805 y^8+0.0000155 y^9)$

$F_{\Sigma' \Delta}^{(n,p)}= e^{-2 y} (-0.0565+0.293 y-0.522 y^2+0.482 y^3-0.256 y^4+0.0777 y^5-0.0130 y^6+0.00107 y^7-0.0000293 y^8+8.89\times10^{-9} y^9)$

\vspace{5mm}

$F_{M}^{(n,n)}= e^{-2 y} (5450.-22900. y+38000. y^2-32600. y^3+15900. y^4-4620. y^5+792. y^6-76.8 y^7+3.76 y^8-0.0710 y^9+0.000442 y^{10}) $

$F_{\Sigma''}^{(n,n)}= e^{-2 y} (0.00278-0.0128 y+0.0234 y^2-0.0218 y^3+0.0117 y^4-0.00384 y^5+0.000786 y^6-0.0000980 y^7+6.83\times10^{-6} y^8-2.00\times10^{-7} y^9+2.12\times10^{-9} y^{10})$

$F_{\Sigma'}^{(n,n)}= e^{-2 y} (0.00557-0.0444 y+0.130 y^2-0.178 y^3+0.130 y^4-0.0552 y^5+0.0142 y^6-0.00220 y^7+0.000195 y^8-8.36\times10^{-6} y^9+1.35\times10^{-7} y^{10}) $

$F_{\Phi''}^{(n,n)}= e^{-2 y} (198.-603. y+725. y^2-445. y^3+151. y^4-28.5 y^5+2.92 y^6-0.146 y^7+0.00276 y^8) $

$F_{\tilde{\Phi}'}^{(n,n)}= e^{-2 y} (0.332-0.841 y+0.845 y^2-0.439 y^3+0.130 y^4-0.0224 y^5+0.00212 y^6-0.0000928 y^7+1.48\times10^{-6} y^8)$

$F_{\Delta}^{(n,n)}= e^{-2 y} (0.0771-0.232 y+0.278 y^2-0.170 y^3+0.0574 y^4-0.0108 y^5+0.00110 y^6-0.0000537 y^7+9.87\times10^{-7} y^8) $

$F_{M \Phi''}^{(n,n)}= e^{-2 y} (-1040.+3760. y-5350. y^2+3910. y^3-1600. y^4+380. y^5-51.3 y^6+3.67 y^7-0.118 y^8+0.00111 y^9) $

$F_{\Sigma' \Delta}^{(n,n)}=e^{-2 y} (-0.0207+0.114 y-0.215 y^2+0.197 y^3-0.0991 y^4+0.0289 y^5-0.00491 y^6+0.000463 y^7-0.0000212 y^8+3.65\times10^{-7} y^9) $

\vspace{5mm}

\subsection{$^{128, 129, 130, 131, 132, 134, 136}$Xe}

\subsection*{$^{128}$Xe}

$F_{M}^{(p,p)}= e^{-2 y} (2920.-10500. y+14700. y^2-10200. y^3+3850. y^4-810. y^5+92.9 y^6-5.31 y^7+0.123 y^8-0.000385 y^9+3.24\times10^{-7} y^{10}) $

$F_{\Phi''}^{(p,p)}=e^{-2 y} (93.8-228. y+206. y^2-87.8 y^3+19.0 y^4-2.04 y^5+0.0921 y^6-0.000816 y^7+2.03\times10^{-6} y^8) $

$F_{M \Phi''}^{(p,p)}= e^{-2 y} (-523.+1580. y-1800. y^2+995. y^3-290. y^4+45.0 y^5-3.49 y^6+0.110 y^7-0.000645 y^8+8.10\times10^{-7} y^9) $

\vspace{5mm}

$F_{M}^{(p,n)}= e^{-2 y} (4000.-15600. y+23800. y^2-18400. y^3+7920. y^4-1960. y^5+277. y^6-21. y^7+0.748 y^8-0.00895 y^9+0.0000135 y^{10}) $

$F_{\Phi''}^{(p,n)}= e^{-2 y} (148.-407. y+432. y^2-227. y^3+63.1 y^4-9.32 y^5+0.676 y^6-0.0192 y^7+0.0000844 y^8) $

$F_{M \Phi''}^{(p,n)}= e^{-2 y} (-825.+2760. y-3590. y^2+2360. y^3-854. y^4+172. y^5-18.7 y^6+1. y^7-0.021 y^8+0.0000337 y^9) $

\vspace{5mm}

$F_{M \Phi''}^{(n,p)}= e^{-2 y} (-717.+2380. y-3010. y^2+1880. y^3-640. y^4+121. y^5-12.2 y^6+0.581 y^7-0.00911 y^8+0.0000337 y^9)$

\vspace{5mm}

$F_{M}^{(n,n)}= e^{-2 y} (5480.-23100. y+38300. y^2-32600. y^3+15800. y^4-4510. y^5+763. y^6-73.5 y^7+3.66 y^8-0.0774 y^9+0.000562 y^{10}) $

$F_{\Phi''}^{(n,n)}=e^{-2 y} (234.-717. y+876. y^2-547. y^3+189. y^4-36.4 y^5+3.77 y^6-0.188 y^7+0.00351 y^8) $

$F_{M \Phi''}^{(n,n)}= e^{-2 y} (-1130.+4120. y-5890. y^2+4320. y^3-1780. y^4+422. y^5-57.0 y^6+4.09 y^7-0.134 y^8+0.00140 y^9)$

\vspace{5mm}

\subsection*{$^{129}$Xe}
  
$F_{M}^{(p,p)}=e^{-2 y} (2920.-10500. y+14700. y^2-10200. y^3+3860. y^4-816. y^5+94. y^6-5.42 y^7+0.127 y^8-0.000405 y^9+3.47\times10^{-7} y^{10}) $

$F_{\Sigma''}^{(p,p)}= e^{-2 y} (5.81\times10^{-8}+4.81\times10^{-6} y+0.0000936 y^2-0.000254 y^3+0.000262 y^4-0.000137 y^5+0.000038 y^6-5.24\times10^{-6} y^7+2.84\times10^{-7} y^8-3.82\times10^{-10} y^9+1.3\times10^{-13} y^{10}) $

$F_{\Sigma'}^{(p,p)}= e^{-2 y} (1.16\times10^{-7}-5.25\times10^{-6} y+0.0000394 y^2+0.000466 y^3+0.000575 y^4-0.00106 y^5+0.000461 y^6-0.0000728 y^7+3.91\times10^{-6} y^8-1.19\times10^{-8} y^9+ 9.35\times10^{-12} y^{10}) $

$F_{\Phi''}^{(p,p)}=e^{-2 y} (97.8-237. y+215. y^2-92.3 y^3+20.2 y^4-2.18 y^5+0.0995 y^6-0.000878 y^7+2.17\times10^{-6} y^8) $

$F_{\Delta}^{(p,p)}= e^{-2 y} (0.0106-0.0256 y+0.025 y^2-0.0127 y^3+0.00346 y^4-0.000482 y^5+0.0000272 y^6-8.52\times10^{-8} y^7+6.86\times10^{-11} y^8) $

$F_{M \Phi''}^{(p,p)}= e^{-2 y} (-534.+1610. y-1840. y^2+1020. y^3-299. y^4+46.7 y^5-3.64 y^6+0.116 y^7-0.000682 y^8+8.68\times10^{-7} y^9)$

$F_{\Sigma' \Delta}^{(p,p)}= e^{-2 y} (0.0000351-0.000836 y-0.00205 y^2+0.00521 y^3-0.00385 y^4+0.00126 y^5-0.000187 y^6+0.0000103 y^7-3.19\times10^{-8} y^8+2.53\times10^{-11} y^9)$

\vspace{5mm}

$F_{M}^{(p,n)}= e^{-2 y} (4050.-15900. y+24300. y^2-18900. y^3+8230. y^4-2060. y^5+295. y^6-22.7 y^7+0.816 y^8-0.00964 y^9+0.0000147 y^{10}) $

$F_{\Sigma''}^{(p,n)}= e^{-2 y} (-0.000187-0.00728 y+0.0282 y^2-0.0476 y^3+0.0423 y^4-0.0208 y^5+0.00568 y^6-0.000811 y^7+0.0000471 y^8-4.24\times10^{-8} y^9+7.22\times10^{-12} y^{10}) $

$F_{\Sigma'}^{(p,n)}= e^{-2 y} (-0.000373+0.00949 y+0.00719 y^2-0.085 y^3+0.148 y^4-0.102 y^5+0.0332 y^6-0.00509 y^7+0.000293 y^8-7.8\times10^{-7} y^9+5.2\times10^{-10} y^{10})$

$F_{\Phi''}^{(p,n)}= e^{-2 y} (161.-442. y+467. y^2-243. y^3+67.1 y^4-9.9 y^5+0.724 y^6-0.021 y^7+0.0000922 y^8) $

$F_{\Delta}^{(p,n)}= e^{-2 y} (-0.00238+0.00162 y-0.00174 y^2+0.00254 y^3-0.00131 y^4+0.000216 y^5+3.22\times10^{-6} y^6-2.34\times10^{-6} y^7+3.81\times10^{-9} y^8) $

$F_{M \Phi''}^{(p,n)}= e^{-2 y} (-879.+2930. y-3810. y^2+2480. y^3-890. y^4+178. y^5-19.4 y^6+1.05 y^7-0.0225 y^8+0.0000369 y^9) $

$F_{\Sigma' \Delta}^{(p,n)}= e^{-2 y} (-7.87\times10^{-6}+0.000174 y+0.000765 y^2+0.0000913 y^3+0.000412 y^4-0.000477 y^5+0.0000931 y^6+1.65\times10^{-6} y^7-8.87\times10^{-7} y^8+1.41\times10^{-9} y^9) $

\vspace{5mm}

$F_{M \Phi''}^{(n,p)}= e^{-2 y} (-741.+2470. y-3140. y^2+1990. y^3-685. y^4+131. y^5-13.5 y^6+0.65 y^7-0.0101 y^8+0.0000369 y^9) $

$F_{\Sigma' \Delta}^{(n,p)}=e^{-2 y} (-0.113+0.454 y-0.779 y^2+0.689 y^3-0.336 y^4+0.0916 y^5-0.0131 y^6+0.000771 y^7-2.09\times10^{-6} y^8+1.41\times10^{-9} y^9)$

\vspace{5mm}

$F_{M}^{(n,n)}= e^{-2 y} (5620.-23800. y+39800. y^2-34400. y^3+16900. y^4-4930. y^5+851. y^6-83.7 y^7+4.24 y^8-0.0885 y^9+0.000627 y^{10}) $

$F_{\Sigma''}^{(n,n)}= e^{-2 y} (0.599-2.91 y+6.61 y^2-8.58 y^3+6.77 y^4-3.15 y^5+0.852 y^6-0.125 y^7+0.00781 y^8-3.53\times10^{-6} y^9+4.02\times10^{-10} y^{10})$

$F_{\Sigma'}^{(n,n)}= e^{-2 y} (1.2-6.77 y+16.9 y^2-23.3 y^3+18.9 y^4-8.95 y^5+2.43 y^6-0.354 y^7+0.0219 y^8-0.0000499 y^9+2.89\times10^{-8} y^{10}) $

$F_{\Phi''}^{(n,n)}= e^{-2 y} (265.-813. y+982. y^2-602. y^3+204. y^4-38.7 y^5+4. y^6-0.203 y^7+0.00392 y^8) $

$F_{\Delta}^{(n,n)}= e^{-2 y} (0.000534+0.000562 y+0.00112 y^2+0.000351 y^3+0.000339 y^4-0.00016 y^5-7.07\times10^{-6} y^6+3.24\times10^{-6} y^7+2.12\times10^{-7} y^8)$

$F_{M \Phi''}^{(n,n)}= e^{-2 y} (-1220.+4460. y-6380. y^2+4670. y^3-1920. y^4+458. y^5-62.3 y^6+4.53 y^7-0.151 y^8+0.00157 y^9) $

$F_{\Sigma' \Delta}^{(n,n)}= e^{-2 y} (0.0253-0.0581 y+0.0627 y^2-0.0563 y^3+0.0697 y^4-0.034 y^5+0.00579 y^6+0.0000349 y^7-0.0000669 y^8+7.83\times10^{-8} y^9) $

\vspace{5mm}

\subsection*{$^{130}$Xe}

$F_{M}^{(p,p)}= e^{-2 y} (2920.-10500. y+14700. y^2-10200. y^3+3830. y^4-804. y^5+91.6 y^6-5.21 y^7+0.122 y^8-0.000468 y^9+4.94\times10^{-7} y^{10}) $

$F_{\Phi''}^{(p,p)}= e^{-2 y} (95.3-232. y+211. y^2-91.3 y^3+20.2 y^4-2.22 y^5+0.105 y^6-0.00107 y^7+3.09\times10^{-6} y^8)$

$F_{M \Phi''}^{(p,p)}= e^{-2 y} (-527.+1590. y-1820. y^2+1010. y^3-297. y^4+46.4 y^5-3.63 y^6+0.117 y^7-0.000798 y^8+1.23\times10^{-6} y^9)$

\vspace{5mm}

$F_{M}^{(p,n)}= e^{-2 y} (4100.-16100. y+24800. y^2-19400. y^3+8440. y^4-2120. y^5+304. y^6-23.4 y^7+0.841 y^8-0.0102 y^9+0.0000189 y^{10}) $

$F_{\Phi''}^{(p,n)}= e^{-2 y} (166.-457. y+484. y^2-252. y^3+70. y^4-10.4 y^5+0.774 y^6-0.0233 y^7+0.000118 y^8) $

$F_{M \Phi''}^{(p,n)}= e^{-2 y} (-917.+3070. y-3990. y^2+2590. y^3-926. y^4+185. y^5-20.1 y^6+1.09 y^7-0.0235 y^8+0.0000472 y^9)$

\vspace{5mm}

$F_{M \Phi''}^{(n,p)}= e^{-2 y} (-742.+2480. y-3170. y^2+2020. y^3-703. y^4+136. y^5-14.2 y^6+0.7 y^7-0.0114 y^8+0.0000472 y^9) $

\vspace{5mm}

$F_{M}^{(n,n)}= e^{-2 y} (5770.-24600. y+41500. y^2-36000. y^3+17900. y^4-5250. y^5+917. y^6-91.3 y^7+4.69 y^8-0.0997 y^9+0.000721 y^{10}) $

$F_{\Phi''}^{(n,n)}= e^{-2 y} (288.-889. y+1080. y^2-659. y^3+223. y^4-42.3 y^5+4.4 y^6-0.227 y^7+0.00451 y^8) $

$F_{M \Phi''}^{(n,n)}= e^{-2 y} (-1290.+4740. y-6820. y^2+5010. y^3-2070. y^4+495. y^5-67.8 y^6+5. y^7-0.17 y^8+0.0018 y^9) $

\vspace{5mm}

\subsection*{$^{131}$Xe}

$F_{M}^{(p,p)}= e^{-2 y} (2920.-10500. y+14600. y^2-10100. y^3+3780. y^4-782. y^5+87.6 y^6-4.86 y^7+0.112 y^8-0.000509 y^9+6.51\times10^{-7} y^{10})$

$F_{\Sigma''}^{(p,p)}= e^{-2 y} (0.000114-0.000144 y+0.0000326 y^2+0.0000504 y^3+4.32\times10^{-7} y^4-0.0000235 y^5+9.65\times10^{-6} y^6-1.45\times10^{-6} y^7+7.7\times10^{-8} y^8-2.18\times10^{-10} y^9+2.61\times10^{-13} y^{10}) $

$F_{\Sigma'}^{(p,p)}= e^{-2 y} (0.000227-0.00165 y+0.00383 y^2-0.0031 y^3+0.00103 y^4-0.000128 y^5+3.37\times10^{-6} y^6+3.47\times10^{-7} y^7-1.3\times10^{-8} y^8-3.08\times10^{-10} y^9+1.32\times10^{-11} y^{10})$

$F_{\Phi''}^{(p,p)}= e^{-2 y} (86.5-211. y+192. y^2-82.8 y^3+18.2 y^4-2. y^5+0.0959 y^6-0.00116 y^7+4.07\times10^{-6} y^8)$

$F_{\tilde{\Phi}'}^{(p,p)}= e^{-2 y} (0.000333+0.0000342 y+0.0000816 y^2-0.0000194 y^3+3.28\times10^{-6} y^4-2.87\times10^{-6} y^5+3.66\times10^{-7} y^6+1.43\times10^{-8} y^7+1.24\times10^{-10} y^8)$

$F_{\Delta}^{(p,p)}=  e^{-2 y} (0.00479-0.0116 y+0.0111 y^2-0.00532 y^3+0.00134 y^4-0.000172 y^5+9.13\times10^{-6} y^6-5.71\times10^{-8} y^7+9.55\times10^{-11} y^8)$

$F_{M \Phi''}^{(p,p)}= e^{-2 y} (-502.+1520. y-1730. y^2+958. y^3-279. y^4+43.3 y^5-3.35 y^6+0.108 y^7-0.000867 y^8+1.63\times10^{-6} y^9)$

$F_{\Sigma' \Delta}^{(p,p)}= e^{-2 y} (0.00104-0.00505 y+0.00695 y^2-0.00414 y^3+0.00117 y^4-0.000147 y^5+5.61\times10^{-6} y^6+2.17\times10^{-7} y^7-1.1\times10^{-8} y^8+3.55\times10^{-11} y^9) $

\vspace{5mm}

$F_{M}^{(p,n)}= e^{-2 y} (4160.-16400. y+25300. y^2-19800. y^3+8600. y^4-2150. y^5+307. y^6-23.6 y^7+0.859 y^8-0.0112 y^9+0.0000247 y^{10}) $

$F_{\Sigma''}^{(p,n)}= e^{-2 y} (0.00377+0.00113 y-0.00909 y^2+0.0104 y^3-0.000659 y^4-0.00372 y^5+0.00188 y^6-0.000355 y^7+0.0000232 y^8-4.11\times10^{-8} y^9+5.39\times10^{-12} y^{10}) $

$F_{\Sigma'}^{(p,n)}=e^{-2 y} (0.00753-0.0607 y+0.171 y^2-0.208 y^3+0.125 y^4-0.0381 y^5+0.00537 y^6-0.000266 y^7-4.9\times10^{-6} y^8+3.89\times10^{-7} y^9+2.63\times10^{-10} y^{10})$

$F_{\Phi''}^{(p,n)}=e^{-2 y} (171.-476. y+505. y^2-263. y^3+72.7 y^4-10.8 y^5+0.817 y^6-0.0257 y^7+0.000154 y^8) $

$F_{\tilde{\Phi}'}^{(p,n)}= e^{-2 y} (0.00234-0.00346 y+0.0015 y^2-0.000686 y^3+0.000286 y^4-0.0000762 y^5+8.16\times10^{-6} y^6+3.53\times10^{-9} y^7-2.48\times10^{-9} y^8) $

$F_{\Delta}^{(p,n)}=  e^{-2 y} (0.0588-0.143 y+0.156 y^2-0.0918 y^3+0.0293 y^4-0.00466 y^5+0.000298 y^6-1.53\times10^{-6} y^7+1.9\times10^{-9} y^8) $

$F_{M \Phi''}^{(p,n)}= e^{-2 y} (-994.+3350. y-4350. y^2+2830. y^3-1000. y^4+199. y^5-21.6 y^6+1.17 y^7-0.0256 y^8+0.0000617 y^9) $

$F_{\Sigma' \Delta}^{(p,n)}=e^{-2 y} (0.0128-0.0621 y+0.0897 y^2-0.0663 y^3+0.025 y^4-0.00421 y^5+0.000227 y^6+5.11\times10^{-6} y^7-3.68\times10^{-7} y^8+7.08\times10^{-10} y^9) $

\vspace{5mm}

$F_{M \Phi''}^{(n,p)}= e^{-2 y} (-716.+2410. y-3090. y^2+1980. y^3-687. y^4+133. y^5-13.9 y^6+0.705 y^7-0.0127 y^8+0.0000617 y^9)$

$F_{\Sigma' \Delta}^{(n,p)}= e^{-2 y} (0.0346-0.195 y+0.367 y^2-0.335 y^3+0.164 y^4-0.0431 y^5+0.00575 y^6-0.000309 y^7+8.3\times10^{-7} y^8+7.08\times10^{-10} y^9) $

\vspace{5mm}

$F_{M}^{(n,n)}= e^{-2 y} (5920.-25400. y+43200. y^2-37700. y^3+18800. y^4-5540. y^5+972. y^6-97.9 y^7+5.17 y^8-0.118 y^9+0.000937 y^{10}) $

$F_{\Sigma''}^{(n,n)}= e^{-2 y} (0.125+0.234 y-0.309 y^2+0.0503 y^3+0.766 y^4-0.896 y^5+0.416 y^6-0.0883 y^7+0.00733 y^8-1.65\times10^{-6} y^9+1.12\times10^{-10} y^{10}) $

$F_{\Sigma'}^{(n,n)}= e^{-2 y} (0.25-2.22 y+7.34 y^2-11.9 y^3+10.8 y^4-5.68 y^5+1.67 y^6-0.253 y^7+0.0154 y^8+0.0000159 y^9+5.26\times10^{-9} y^{10}) $

$F_{\Phi''}^{(n,n)}= e^{-2 y} (339.-1060. y+1290. y^2-788. y^3+266. y^4-50.7 y^5+5.34 y^6-0.283 y^7+0.00586 y^8) $

$F_{\tilde{\Phi}'}^{(n,n)}= e^{-2 y} (0.0164-0.0503 y+0.0582 y^2-0.0335 y^3+0.0111 y^4-0.00211 y^5+0.000209 y^6-5.88\times10^{-6} y^7+4.95\times10^{-8} y^8) $

$F_{\Delta}^{(n,n)}= e^{-2 y} (0.722-1.76 y+2.15 y^2-1.48 y^3+0.596 y^4-0.123 y^5+0.0103 y^6-0.0000387 y^7+3.8\times10^{-8} y^8) $

$F_{M \Phi''}^{(n,n)}= e^{-2 y} (-1420.+5250. y-7610. y^2+5610. y^3-2320. y^4+557. y^5-76.8 y^6+5.77 y^7-0.204 y^8+0.00234 y^9) $

$F_{\Sigma' \Delta}^{(n,n)}= e^{-2 y} (0.425-2.4 y+4.66 y^2-4.84 y^3+2.9 y^4-0.979 y^5+0.17 y^6-0.0118 y^7+0.0000139 y^8+1.41\times10^{-8} y^9) $

\vspace{5mm}

\subsection*{$^{132}$Xe}

$F_{M}^{(p,p)}= e^{-2 y} (2920.-10500. y+14600. y^2-10100. y^3+3770. y^4-781. y^5+87.2 y^6-4.83 y^7+0.111 y^8-0.000539 y^9+7.37\times10^{-7} y^{10})$

$F_{\Phi''}^{(p,p)}= e^{-2 y} (87.4-213. y+195. y^2-84.5 y^3+18.8 y^4-2.09 y^5+0.102 y^6-0.00126 y^7+4.61\times10^{-6} y^8) $

$F_{M \Phi''}^{(p,p)}= e^{-2 y} (-505.+1530. y-1740. y^2+967. y^3-283. y^4+44. y^5-3.43 y^6+0.112 y^7-0.000926 y^8+1.84\times10^{-6} y^9)$

\vspace{5mm}

$F_{M}^{(p,n)}=e^{-2 y} (4210.-16700. y+25800. y^2-20200. y^3+8830. y^4-2220. y^5+319. y^6-24.6 y^7+0.896 y^8-0.0117 y^9+0.0000274 y^{10}) $

$F_{\Phi''}^{(p,n)}= e^{-2 y} (175.-487. y+518. y^2-269. y^3+74.8 y^4-11.2 y^5+0.855 y^6-0.0275 y^7+0.000171 y^8) $

$F_{M \Phi''}^{(p,n)}= e^{-2 y} (-1010.+3400. y-4430. y^2+2870. y^3-1020. y^4+202. y^5-22. y^6+1.2 y^7-0.0266 y^8+0.0000684 y^9) $

\vspace{5mm}

$F_{M \Phi''}^{(n,p)}= e^{-2 y} (-729.+2460. y-3180. y^2+2040. y^3-715. y^4+140. y^5-14.8 y^6+0.758 y^7-0.0137 y^8+0.0000684 y^9) $

\vspace{5mm}

$F_{M}^{(n,n)}= e^{-2 y} (6080.-26200. y+44800. y^2-39400. y^3+19800. y^4-5890. y^5+1040. y^6-106. y^7+5.65 y^8-0.129 y^9+0.00102 y^{10}) $

$F_{\Phi''}^{(n,n)}= e^{-2 y} (349.-1090. y+1330. y^2-814. y^3+274. y^4-52.2 y^5+5.53 y^6-0.299 y^7+0.00635 y^8) $

$F_{M \Phi''}^{(n,n)}= e^{-2 y} (-1460.+5430. y-7900. y^2+5840. y^3-2420. y^4+584. y^5-81.1 y^6+6.15 y^7-0.221 y^8+0.00254 y^9) $

\vspace{5mm}

\subsection*{$^{134}$Xe}

$F_{M}^{(p,p)}=e^{-2 y} (2920.-10500. y+14600. y^2-10100. y^3+3730. y^4-767. y^5+84.5 y^6-4.59 y^7+0.104 y^8-0.000515 y^9+7.25\times10^{-7} y^{10}) $

$F_{\Phi''}^{(p,p)}= e^{-2 y} (82.-200. y+183. y^2-79.5 y^3+17.7 y^4-1.97 y^5+0.0965 y^6-0.00122 y^7+4.53\times10^{-6} y^8) $

$F_{M \Phi''}^{(p,p)}= e^{-2 y} (-489.+1480. y-1690. y^2+935. y^3-273. y^4+42.2 y^5-3.26 y^6+0.105 y^7-0.000887 y^8+1.81\times10^{-6} y^9)$

\vspace{5mm}

$F_{M}^{(p,n)}= e^{-2 y} (4320.-17200. y+26800. y^2-21100. y^3+9250. y^4-2330. y^5+336. y^6-25.9 y^7+0.943 y^8-0.0125 y^9+0.0000299 y^{10}) $

$F_{\Phi''}^{(p,n)}= e^{-2 y} (180.-504. y+537. y^2-280. y^3+77.7 y^4-11.7 y^5+0.9 y^6-0.0295 y^7+0.000187 y^8) $

$F_{M \Phi''}^{(p,n)}= e^{-2 y} (-1070.+3630. y-4740. y^2+3070. y^3-1090. y^4+215. y^5-23.3 y^6+1.27 y^7-0.0283 y^8+0.0000748 y^9) $

\vspace{5mm}

$F_{M \Phi''}^{(n,p)}= e^{-2 y} (-724.+2460. y-3200. y^2+2070. y^3-731. y^4+144. y^5-15.5 y^6+0.799 y^7-0.0147 y^8+0.0000748 y^9) $

\vspace{5mm}

$F_{M}^{(n,n)}= e^{-2 y} (6400.-27900. y+48100. y^2-42800. y^3+21800. y^4-6570. y^5+1180. y^6-122. y^7+6.59 y^8-0.153 y^9+0.00124 y^{10}) $

$F_{\Phi''}^{(n,n)}=e^{-2 y} (394.-1250. y+1530. y^2-934. y^3+314. y^4-60.1 y^5+6.43 y^6-0.353 y^7+0.00773 y^8) $

$F_{M \Phi''}^{(n,n)}= e^{-2 y} (-1590.+5980. y-8770. y^2+6530. y^3-2730. y^4+663. y^5-93.1 y^6+7.16 y^7-0.262 y^8+0.00309 y^9) $

\vspace{5mm}

\subsection*{$^{136}$Xe}

$F_{M}^{(p,p)}= e^{-2 y} (2920.-10500. y+14600. y^2-10100. y^3+3740. y^4-768. y^5+84.7 y^6-4.61 y^7+0.105 y^8-0.000533 y^9+7.76\times10^{-7} y^{10}) $

$F_{\Phi''}^{(p,p)}= e^{-2 y} (83.7-204. y+187. y^2-81.7 y^3+18.3 y^4-2.06 y^5+0.102 y^6-0.0013 y^7+4.85\times10^{-6} y^8) $

$F_{M \Phi''}^{(p,p)}= e^{-2 y} (-494.+1500. y-1710. y^2+947. y^3-277. y^4+43.1 y^5-3.35 y^6+0.109 y^7-0.000926 y^8+1.94\times10^{-6} y^9)$

\vspace{5mm}

$F_{M}^{(p,n)}= e^{-2 y} (4430.-17700. y+27700. y^2-22000. y^3+9710. y^4-2470. y^5+358. y^6-27.8 y^7+1.01 y^8-0.0133 y^9+0.0000325 y^{10}) $

$F_{\Phi''}^{(p,n)}= e^{-2 y} (183.-516. y+551. y^2-287. y^3+79.8 y^4-12.1 y^5+0.943 y^6-0.0317 y^7+0.000203 y^8) $

$F_{M \Phi''}^{(p,n)}= e^{-2 y} (-1080.+3680. y-4810. y^2+3110. y^3-1100. y^4+217. y^5-23.7 y^6+1.32 y^7-0.0298 y^8+0.0000813 y^9) $

\vspace{5mm}

$F_{M \Phi''}^{(n,p)}= e^{-2 y} (-750.+2560. y-3350. y^2+2190. y^3-781. y^4+156. y^5-16.9 y^6+0.879 y^7-0.016 y^8+0.0000813 y^9) $

\vspace{5mm}

$F_{M}^{(n,n)}= e^{-2 y} (6720.-29500. y+51400. y^2-46300. y^3+23900. y^4-7300. y^5+1330. y^6-139. y^7+7.57 y^8-0.174 y^9+0.00136 y^{10}) $

$F_{\Phi''}^{(n,n)}= e^{-2 y} (400.-1280. y+1570. y^2-959. y^3+322. y^4-61.6 y^5+6.66 y^6-0.375 y^7+0.00853 y^8) $

$F_{M \Phi''}^{(n,n)}= e^{-2 y} (-1640.+6220. y-9200. y^2+6900. y^3-2900. y^4+710. y^5-101. y^6+7.85 y^7-0.292 y^8+0.00341 y^9) $

\vspace{10mm}

\section{Occupation Number Tables}\label{occupation numbers}

The occupation numbers $n^N_{nlj}$ of each valence single-nucleon orbital are given in order of increasing orbit, with $N=\{p,n\}$ denoting protons and neutrons respectively, and $n$ begins at $1$.  ``Max Occ" refers to the maximum occupation number allowed for each orbital.\\

\twocolumngrid

\subsection{$^{19}$F}

\begin{table}[H]
\centering
\begin{tabular}{lccc|cc}
\hline \hline 
$n$  & $l$ & $j$ & {\footnotesize Max Occ} & $n^p_{nlj}$ & $n^n_{nlj}$ \\
\hline 
1   & 2  &  5/2 & 6 & 0.47 & 1.08 \\
2  &  0 &  1/2 &  2 & 0.41 & 0.77 \\
1   & 2 &  3/2 &  4 & 0.12 & 0.16 \\
\hline \hline
\end{tabular}
\end{table}

\subsection{$^{23}$Na}

\begin{table}[H]
\centering
\begin{tabular}{lccc|cc}
\hline \hline 
$n$  & $l$ & $j$ & {\footnotesize Max Occ} & $n^p_{nlj}$ & $n^n_{nlj}$ \\
\hline 
1   & 2  &  5/2 & 6 & 2.30 & 3.11 \\
2  &  0 &  1/2 &  2 & 0.45 & 0.44 \\
1   & 2 &  3/2 &  4 & 0.25 & 0.46 \\
\hline \hline
\end{tabular}
\end{table}

\subsection{$^{28, 29, 30}$Si}

\subsection*{$^{28}$Si}

\begin{table}[H]
\centering
\begin{tabular}{lccc|cc}
\hline \hline 
$n$  & $l$ & $j$ & {\footnotesize Max Occ} & $n^p_{nlj}$ & $n^n_{nlj}$ \\
\hline 
1   & 2  &  5/2 & 6 & 4.66 & 4.66 \\
2  &  0 &  1/2 &  2 & 0.71 & 0.71 \\
1   & 2 &  3/2 &  4 & 0.63  & 0.63 \\
\hline \hline
\end{tabular}
\end{table}

\subsection*{$^{29}$Si}

\begin{table}[H]
\centering
\begin{tabular}{lccc|cc}
\hline \hline 
$n$  & $l$ & $j$ & {\footnotesize Max Occ} & $n^p_{nlj}$ & $n^n_{nlj}$ \\
\hline 
1   & 2  &  5/2 & 6 & 4.94 & 5.28 \\
2  &  0 &  1/2 &  2 & 0.59 & 1.09 \\
1   & 2 &  3/2 &  4 & 0.47 & 0.63 \\
\hline \hline
\end{tabular}
\end{table}

\subsection*{$^{30}$Si}

\begin{table}[H]
\centering
\begin{tabular}{lccc|cc}
\hline \hline 
$n$  & $l$ & $j$ & {\footnotesize Max Occ} & $n^p_{nlj}$ & $n^n_{nlj}$ \\
\hline 
1   & 2  &  5/2 & 6 & 5.09 & 5.49 \\
2  &  0 &  1/2 &  2 & 0.53 & 1.37 \\
1   & 2 &  3/2 &  4 & 0.37  & 1.15 \\
\hline \hline
\end{tabular}
\end{table}

\onecolumngrid

\vspace{5mm}

\subsection{$^{40}$Ar}

\begin{table}[H]
\centering
\begin{tabular}{lccc|ccc|ccc|ccc|cc}
\hline \hline 
 &  &  & & \multicolumn{2}{c}{SDPF-NR} & & \multicolumn{2}{c}{SDPF-U} & & \multicolumn{2}{c}{EPQQM} & & \multicolumn{2}{c}{SDPF-MU} \\
\cline{5-6}
\cline{8-9}
\cline{11-12}
\cline{14-15}
$n$  & $l$ & $j$ & {\footnotesize Max Occ} & $n^p_{nlj}$ & $n^n_{nlj}$ & & $n^p_{nlj}$ & $n^n_{nlj}$ & & $n^p_{nlj}$ & $n^n_{nlj}$ & & $n^p_{nlj}$ & $n^n_{nlj}$\\
\hline 
1   & 2  &  5/2 & 6 & 5.92 & 6.00 & & 5.91 & 6.00 & & 5.92 & 6.00 & & 5.92 & 6.00 \\
2  &  0 &  1/2 &  2 & 1.83 & 2.00  & & 1.82 & 2.00 & & 1.85 & 2.00 & & 1.78 & 2.00\\
1   & 2 &  3/2 &  4 & 2.25  & 4.00  & & 2.27 & 4.00 & & 2.23 & 4.00 & & 2.30 & 4.00\\
1   & 3 &  7/2 &  8 & 0 & 1.79  & & 0 & 1.83 & & 0 & 1.47 & & 0 & 1.91 \\
2   & 1  &  3/2 &  4 & 0 & 0.11  & & 0 & 0.08 & & 0 & 0.40 & & 0 & 0.06\\
1  &  3 &  5/2 &  6 & 0 & 0.07  & & 0 & 0.07  & & 0 & 0.06 & & 0 & 0.02 \\
2   & 1 &  1/2 &  2 &  0 & 0.02  & & 0 & 0.02 & & 0 & 0.07 & & 0 & 0.01 \\
\hline \hline
\end{tabular}
\end{table}

\newpage

\twocolumngrid

\subsection{$^{70,72,73,74,76}$Ge}

\subsection*{$^{70}$Ge}

\begin{table}[H]
\centering
\begin{tabular}{lccc|ccc|cc}
\hline \hline 
 &  &  & & \multicolumn{2}{c}{JUN45} & & \multicolumn{2}{c}{jj44b}  \\
\cline{5-6}
\cline{8-9}
$n$  & $l$ & $j$ & {\footnotesize Max Occ} & $n^p_{nlj}$ & $n^n_{nlj}$ & & $n^p_{nlj}$ & $n^n_{nlj}$\\
\hline 
2   & 1  &  3/2 &  4 & 2.28 & 3.23 & & 1.58 & 3.04 \\
1  &  3 &  5/2 &  6 & 0.96 & 4.00 & & 1.66 &3.41 \\
2   & 1 &  1/2 &  2 &  0.49 & 1.07 & & 0.40 &0.93 \\
1  &  4 &  9/2 &  10 & 0.27 & 1.70 & & 0.35 & 2.62 \\
\hline \hline
\end{tabular}
\end{table}

\subsection*{$^{72}$Ge}

\begin{table}[H]
\centering
\begin{tabular}{lccc|ccc|cc}
\hline \hline 
 &  &  & & \multicolumn{2}{c}{JUN45} & & \multicolumn{2}{c}{jj44b}  \\
\cline{5-6}
\cline{8-9}
$n$  & $l$ & $j$ & {\footnotesize Max Occ} & $n^p_{nlj}$ & $n^n_{nlj}$ & & $n^p_{nlj}$ & $n^n_{nlj}$\\
\hline 
2   & 1  &  3/2 &  4 & 2.09 & 3.35 & & 1.53 & 3.22 \\
1  &  3 &  5/2 &  6 & 1.27 & 4.49 & & 1.80 & 3.95 \\
2   & 1 &  1/2 &  2 &  0.36 & 1.25 & & 0.37 & 1.03 \\
1  &  4 &  9/2 &  10 & 0.27 & 2.90 & & 0.30 & 3.79 \\
\hline \hline
\end{tabular}
\end{table}

\subsection*{$^{73}$Ge}

\begin{table}[H]
\centering
\begin{tabular}{lccc|ccc|cc}
\hline \hline 
 &  &  & & \multicolumn{2}{c}{JUN45} & & \multicolumn{2}{c}{jj44b}  \\
\cline{5-6}
\cline{8-9}
$n$  & $l$ & $j$ & {\footnotesize Max Occ} & $n^p_{nlj}$ & $n^n_{nlj}$ & & $n^p_{nlj}$ & $n^n_{nlj}$\\
\hline 
2   & 1  &  3/2 &  4 & 1.95 & 3.46 & & 1.48 & 3.31 \\
1  &  3 &  5/2 &  6 & 1.45 & 4.70 & & 1.89 & 4.13 \\
2   & 1 &  1/2 &  2 &  0.35 & 1.30 & & 0.39 & 1.04 \\
1  &  4 &  9/2 &  10 & 0.25 & 3.52 & & 0.25 & 4.51 \\
\hline \hline
\end{tabular}
\end{table}

\subsection*{$^{74}$Ge}

\begin{table}[H]
\centering
\begin{tabular}{lccc|ccc|cc}
\hline \hline 
 &  &  & & \multicolumn{2}{c}{JUN45} & & \multicolumn{2}{c}{jj44b}  \\
\cline{5-6}
\cline{8-9}
$n$  & $l$ & $j$ & {\footnotesize Max Occ} & $n^p_{nlj}$ & $n^n_{nlj}$ & & $n^p_{nlj}$ & $n^n_{nlj}$\\
\hline 
2   & 1  &  3/2 &  4 & 1.74 & 3.44 & & 1.49 & 3.40 \\
1  &  3 &  5/2 &  6 & 1.70 & 4.76 & & 1.90 & 4.34 \\
2   & 1 &  1/2 &  2 & 0.31  & 1.32 & & 0.35 & 1.23 \\
1  &  4 &  9/2 &  10 & 0.25 & 4.48 & & 0.26 & 5.01 \\
\hline \hline
\end{tabular}
\end{table}

\subsection*{$^{76}$Ge}

\begin{table}[H]
\centering
\begin{tabular}{lccc|ccc|cc}
\hline \hline 
 &  &  & & \multicolumn{2}{c}{JUN45} & & \multicolumn{2}{c}{jj44b}  \\
\cline{5-6}
\cline{8-9}
$n$  & $l$ & $j$ & {\footnotesize Max Occ} & $n^p_{nlj}$ & $n^n_{nlj}$ & & $n^p_{nlj}$ & $n^n_{nlj}$\\
\hline 
2   & 1  &  3/2 &  4 & 1.59 & 3.54 & & 1.44 & 3.62 \\
1  &  3 &  5/2 &  6 & 1.91 & 5.18 & & 2.00 & 4.74 \\
2   & 1 &  1/2 &  2 & 0.27  & 1.43 & & 0.34 & 1.48 \\
1  &  4 &  9/2 &  10 & 0.24 & 5.85 & & 0.22 & 6.16 \\
\hline \hline
\end{tabular}
\end{table}

\subsection{$^{127}$I}

\begin{table}[H]
\centering
\begin{tabular}{lccc|ccc|cc}
\hline \hline 
 &  &  & & \multicolumn{2}{c}{SN100PN} & & \multicolumn{2}{c}{GCN5082}  \\
\cline{5-6}
\cline{8-9}
$n$  & $l$ & $j$ & {\footnotesize Max Occ} & $n^p_{nlj}$ & $n^n_{nlj}$ & & $n^p_{nlj}$ & $n^n_{nlj}$\\
\hline 
1   & 4  &  7/2 & 8 & 2.03 & 7.35 & & 1.43 & 7.54 \\
2  &  2 &  5/2 &  6 & 0.62 & 5.99 & & 0.98 &  5.99 \\
2   & 2 &  3/2 &  4 &  0.22 & 1.94 & & 0.39 &  2.10 \\
3  &  0 &  1/2 & 2 & 0.08 & 1.08 & & 0.12 &  1.52 \\
1  &  5 &  11/2 &  12 & 0.05 & 7.62 & & 0.10 &  6.84 \\
\hline \hline
\end{tabular}
\end{table}

\subsection{$^{128, 129, 130, 131, 132, 134, 136}$Xe}

\subsection*{$^{128}$Xe}

\begin{table}[H]
\centering
\begin{tabular}{lccc|cc}
\hline \hline 
$n$  & $l$ & $j$ & {\footnotesize Max Occ} & $n^p_{nlj}$ & $n^n_{nlj}$ \\
\hline 
1   & 4  &  7/2 & 8 & 2.49 & 8.00 \\
2  &  2 &  5/2 &  6 & 0.80 & 6.00 \\
2   & 2 &  3/2 &  4 & 0.34 & 1.55 \\
3  &  0 &  1/2 & 2 & 0.18 & 0.75 \\
1  &  5 &  11/2 &  12 & 0.19 & 7.70 \\
\hline \hline
\end{tabular}
\end{table}

\subsection*{$^{129}$Xe}

\begin{table}[H]
\centering
\begin{tabular}{lccc|cc}
\hline \hline 
$n$  & $l$ & $j$ & {\footnotesize Max Occ} & $n^p_{nlj}$ & $n^n_{nlj}$ \\
\hline 
1   & 4  &  7/2 & 8 & 2.41 & 7.53 \\
2  &  2 &  5/2 &  6 & 0.88 & 6.00 \\
2   & 2 &  3/2 &  4 & 0.33 & 2.06 \\
3  &  0 &  1/2 & 2 & 0.19 & 1.26 \\
1  &  5 &  11/2 &  12 & 0.19 & 8.14 \\
\hline \hline
\end{tabular}
\end{table}

\subsection*{$^{130}$Xe}

\begin{table}[H]
\centering
\begin{tabular}{lccc|cc}
\hline \hline 
$n$  & $l$ & $j$ & {\footnotesize Max Occ} & $n^p_{nlj}$ & $n^n_{nlj}$ \\
\hline 
1   & 4  &  7/2 & 8 & 2.55 & 7.55 \\
2  &  2 &  5/2 &  6 & 0.83 & 6.00 \\
2   & 2 &  3/2 &  4 & 0.26 & 2.31 \\
3  &  0 &  1/2 & 2 & 0.14 & 1.41 \\
1  &  5 &  11/2 &  12 & 0.23 & 8.73 \\
\hline \hline
\end{tabular}
\end{table}

\subsection*{$^{131}$Xe}

\begin{table}[H]
\centering
\begin{tabular}{lccc|cc}
\hline \hline 
$n$  & $l$ & $j$ & {\footnotesize Max Occ} & $n^p_{nlj}$ & $n^n_{nlj}$ \\
\hline 
1   & 4  &  7/2 & 8 & 2.80 & 7.71 \\
2  &  2 &  5/2 &  6 & 0.62 & 5.60 \\
2   & 2 &  3/2 &  4 &  0.22 & 2.39 \\
3  &  0 &  1/2 & 2 & 0.09  & 1.33 \\
1  &  5 &  11/2 &  12 & 0.26 & 9.95 \\
\hline \hline
\end{tabular}
\end{table}

\newpage

\onecolumngrid

\subsection*{$^{132}$Xe}

\begin{table}[H]
\centering
\begin{tabular}{lccc|cc}
\hline \hline 
$n$  & $l$ & $j$ & {\footnotesize Max Occ} & $n^p_{nlj}$ & $n^n_{nlj}$ \\
\hline 
1   & 4  &  7/2 & 8 & 2.82 & 7.72 \\
2  &  2 &  5/2 &  6 & 0.64 & 5.61 \\
2   & 2 &  3/2 &  4 &  0.18 & 2.79 \\
3  &  0 &  1/2 & 2 &  0.08 & 1.52 \\
1  &  5 &  11/2 &  12 & 0.28 & 10.36 \\
\hline \hline
\end{tabular}
\end{table}

\hfill

\subsection*{$^{134}$Xe}

\begin{table}[H]
\centering
\begin{tabular}{lccc|cc}
\hline \hline 
$n$  & $l$ & $j$ & {\footnotesize Max Occ} & $n^p_{nlj}$ & $n^n_{nlj}$ \\
\hline 
1   & 4  &  7/2 & 8 & 2.99 & 7.88 \\
2  &  2 &  5/2 &  6 & 0.56 & 5.79 \\
2   & 2 &  3/2 &  4 & 0.13 & 3.24 \\
3  &  0 &  1/2 & 2 &  0.05 & 1.67 \\
1  &  5 &  11/2 &  12 & 0.28 & 11.42 \\
\hline \hline
\end{tabular}
\end{table}

\subsection*{$^{136}$Xe}

\begin{table}[H]
\centering
\begin{tabular}{lccc|cc}
\hline \hline 
$n$  & $l$ & $j$ & {\footnotesize Max Occ} & $n^p_{nlj}$ & $n^n_{nlj}$ \\
\hline 
1   & 4  &  7/2 & 8 & 2.97 & 8.00 \\
2  &  2 &  5/2 &  6 & 0.59 & 6.00 \\
2   & 2 &  3/2 &  4 &  0.11 & 4.00 \\
3  &  0 &  1/2 & 2 &  0.04 & 2.00 \\
1  &  5 &  11/2 &  12 & 0.29 & 12.00 \\
\hline \hline
\end{tabular}
\end{table}

\end{document}